\documentclass[10pt,journal]{IEEEtran}

\IEEEoverridecommandlockouts

\usepackage{graphicx,subfigure}
\usepackage{amsmath,amsthm,amsfonts,amssymb}
\usepackage{cite}
\usepackage{bm}
\usepackage{bbm}
\usepackage{url}
\usepackage{array}
\usepackage{color,soul}
\usepackage{multirow}
\usepackage{booktabs}
\usepackage[table,xcdraw]{xcolor}
\usepackage{enumerate}
\usepackage{enumitem}
\usepackage{makecell}

\usepackage[english]{babel}
\usepackage{algorithm}
\usepackage[noend]{algpseudocode}

\theoremstyle{plain}
\newtheorem{thm}{Theorem}[section]

\newtheorem{prop}[thm]{Proposition}

\newtheorem{rem}{Remark}
\newtheorem{defi}{Definition}[section]

\newtheorem{sty2}{Theorem}
\newtheorem{example}[sty2]{Example}
\newtheorem{sty3}{Theorem}
\newtheorem{problem}[sty3]{Problem}

\theoremstyle{plain}
\newtheorem{thmApp}{Theorem}[section]
\newtheorem{lemApp}[thmApp]{Lemma}
\newtheorem{propApp}[thmApp]{Proposition}
\newtheorem{corApp}[thmApp]{Corollary}
\newtheorem{defiApp}{Definition}[section]

\newenvironment{NewProof}{{\noindent\it Proof.}}{\hfill $\blacksquare$\par}
\newcommand{\RR}{I\!\!R}

\makeatletter
\newcommand{\algrule}[1][.2pt]{\par\vskip.5\baselineskip\hrule height #1\par\vskip.5\baselineskip}
\makeatother

\makeatletter
\newcommand{\ssymbol}[1]{^{\@fnsymbol{#1}}}
\newcommand{\algmargin}{\the\ALG@thistlm}
\makeatother

\DeclareMathOperator*{\argmax}{arg\,max}
\DeclareMathOperator*{\argmin}{arg\,min}

\usepackage{mathtools}

\usepackage[english]{babel}
\usepackage{algorithm}
\usepackage[noend]{algpseudocode}

\allowdisplaybreaks[4]

\begin{document}
\title{A Theory of Semantic Communication}

\author{
Yulin~Shao,~\IEEEmembership{Member, IEEE},
Qi~Cao,
Deniz~G\"und\"uz,~\IEEEmembership{Fellow,~IEEE}
\thanks{Y. Shao is with the Department of Electrical and Electronic Engineering, University of Hong Kong, Hong Kong S.A.R., China. (e-mail: ylshao@hku.hk).}
\thanks{Qi Cao is with Xidian-Guangzhou Research Institute, Xidian University, Guangzhou, China (e-mail: caoqi@xidian.edu.cn).
}
\thanks{D. G\"und\"uz is with the Department of Electrical and Electronic Engineering, Imperial College London, London SW7 2AZ, U.K. (e-mail: d.gunduz@imperial.ac.uk).}
}

\IEEEtitleabstractindextext{%
\begin{abstract}
Semantic communication is an emerging research area that has gained a wide range of attention recently. Despite this growing interest, there remains a notable absence of a comprehensive and widely-accepted framework for characterizing semantic communication.
This paper introduces a new conceptualization of semantic communication and formulates two fundamental problems, which we term \textit{language exploitation} and \textit{language design}.
Our contention is that the challenge of language design can be effectively situated within the broader framework of joint source-channel coding theory, underpinned by a comprehensive end-to-end distortion metric.
To tackle the language exploitation problem, we put forth three approaches: semantic encoding, semantic decoding, and a synergistic combination of both in the form of combined semantic encoding and decoding. 
Furthermore, we establish the semantic distortion-cost region as a critical framework for assessing the language exploitation problem. For each of the three proposed approaches, the achievable distortion-cost region is characterized. Overall, this paper aims to shed light on the intricate dynamics of semantic communication, paving the way for a deeper understanding of this evolving field.
\end{abstract}

}

\maketitle
\IEEEdisplaynontitleabstractindextext
\IEEEpeerreviewmaketitle

\section{Introduction}\label{sec:intro}
\subsection{What is semantic communication?}
One year after the birth of Shannon's information theory \cite{shannon1948}, Weaver classified a broad communication problem into three levels \cite{weaver1953}:
\begin{itemize}
\item \textbf{Level A} (The technical problem): How accurately can the symbols of communication be transmitted?
\item \textbf{Level B} (The semantic problem): How precisely do the transmitted symbols convey the desired meaning?
\item \textbf{Level C} (The effectiveness problem): How effectively does the received meaning affect conduct in the desired way?
\end{itemize}

\begin{figure}[t]
  \centering
  \includegraphics[width=0.75\linewidth]{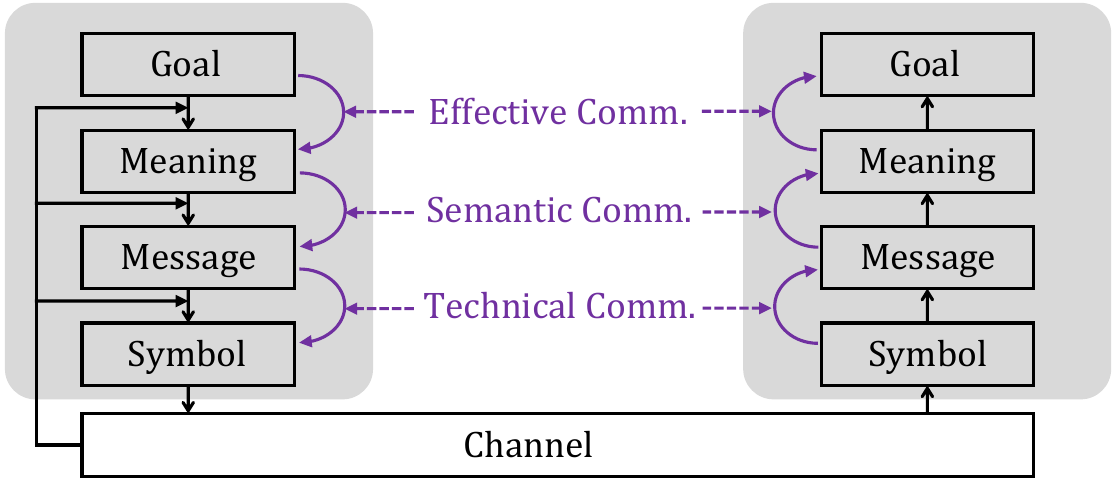}\\
  \caption{A broad communication process with three levels of problems.}
\label{fig:1}
\end{figure}

Following Weaver's formulation, we depict a communication process in Fig.~\ref{fig:1} and expound on some key concepts to be used throughout this paper. We will provide general descriptions of these concepts here, and relegate their more precise mathematical definitions in our context to Section \ref{sec:problem}.
\begin{itemize}
\item \textbf{Goal}. Communication serves for a cooperative task between the transmitter and receiver. The task is the ultimate goal of communication.
\item \textbf{Meaning}. At different phases of the task, the transmitter can generate different meanings to be conveyed to the receiver for achieving the ultimate goal. The intended meaning can be an observation of the environment, an instruction, a state update, etc.
\item \textbf{Message}. Messages are the carriers of meaning and are constructed following an agreed-upon language between the transmitter and receiver.
\item \textbf{Symbol}. Symbols refer to the coded channel symbols to be transmitted to the receiver via the physical channel.
\end{itemize}

As shown in Fig. \ref{fig:1}, effective/pragmatic communication deals with the problem of generating the right meaning for achieving the ultimate goal, considering the current states of the transmitter, receiver, and the progress of the task; semantic communication deals with the problem of how to construct the right message to accurately convey the meaning based on the agreed language; technical communication studies how to design channel symbols for different messages such that the messages can be reconstructed at the receiver as accurately as possible.

Shannon provided an elegant solution to the technical problem \cite{shannon1948}, which further spawned the field of information theory. In his solution, the amount of information in a message is measured by the notion of ``uncertainty'', which is inversely proportional to the statistical probability of the message. Thus, it is the probabilities, but not the meanings of messages matter in Shannon's theory. In contrast, semantic communication is more concerned with the meaning that can be conveyed by a message.
We give two examples below to show the difference between semantic and technical communications.

\begin{example}[Successful technical communication but failed semantic communication]\label{exp:1}
Consider the communication between a farmer and his grandchild. The farmer says: ``The apple looks good''. The intended meaning of the farmer is that the fruit ``apple'' looks good. His grandchild, however, may interpret ``apple'' as the mobile phone. In this example, technical communication succeeds since the message has been perfectly transmitted to the receiver. Semantic communication, however, fails due to misinterpretation -- the same message can convey different meanings in a language.
\end{example}
\begin{example}[Failed technical communication but successful semantic communication]\label{exp:2}
Consider the message transmission from Bob to Ted, and then to Alice. Bob says: ``Carol does not like carrots''. Ted relays the message to Alice: ``Carol dislikes carrots''.
In this example, technical communication fails since the received message of Alice is different from the original message transmitted by Bob. Nevertheless, semantic communication succeeds because Alice knows ``does not like'' is the same as ``dislikes'' in English -- the intended meaning of Bob is successfully conveyed.
\end{example}

\subsection{Fundamental problems of semantic communication}
The cornerstone of communication is an agreed language among the parties involved. In Shannon's theory, for example, two key procedures for transmitting a message are source coding and channel coding. In source coding, each message is assigned a unique bit sequence (more generally, a sequence of alphabet symbols) according to the statistical probability of the message. In channel coding, structured redundancies are added to the source-coded bit sequences to combat physical channel impairments, yielding channel symbols to be transmitted to the receiver. At the receiver, first the channel input is resolved then the message is decoded. 

The successful operation in Shannon's approach relies on the transmitter and receiver's consensuses on 
1) the statistical distribution of messages;
2) the source code, e.g., the one-to-one mappings between messages and bit sequences in lossless source coding; and 3) the channel code, e.g., the algebraic structures among coded symbols, before transmission. These serve as the common ground for technical communication, and we refer to them as {\it technical language} in the sequel.

Likewise, semantic communication relies on a common {\it semantic language} between the transmitter and receiver for conveying the meaning. A typical example of semantic language is the human language, e.g., English, as in Examples \ref{exp:1} and \ref{exp:2}. Unlike hand-crafted technical language, semantic language is often formed naturally over the course of interactions and is much richer in expression and interpretation: a meaning can be expressed by multiple messages and a message can be interpreted as multiple meanings. The reason for such richness can be attributed to:
\begin{enumerate}
\item Semantic communication pursues not only communication efficiency -- which is the exclusive pursuit of technical communication -- but also elegance, charm, comprehensibility, politeness, etc \cite{feist2022}.
\item Semantic language has to be robust to the mismatched or unagreed prior information between the transmitter and receiver \cite{juba2011}. That is, the prior distribution of the intended meaning may not be known to the receiver before transmission.
\end{enumerate}

Based on the above understanding of languages, we formulate two fundamental problems of semantic communication.

\begin{problem}[Language exploitation]\label{problem:1}
The transmitter and receiver have agreed on semantic and technical languages. 
When conveying an intended meaning, how to minimize the misinterpretations of the receiver from the transmitter's perspective (semantic encoding) or the receiver's perspective (semantic decoding)?
\begin{itemize}
\item \textbf{Semantic encoding}: How can the transmitter generate the message such that the intended meaning can be recovered at the receiver as accurately as possible while the communication cost is minimized?
\item \textbf{Semantic decoding}: Given a received message, how can the receiver decode the intended meaning of the transmitter without accurate prior information about the meaning?
\end{itemize}
\end{problem}

As noted by Weaver, ``the semantic problems are concerned with the interpretation of meaning by the receiver, as compared with the intended meaning of the sender'', the first fundamental problem emphasizes how to leverage the agreed language to reduce misinterpretations of the receiver. An implication of the language exploitation problem is that there is no negotiation before the transmission. The only things that the transmitter and receiver agree on are the semantic and technical languages. 

Note that semantic encoding and decoding can also be used at the same time, which we refer to as combined semantic encoding and decoding (CSED). The language exploitation problem resembles human communications.

\begin{problem}[Language design]
Assuming that the transmitter and receiver are allowed to negotiate before the transmission, how can the semantic and technical languages be designed to efficiently convey the meaning of a semantic source?
\end{problem}

The second fundamental problem concerns how to design common languages or codebooks between the transmitter and receiver to efficiently convey the meaning. 
Since the transmitter and receiver are given full freedom to design the languages, we can directly design the mappings between meanings and channel symbols, to which joint source-channel coding (JSCC) theory naturally applies.
In other words, language design in semantic communication falls into the scope of the classical communication problem that researchers have been striving to solve for decades -- designing the communication system to minimize a prescribed distortion measure between the transmitted and reconstructed data.
In technical communication, we focus on sequences of equal likely bits and consider the block error rate (BLER) as the measure of reconstruction quality. 
Semantic communication, on the other hand, considers more general sources (e.g., text \cite{xie2021}, image \cite{bourtsoulatze2019}, video \cite{tung2022}, point cloud \cite{PC}) and end-to-end distortion measures (e.g., classification loss \cite{xie2021,shao2021federated}, perceptual loss \cite{blau2019}, goal-oriented distortions \cite{beck2022,noisynn}).

As will be detailed in Section \ref{sec:prior}, the language design problem can be further divided into two classes of subproblems: joint semantic and technical language design, and technical language design under a given semantic language.
A popular trend nowadays is leveraging data-driven techniques, in particular deep learning (DL), to learn a common language \cite{bourtsoulatze2019,xu2023deep,xie2021,shao2022semantic}. In this approach, there is a training phase and an evaluation phase. The training phase is essentially a process of language negotiation/design between the transmitter and receiver for a specific kind of source. When the training is done, the agreed language, manifested as the neural encoder and decoder, is used in the evaluation phase for semantic transmission. In general, data-driven solutions to the language design problem are particularly useful in the case of analytically intractable distortion measures,  a semantic source with memory, and resource-constrained channels, to name a few.

\begin{figure}[t]
  \centering
  \includegraphics[width=0.9\linewidth]{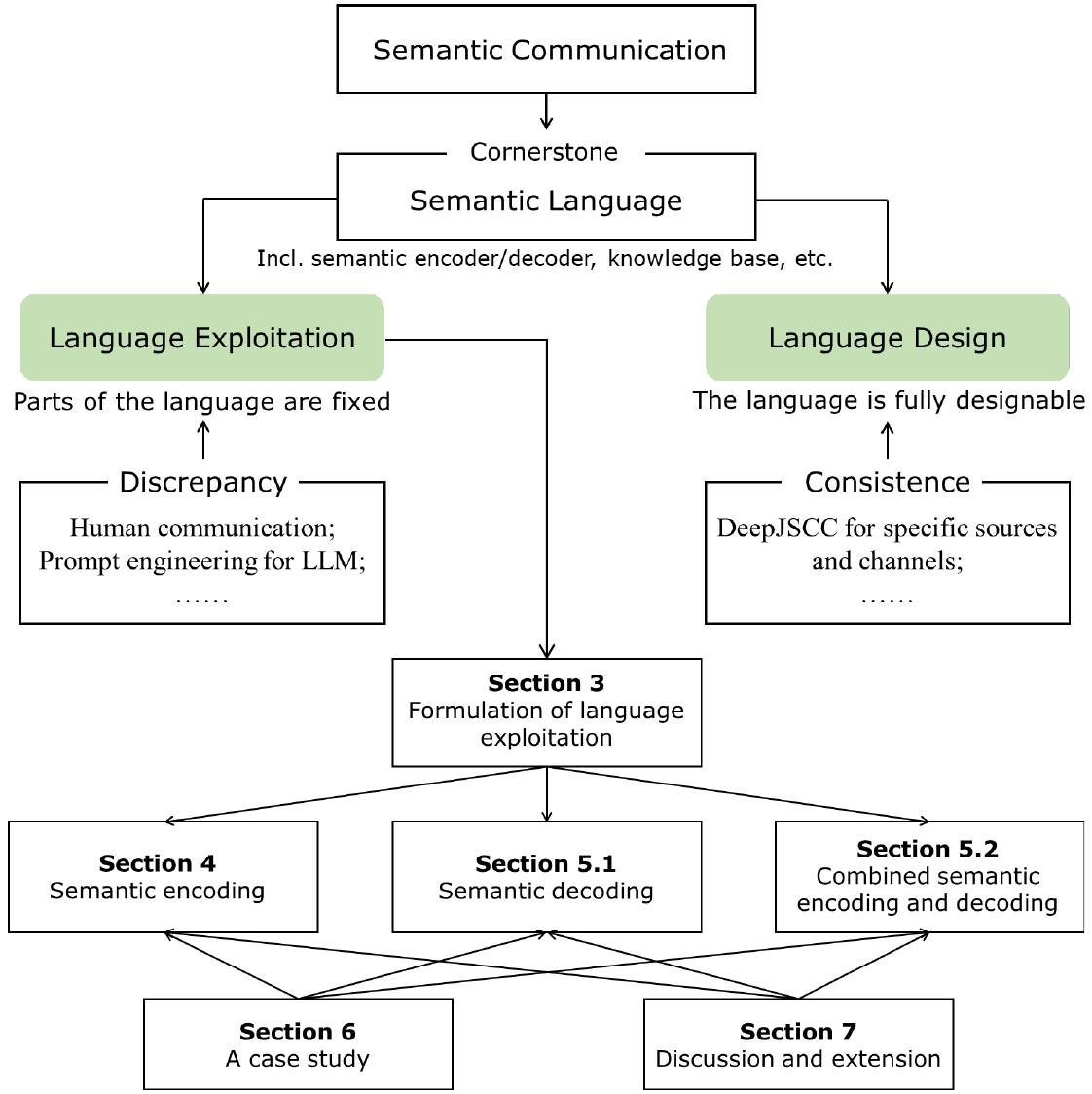}\\
  \caption{{The framework of semantic communications. The language design problem focuses on crafting a new set of languages for efficient communication between the transmitter and receiver for specific sources, aligning with classical communication principles. In contrast, the language exploitation problem addresses the nuanced challenges arising from discrepancies between the transmitter and receiver, extending beyond the traditional communication framework.}}
\label{fig:1a}
\end{figure}

\subsection{Contributions and roadmap}

In this paper, we provide a comprehensive exploration of semantic communication, shedding light on its essence and framing it within a general framework, as outlined in Fig.~\ref{fig:1a}. We emphasize that the core of semantic communication lies in what we term the ``semantic language'', which encompasses the semantic encoder of the transmitter, the semantic decoder of the receiver, and the collective knowledge base shared by all communication parties.

Within the framework, we formulate two fundamental challenges intrinsic to semantic communication: language exploitation and language design.
\begin{itemize}[leftmargin=0.5cm]
    \item Language exploitation pertains to scenarios where certain elements of the semantic language are fixed and unalterable, necessitating the strategic design of other components to efficiently convey intended meanings. This challenge captures situations characterized by disparities between the transmitter and receiver, with the primary objective being to mitigate these disparities. A typical example of this problem can be found in prompt engineering for large language models (LLMs). In this context, the semantic decoder of LLMs remains unchangeable, and the goal is carefully engineer the semantic encoder of human beings to effectively interact with the LLM \cite{llmind}.
    \item Language Design, on the other hand, ventures into the optimal crafting of the semantic language, striking a balance between data transmission efficiency -- often quantified in terms of bits per sample or channel uses per sample -- and a versatile distortion metric. Here, the entire semantic language, inclusive of the semantic encoder and decoder, is open to design, allowing for their joint optimization tailored to specific sources or channels. This approach ensures consistency between the transmitter and receiver. Applications of the Language Design problem encompass a wide range of data sources, spanning text, images, video, and point clouds, under the DeepJSCC paradigm.
\end{itemize}

Importantly, language design aligns with the classical principles of communication and is within the scope of traditional communication problems \cite{gunduz2022survey,qin2021survey,lan2021survey}. In contrast, language exploitation addresses the nuanced challenges that arise from discrepancies between the transmitter and receiver, extending beyond the boundaries of conventional communication frameworks.

In this paper, we delve into the challenge of language exploitation, exploring three key
approaches to address it: semantic encoding, semantic decoding, and a combined strategy of semantic encoding and decoding. Moreover, we introduce the semantic distortion-cost region as a pivotal metric for assessing semantic communication performance. For each of the three proposed approaches, the achievable regions are meticulously characterized.

{\it Roadmap} -- The remainder of this paper is articulated in the following manner.
Section \ref{sec:prior} reviews the prior art on semantic communication according to the problem of language exploitation and language design.
Section \ref{sec:problem} defines some basic semantic-related concepts and rigorously formulates the language exploitation problem.
Sections \ref{sec:enc}, \ref{sec:decA}, and \ref{sec:decB} solve the language exploitation problem by semantic encoding, semantic decoding, and CSED, respectively.
A concrete example to illustrate the essence of our formulation and solution is given in Section \ref{sec:example}.
Discussion and extension are given in Section \ref{sec:ext}.
Section \ref{sec:con} concludes this paper.

\section{Related Works}\label{sec:prior}
\subsection{Language utilization}
In the literature, there are relatively few research efforts \cite{carnap1952,bao2011,juba2008,zhong2017} devoted to formulating the theory of semantic communication, perhaps owing to the lack of clear definitions of semantics and languages. Among these few studies, the most notable works are Carnap and Bar-Hillel's characterization of semantic information \cite{carnap1952} and the follow-up attempt at formulating semantic communication \cite{bao2011}.

Carnap and Bar-Hillel's work focused on how to define semantic information and how to measure the amount of semantic information contained in a message. To this end,
\begin{enumerate}
\item They defined a language system with messages being limited to inductive logic. That is, messages can only be declarative statements or propositions.
\item They measured the amount of semantic information carried in a message by the probability that the message is logically true.
\end{enumerate}

In this context, the semantic entropy of a message $s$ is defined as
\begin{equation}
H_{\text{sem}}(s)=-\log_2 {\sum_{w\models s} p(w)},
\end{equation}
where $w$ denotes a world model (e.g., a meaning under Weaver's formulation), $\models$ denotes the propositional satisfaction relation, and ${\sum_{w\models s} p(w)}$ is the probability of all world models in which $s$ is true.

As can be seen, in Shannon's theory, the information of a message is governed by its statistical probability but not its meaning (e.g., whether the message itself is true or false). By contrast, in Carnap and Bar-Hillel's formulation, the statistical probability $p(s)$ is irrelevant, and the semantic information of $s$ is determined by its logical probability ${\sum_{w\models s} p(w)}$ under their language system.

Following Carnap and Bar-Hillel's characterization of semantic information, the authors of \cite{bao2011} defined a complete procedure of semantic communication and established semantic source coding and semantic channel coding by mimicking Shannon's theory. They assumed the same logical language system as \cite{carnap1952}. In source coding, for example, the authors proposed to merge semantically equivalent messages in the language into a single message, thereby constructing a minimum subset of the original set of messages with no semantic loss. In particular, the minimum subset exhibits the minimum entropy, and hence, the minimum expected coding length.
 
Although established under the specific logical language system, the formulations and insights obtained in \cite{carnap1952,bao2011} shed light on forming a generic theory of semantic communication, and also inspire our work.

\subsection{Language design}
In the language design problem, the transmitter and receiver are allowed to negotiate both semantic and technical languages. For now, the problem that receives wide attention is the joint design of semantic and technical languages using DL techniques.

\subsubsection{Joint semantic and technical language design via DL}
In the basic form, the objective of DL-based semantic communication is transmitting a specific kind of source beyond a simple bitstream (e.g., text \cite{xie2021}, image \cite{bourtsoulatze2019}, video \cite{tung2022}), for which DL techniques are utilized to design a common language. As stated in the introduction, DL-based approaches consist of a training phase for language design and an evaluation phase for data transmission. The main research challenge is how to learn an efficient language in the training phase tailored for the considered source. To this end, various schemes, such as DeepJSCC \cite{bourtsoulatze2019,xie2021}, discrete-time analog transmission \cite{shao2022semantic}, the information bottleneck approach \cite{beck2022}, the goal-oriented principle \cite{farshbafan2022}, the nonlinear transform \cite{dai2022}, etc., have been proposed. We refer readers to survey papers \cite{gunduz2022survey,qin2021survey,lan2021survey} for a more detailed treatment of this line of work.

\subsubsection{Technical language design under a given semantic language}
In addition to the joint design of semantic and technical languages, another interesting problem is the design of technical language under a given semantic language. This line of work is relatively few.

In \cite{guler2018}, the authors proposed to optimize the communication system performance by taking the semantic content of messages into account. There are three parties in their framework: a transmitter, a receiver, and an influencer. The transmitter aims to transmit a message (more precisely, the meaning of the message) to the receiver, which, in turn, decodes the message with the side information provided by the influencer. To quantify the semantic distortion in the system, the authors defined a semantic measure that quantifies the similarity of messages. The transmitter and receiver have a fixed set of encoding and decoding functions, respectively. In this semantic communication system, the semantic language and the semantic distortion among messages are given. To deliver the meaning of a message, the transmitter and receiver design the technical language, i.e., choose between the set of encoding and decoding functions, thereby minimizing the semantic distortion in the presence of the influencer. 

In \cite{liu2022}, the authors considered a semantic source with an intrinsic state and an extrinsic observation, where the transitions between the state and observation are determined. The receiver is interested in both state and observation, thus there are two distortions at the receiver with respect to the reconstructed state and observation, respectively. The authors characterized the trade-off between the two distortions under a given code rate. Translating into our formulation, the intrinsic state and extrinsic observation can be viewed as meaning and message, respectively, and the semantic language is the fixed (and known) mapping between state and observation. The authors designed the technical language (codebook) to strike a balance between the two distortions using techniques from indirect rate-distortion theory and rate-distortion under multiple distortion measures.

\section{Problem Formulation}\label{sec:problem}
This section formulates the language utilization problem rigorously. We shall go through some basic semantic-related definitions, such as language, channel, distortion, cost, etc., and define the problems of semantic encoding and decoding, respectively. Let us set out to define semantic language, the prerequisite of semantic communication.

\begin{figure}[t]
  \centering  \includegraphics[width=0.7\linewidth]{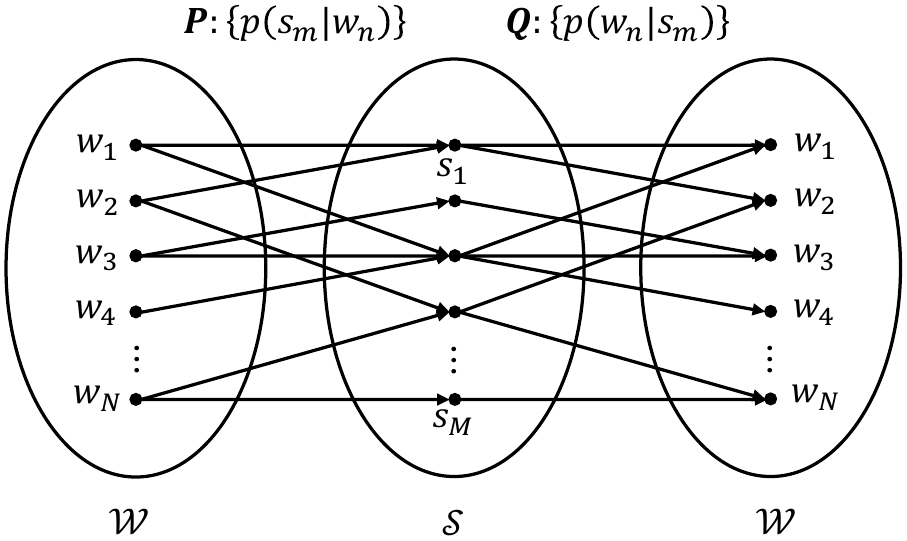}\\
  \caption{The main ingredients of a semantic language $(\mathcal{W},\mathcal{S},\bm{P},\bm{Q})$.}
\label{fig:language}
\end{figure}

A semantic language is a structured system that defines the transformation between meanings and messages. In this paper, we characterize a semantic language by four main ingredients: word, syntax, expression, and interpretation. To ease exposition, henceforth ``language'' refers specifically to semantic language, unless specified otherwise.

\begin{defi}[Words and syntax]
Words are the smallest elements of a message in a language. Words can be used on their own or together to form a message. Syntax is a set of rules that determine the arrangement of words in a message. 
\end{defi}

\begin{defi}[The set of messages]
The word and syntax of a language determine the set of all possible messages, denoted by $\mathcal{S}$. Suppose $\mathcal{S}$ is finite or countably infinite, we define $\mathcal{S}\triangleq \{s_m: m\in [M]\}$, where $s_m$ denotes a message, $M$ the number of all possible messages, and we define $[M] \triangleq \{1,2,\ldots,M\}$. 
\end{defi}

\begin{defi}[The set of meanings]
Suppose the messages in $\mathcal{S}$ can convey a finite or countably infinite number of meanings. We define the set of all possible meanings as $\mathcal{W}\triangleq \{w_n:n \in [N]\}$, where $w_n$ denotes one meaning and $N$ denotes the number of meanings. Furthermore, we denote by $p(w_n)$ the probability that the intended meaning of the transmitter is $w_n$.
\end{defi}

\begin{defi}[Expression]
The \textit{expression} of a language defines a mapping from the set of meanings to the set of messages, denoted by 
\begin{equation}
\left\{p(s|w)\in[0,1]: w\in\mathcal{W},s\in\mathcal{S},\sum_s p(s|w) = 1 \right\}.
\end{equation}
To simplify the notation, we write the mappings into a matrix form as $\bm{P}\in \RR^{N\times M}$: the element on the $n$-th row and $m$-th column of $\bm{P}$ is $p(s_m| w_n)$ and each row of $\bm{P}$ sums to $1$.
\end{defi}

\begin{defi}[Interpretation]
The \textit{interpretation} of a language defines a mapping from the set of messages to the set of meanings, denoted by 
\begin{equation}
\left\{q(w|s)\in[0,1]: w\in\mathcal{W},s\in\mathcal{S},\sum_w q(w|s) = 1 \right\}
\end{equation}
or the matrix form $\bm{Q}\in \RR^{M\times N}$. Each column of $\bm{Q}$ sums to $1$.
\end{defi}

Based on the above definitions, we denote a semantic language by a $4$-tuple $(\mathcal{W},\mathcal{S},\bm{P},\bm{Q})$, as illustrated in Fig.~\ref{fig:language}. As can be seen, the words and syntax of the language formulate the message set $\mathcal{S}$. A meaning $w\in\mathcal{W}$ can be expressed by a subset of messages $\{s:s\in\mathcal{S},p(s|w)>0\}$, the cardinality of which reflects the expressive redundancy of the language. On the other hand, a given message $s$ can be interpreted as a subset of meanings $\{w:w\in\mathcal{W},q(w|s)>0\}$, the cardinality of which reflects the interpretation ambiguity of the language.

\begin{figure}[t]
  \centering  \includegraphics[width=0.75\linewidth]{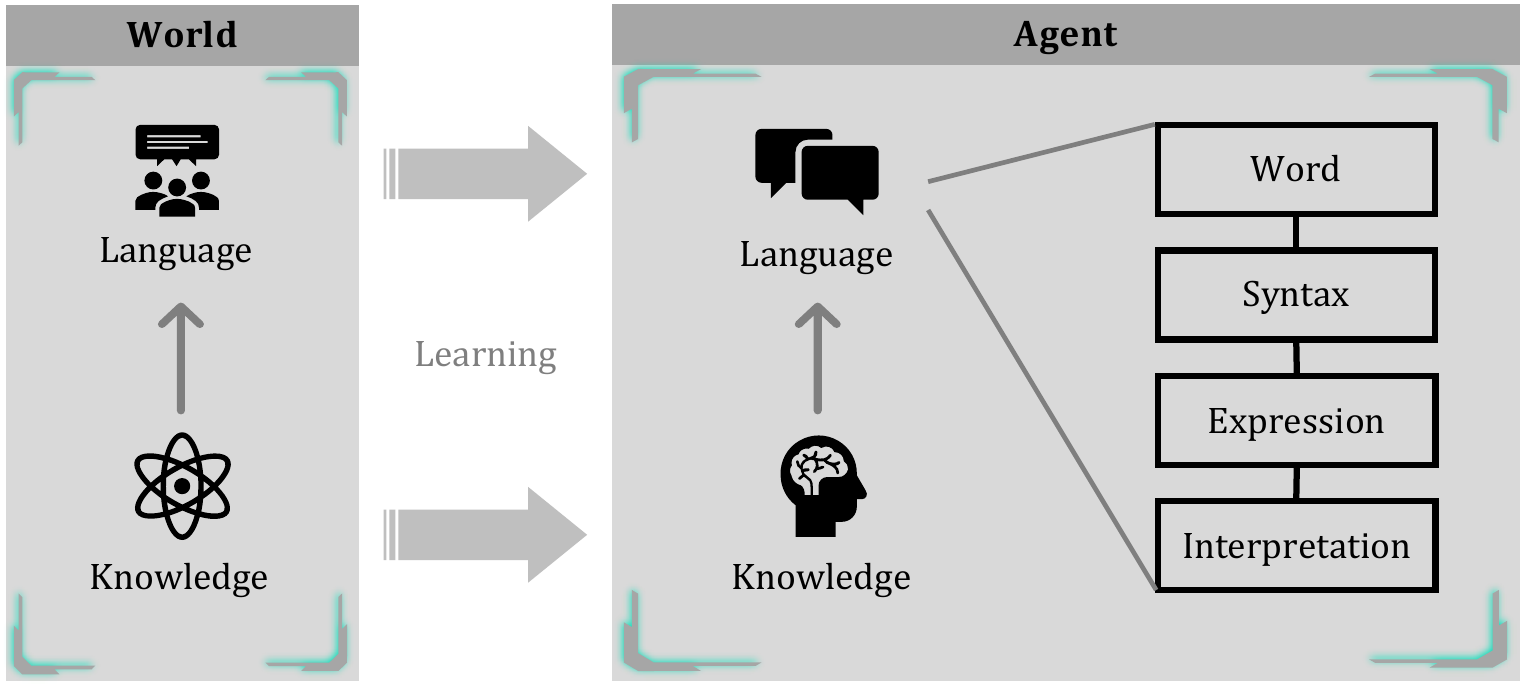}\\
  \caption{A human being develops his/her own language by learning from the world.}
\label{fig:learning}
\end{figure}

\begin{defi}[Logically self-consistent language]
A language is logically self-consistent if and only if
\begin{equation}\label{eq:self}
q(w|s)=\frac{p(s|w)p(w)}{p(s)}.  
\end{equation}
\end{defi}

In this paper, we do not impose any constraint on the logical self-consistency of the language; and hence, $q(w|s)$ does not need to satisfy \eqref{eq:self} and can be any mapping. The advantage of such a formulation is that it encompasses a more practical semantic communication scenario with mismatched expression and interpretation.

Consider human communication as an example. The language of a human being is acquired from the world through interactions, which we refer to as the learning process. As illustrated in Fig.~\ref{fig:learning}, the language of the world is supported by the knowledge of the world. Each person acquires both knowledge and language from the world, thereby developing a language system with its own message set and styles of expression and interpretation. In other words, the language of each agent can be treated as unique, i.e., a distinct pair of mappings $\bm{P}$ and $\bm{Q}$, which rely on the unique personal experience. The communication between two human beings, in this case, faces logical inconsistency, which is allowed in our formulation.

Next, we define the semantic channel.
\begin{defi}[Semantic channel]
Consider the communication between a transmitter and a receiver. Suppose the transmitted message is $s\in\mathcal{S}$ and the received message is $\hat{s}\in\mathcal{S}$.\footnote{More generally, one can define another message set $\hat{\mathcal{S}}$ at the receiver and define $\hat{s}\in\hat{\mathcal{S}}$.}
We characterize the technical channel by the transition probabilities from $s$ to $\hat{s}$ and denote it by $\left\{c(\hat{s}|s)\in[0,1]:s,\hat{s}\in\mathcal{S},\sum_{\hat{s}} c(\hat{s}|s) = 1 \right\}$,
or the matrix form $\bm{C}\in \RR^{M\times M}$. 
Given the process of expression and interpretation within the semantic language, the semantic channel can be written as
\begin{equation}
    \Pr(\hat{w}|w) = \sum_{s,\hat{s}}p(s|w)c(\hat{s}|s)q(w|\hat{s}).
\end{equation}
\end{defi}

Our formulation abstracts technical communication as part of the semantic channel, the quality of which is determined by the manner technical communication is implemented (e.g., coding and modulation schemes, etc.).

As stated in Problem \ref{problem:1}, semantic communication aims to minimize the misinterpretations of the receiver, leveraging the agreed semantic and technical languages between the transmitter and receiver. This can be achieved either from the transmitter's perspective (semantic encoding) or the receiver's perspective (semantic decoding). In the following, we formulate both problems based on the above definitions.

\begin{rem}[One-shot transmission]
Unlike Shannon's information theory, which considers the transmission of a sequence of i.i.d. messages, this paper considers one-shot transmission (i.e., conveying one single meaning) from the transmitter to the receiver. Further discussion on the joint transmission of a sequence of meanings can be found in Section~\ref{sec:ext}.
\end{rem}

\subsection{Semantic encoding}
From the transmitter's perspective, the semantic decoder at the receiver is dictated by the interpretation of the agreed language. The semantic encoding problem is thus how to encode the intended meaning using messages to minimize the misinterpretation at the receiver, considering the semantic channel as well as the given semantic decoder. For a figurative analogy, consider a mother talking to her son. The mother knows the way her son would interpret her message. Therefore, for the mother, semantic encoding is the problem of choosing the best messages to express her meaning such that the meaning can be precisely conveyed to the son.
A practical example of semantic encoding is video compression, wherein international standards typically specify only the decoder \cite{MPEG}.
The manufacturers are free to design the video encoders as long as the encoded video can be decoded by the standardized decoder. While we will assume that the stochastic semantic channel and interpretation mappings are known by the encoder in this work, a data-driven approach where this mapping can be learned through interactions is left as a potential future extension.

\begin{defi}[Semantic encoding schemes]
A semantic encoding scheme is a mapping from the meaning set $\mathcal{W}$ to the message set $\mathcal{S}$, denoted by 
\begin{equation}
\left\{u(s|w)\in[0,1]: w\in\mathcal{W},s\in\mathcal{S},\sum_s u(s|w) = 1 \right\},
\end{equation}
or the matrix form $\bm{U}\in \RR^{N\times M}$.
\end{defi}

In principle, we can impose constraints on $u(s|w)$ such that the semantic encoding scheme complies with the semantic language. For example, if the semantic encoding scheme must comply with the interpretation, we can impose the constraint ``$u(s|w)=0$ if $p(s|w)=0$''; if the transmitter is not allowed to deceive the receiver, we can impose the constraint ``$u(s|w)=0$ if $q(w|s)=0$''; if the transmitter is not allowed to misdirect the receiver, we can impose the constraint ``$u(s|w)\geq u(s^\prime|w)$ if $q(w|s)\geq q(w|s^\prime)$''. In this paper, however, we consider a general setup and do not impose any constraints on  $u(s|w)$.

To evaluate the performance of semantic encoding, we introduce two metrics below, one is semantic distortion and the other is semantic cost.

\begin{defi}[Semantic distortion]
Let $w,\hat{w}\in\mathcal{W}$ be the transmitted and reconstructed meanings at the transmitter and receiver, respectively. We define a semantic distortion measure
$d(w,\hat{w}):\mathcal{W}\times\mathcal{W}\to\RR^+$, where $\RR^+$ is the set of non-negative real numbers. The average distortion achieved by a semantic encoding scheme $\bm{U}$ is given by
\begin{equation}
D_{\bm{U,Q}}=\sum_{w,s,\hat{s},\hat{w}}p(w)u(s|w)c(\hat{s}|s)q(\hat{w}|\hat{s})d(w,\hat{w}).
\end{equation}
\end{defi}

\begin{defi}[Semantic cost]
Each message is associated with a cost. Semantically, the cost of a message can be its conciseness, comprehensibility, or elegance. Technically, the cost of a message can be the length of the corresponding bit sequence. In general, we define a cost function $\ell(s):\mathcal{S}\to\RR^+$, $\forall s\in\mathcal{S}$. The average cost achieved by a semantic encoding scheme $\bm{U}$ is 
\begin{equation}
L_{\bm{U}}=\sum_{w,s}p(w)u(s|w)\ell(s).
\end{equation}
\end{defi}

Without loss of generality, we assume $\mathcal{S}$ is an ordered set throughout the paper such that for any two messages $s_{m_1}$,$s_{m_2}$, $m_1,m_2\in [M]$, we have $\ell(s_{m_1})\leq \ell(s_{m_2})$ if $m_1<m_2$.

\begin{figure}[t]
  \centering  \includegraphics[width=0.95\linewidth]{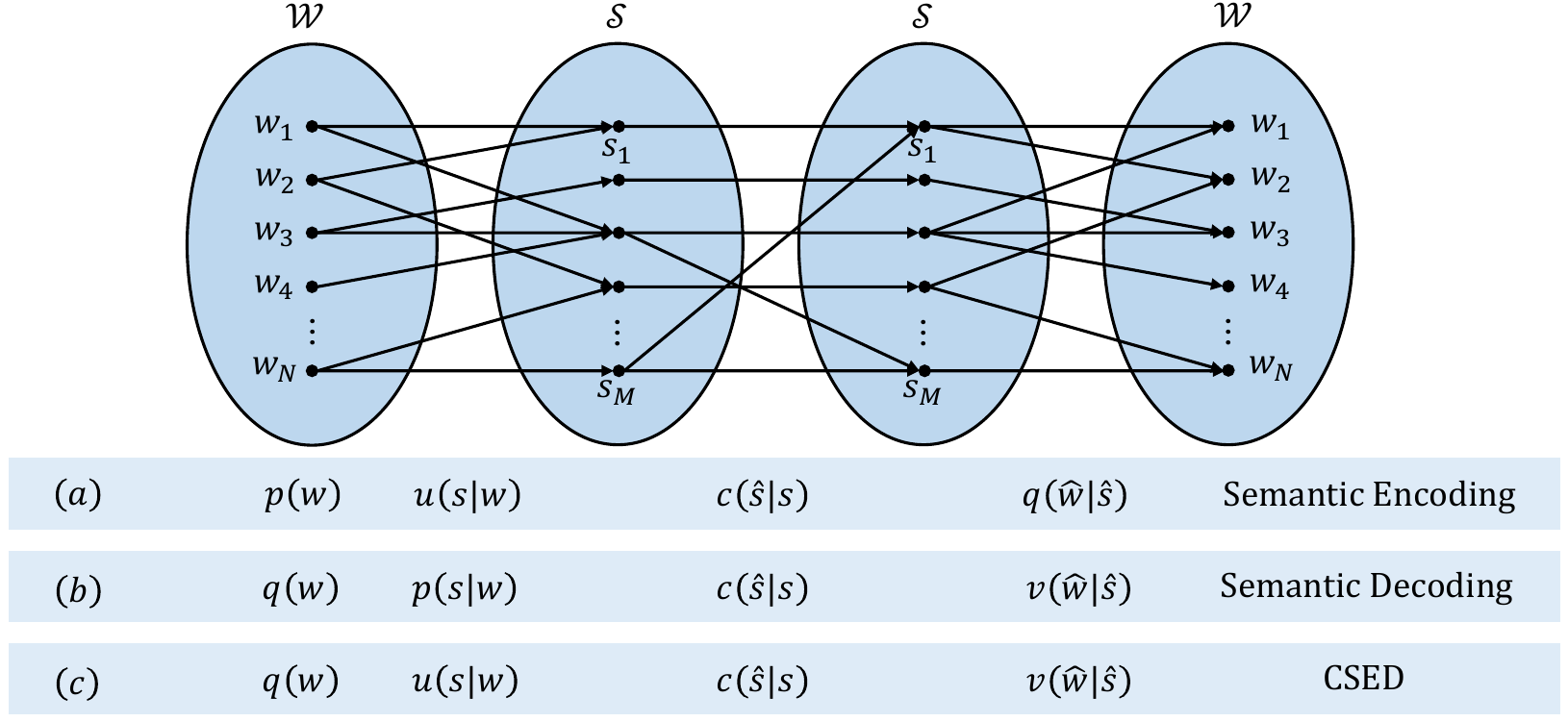}\\
  \caption{A point-to-point semantic communication process with (a) semantic encoding at the transmitter, (b) semantic decoding at the receiver, or (c) combined semantic encoding and decoding at the transmitter and receiver, respectively. We have assumed that the receiver has the same message and meaning sets as the transmitter. More generally, we can define a new message set $\hat{\mathcal{S}}$ and a new meaning set $\hat{\mathcal{W}}$ for the receiver.}
\label{fig:complete}
\end{figure}

Given the above definitions, we are ready to describe a point-to-point semantic communication process with semantic encoding at the transmitter. As shown in Fig.~\ref{fig:complete}(a), the transmitter first generates a meaning $w\sim p(w)$. To convey $w$ to the receiver, the transmitter constructs a message $s\sim u(s|w)$ following its semantic encoding scheme. After passing through the semantic channel, the receiver observes $\hat{s}\sim c(\hat{s}|s)$ and reconstructs a meaning $\hat{w}\sim q(\hat{w}|\hat{s})$ following the interpretation of the language. The main problem of semantic encoding is to characterize the optimal encoding scheme that achieves the best trade-off between the semantic distortion and semantic cost.

\subsection{Semantic decoding and combined semantic encoding and decoding (CSED)}
From the receiver's perspective, the semantic encoder at the transmitter is dictated by the expression of the agreed language. The semantic decoding problem is thus how to decode a received message to minimize semantic distortion, considering the semantic channel as well as the given semantic encoder at the transmitter. An analogy of semantic decoding is a son talking to his mother. The mother knows the way her son expresses his meaning and aims to find the best way to interpret the received messages.
Another example of semantic decoding is reading a text. A reader tries to optimize his/her semantic decoder to minimize the distortion between the meaning intended by the writer and his/her interpretation.


\begin{defi}[Semantic decoding schemes]
A semantic decoding scheme is a mapping from the message set $\mathcal{S}$ to the meaning set $\mathcal{W}$, denoted by 
\begin{equation}
\left\{v(w|s)\in[0,1]: w\in\mathcal{W},s\in\mathcal{S},\sum_w v(w|s) = 1 \right\},
\end{equation}
or the matrix form $\bm{V}\in \RR^{M\times N}$.
\end{defi}

Similar to the semantic encoding problem, we do not impose any constraints on $v(w|s)$. To evaluate the performance of a semantic decoding scheme, we define the semantic distortion below.

\begin{defi}[Semantic distortion of semantic decoding]
Let $w,\hat{w}\in\mathcal{W}$ be the transmitted and reconstructed meanings at the transmitter and receiver, respectively. The average distortion achieved by a  semantic decoding scheme $\bm{V}$ is
\begin{equation}
D_{\bm{P,V}}=\sum_{w,s,\hat{s},\hat{w}}p(w)p(s|w)c(\hat{s}|s)v(\hat{w}|\hat{s})d(w,\hat{w}).
\end{equation}
\end{defi}

The point-to-point semantic communication process with semantic decoding at the receiver is illustrated in Fig.~\ref{fig:complete}(b). Compared with semantic encoding, semantic decoding may consider a potentially inaccurate prior distribution $q(w)$ at the receiver and a fixed semantic encoder $\bm{P}$ dictated by the language. The goal is to find the optimal decoding scheme $\bm{V}^*$ such that the semantic distortion is minimized.

In the language utilization problem, the agreed language is the only common ground between the transmitter and receiver and there is no codebook negotiation. Therefore, semantic encoding and decoding are essentially two decoupled problems -- jointly designing the semantic encoding and decoding falls into the language design problem.

Nevertheless, the transmitter and receiver can perform semantic encoding and decoding, separately, at the same time, which we refer to as the combined semantic encoding and decoding (CSED), as illustrated in Fig.~\ref{fig:complete}(c). CSED will be defined rigorously later in Section \ref{sec:decB}.

\section{Semantic Encoding}\label{sec:enc}
Based on the formulation in Section \ref{sec:problem}, this section focuses on the semantic encoding problem. For a given semantic language, our main goal in this section is to define and characterize the semantic distortion-cost region (and function) of semantic encoding.

\subsection{Semantic distortion-cost region and function}
A semantic encoding scheme can be evaluated from two angles: semantic distortion and semantic cost. We next define the semantic distortion-cost region for semantic encoding.

\begin{defi}[The distortion-cost region of semantic encoding]
Consider a semantic language $(\mathcal{W},\mathcal{S},\bm{P},\bm{Q})$, a semantic channel $\bm{C}$, a message cost function $\ell(s):\mathcal{S}\to\RR^+$, and a distortion measure $d(w,\hat{w}):\mathcal{W}\times\mathcal{W}\to\RR^+$. A distortion-cost pair $(L,D)$ is achievable if there exists a semantic encoding scheme $\bm{U}$ such that $D_{\bm{U}}=D$, $L_{\bm{U}}=L$. The distortion-cost region $R_{\text{enc}}$ is the set of all achievable distortion-cost pairs $(L,D)$.
\end{defi}

A trivial outer bound of the region is
\begin{equation}
\begin{cases}
L_{\text{min}} \leq L \leq L_{\text{max}}, \\
D_{\text{min}} \leq D \leq D_{\text{max}},
\end{cases}
\end{equation}
where
\begin{equation}
L_{\text{min}} \triangleq \min_{(L,D)\in R_{\text{enc}}} L = \min_s \ell(s), 
\end{equation}
\begin{equation}
L_{\text{max}} \triangleq \max_{(L,D)\in R_{\text{enc}}} L = \max_s \ell(s),
\end{equation}
are the minimum and maximum costs, respectively, and
\begin{equation}
D_{\text{min}} \triangleq \min_{(L,D)\in R_{\text{enc}}} D,~~~
D_{\text{max}} \triangleq \max_{(L,D)\in R_{\text{enc}}} D,
\end{equation}
are the minimum and maximum achievable distortions with semantic encoding, respectively.

\begin{figure}[t]
  \centering  \includegraphics[width=0.6\linewidth]{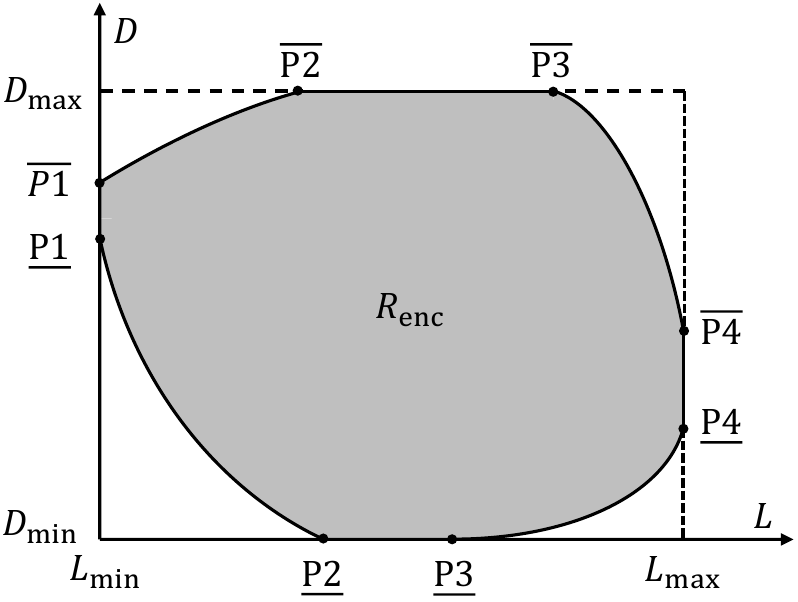}\\
  \caption{An illustration of the distortion-cost region and function of semantic encoding.}
\label{fig:region}
\end{figure}

The exact distortion-cost region, on the other hand, can be more involved. For illustration purposes, we give a schematic plot of the region in Fig.~\ref{fig:region}.

To characterize the region, we first define eight critical points as follows:
\begin{small}
\begin{eqnarray*}
&&\hspace{-0.7cm}
\underline{\text{P1}}:\left(L_{\text{min}},\min_{(L_{\text{min}},D)\in R_{\text{enc}}} D \right),~
\underline{\text{P2}}:\left(\min_{(L,D_{\text{min}})\in R_{\text{enc}}} L, D_{\text{min}} \right), \\
&&\hspace{-0.7cm}
\underline{\text{P3}}:\left(\max_{(L,D_{\text{min}})\in R_{\text{enc}}} L, D_{\text{min}} \right),~
\underline{\text{P4}}:\left(L_{\text{max}},\min_{(L_{\text{max}},D)\in R_{\text{enc}}} D \right),\\
&&\hspace{-0.7cm}
\overline{\text{P1}}:\left(L_{\text{min}},\max_{(L_{\text{min}},D)\in R_{\text{enc}}} D \right),~
\overline{\text{P2}}:\left(\min_{(L,D_{\text{max}})\in R_{\text{enc}}} L, D_{\text{max}} \right), \\
&&\hspace{-0.7cm}
\overline{\text{P3}}:\left(\max_{(L,D_{\text{max}})\in R_{\text{enc}}} L, D_{\text{max}} \right),~
\overline{\text{P4}}:\left(L_{\text{max}},\max_{(L_{\text{max}},D)\in R_{\text{enc}}} D \right),
\end{eqnarray*}
\end{small}

Each of these eight points corresponds to a semantic encoding scheme. For example, $\underline{\text{P1}}$ corresponds to the most ``accurate'' encoding scheme with minimum cost, while $\overline{\text{P1}}$ corresponds to the most ``inaccurate'' encoding scheme with minimum cost. On the other hand, $\underline{\text{P2}}$ corresponds to the most ``cost-efficient'' encoding scheme that achieves the minimum distortion, while $\underline{\text{P3}}$ corresponds to the most ``cost-inefficient'' encoding scheme that achieves the minimum distortion.

Since minimizing the distortion is the main concern of semantic communication, we are particularly interested in the lower-envelope of the region, i.e., the curve connecting $\underline{\text{P1}}$, $\underline{\text{P2}}$, $\underline{\text{P3}}$, $\underline{\text{P4}}$ in Fig.~\ref{fig:region}. As with Shannon's rate-distortion theory, we define the distortion-cost function for semantic encoding.

\begin{defi}[The distortion-cost function of semantic encoding]
With semantic encoding, the minimum achievable semantic distortion under an average message cost $L$ is called the  distortion-cost function of semantic encoding and is given by 
\begin{equation}
D^*_{\bm{U,Q}}(L) = \min \left\{D:(L,D)\in R_{\text{enc}} \right\}.
\end{equation}
\end{defi}

As illustrated in Fig.~\ref{fig:region}, the distortion-cost function of semantic encoding consists of three segments in general, and we denote them by $(\underline{\text{P1}},\underline{\text{P2}})$, $(\underline{\text{P2}},\underline{\text{P3}})$, and $(\underline{\text{P3}},\underline{\text{P4}})$, respectively. In $(\underline{\text{P1}},\underline{\text{P2}})$, minimizing distortion and minimizing cost is a trade-off; in $(\underline{\text{P2}},\underline{\text{P3}})$, the minimum distortion is achieved; in $(\underline{\text{P3}},\underline{\text{P4}})$, minimizing distortion or cost boosts the other one.

Note that, in Shannon's rate-distortion theory \cite{berger2003},
\begin{enumerate}
\item A lower compression rate implies a larger distortion, and hence, there is a trade-off between distortion and rate. This corresponds to the segment $(\underline{\text{P1}},\underline{\text{P2}})$ in the distortion-cost function of semantic encoding.
\item If a rate-distortion pair $(\mathfrak{R},D)$ is achievable, then the rate-distortion pair $(\mathfrak{R}^\prime,D)$ is achievable for any $\mathfrak{R}^\prime>\mathfrak{R}$. Thus, the rate-distortion (or distortion-rate) function is convex and monotonic -- Shannon's rate-distortion function has no `boosting' segment as in $(\underline{\text{P3}},\underline{\text{P4}})$ in Fig.~\ref{fig:region}.
\end{enumerate}

By contrast, the distortion-cost function of semantic encoding is determined by the semantic language as well as the exact forms of semantic cost and distortion. There can be an additional boost segment $(\underline{\text{P3}},\underline{\text{P4}})$ because a larger semantic cost does not necessarily imply a smaller semantic distortion. For an intuitive example, consider the communication between a police officer and a suspect. The police officer wants to know the suspect's activities over a period. As the suspect's description increases, the police officer knows more and more about the suspect's activities, and the semantic distortion -- in which case the uncertainty of the suspect's activities -- decreases. However, when the suspect over-describes some details, the police may doubt the suspect's motive and the semantic distortion starts to increase. A more mathematical example will be given in Section \ref{sec:example}.

\begin{algorithm}[t]
\caption{Discovering the six subsets of messages for a meaning $w$.}
\label{algo:sixsets}
\begin{algorithmic}[1]
\State{\textbf{Input}: $w$; $\mathcal{S}$.}
\State{{\textbf{Output}: $\underline{\mathcal{S}}_w$ and $\overline{\mathcal{S}}_w$;}}
\algrule
\State{$\underline{\mathcal{S}}^\prime_w=\mathcal{S}$, $\underline{s}^l=s_1$; $\underline{\mathcal{S}}^{\prime\prime}_w=\mathcal{S}$, $\underline{s}^r=s_M$;}
\State{$\overline{\mathcal{S}}^\prime_w=\mathcal{S}$, $\overline{s}^l=s_1$;
$\overline{\mathcal{S}}^{\prime\prime}_w=\mathcal{S}$, $\overline{s}^r=s_M$;}
\For{$m=2,3,...,M$}
\If{$\varphi(w,\underline{s}^l)\leq \varphi(w,s_m)$}
\State{$\underline{\mathcal{S}}^\prime_w=\underline{\mathcal{S}}^\prime_w\backslash s_m$;}
\ElsIf{$\ell(\underline{s}^l)==\ell(s_m)$}
\State{$\underline{\mathcal{S}}^\prime_w=\underline{\mathcal{S}}^\prime_w\backslash \underline{s}^l$;}
\State{$\underline{s}^l=s_m$;}
\Else
\State{$\underline{s}^l=s_m$;}
\EndIf
\If{$\varphi(w,\underline{s}^r)\leq \varphi(w,s_{M-m+1})$}
\State{$\underline{\mathcal{S}}^{\prime\prime}_w=\underline{\mathcal{S}}^{\prime\prime}_w\backslash s_{M-m+1}$;}
\ElsIf{$\ell(\underline{s}^r)==\ell(s_{M-m+1})$}
\State{$\underline{\mathcal{S}}^{\prime\prime}_w=\underline{\mathcal{S}}^{\prime\prime}_w\backslash \underline{s}^r$;}
\State{$\underline{s}^r=s_{M-m+1}$;}
\Else
\State{$\underline{s}^r=s_{M-m+1}$;}
\EndIf
\If{$\varphi(w,\overline{s}^l)\leq \varphi(w,s_m)$}
\State{$\overline{\mathcal{S}}^\prime_w=\overline{\mathcal{S}}^\prime_w\backslash s_m$;}
\ElsIf{$\ell(\overline{s}^l)==\ell(s_m)$}
\State{$\overline{\mathcal{S}}^\prime_w=\overline{\mathcal{S}}^\prime_w\backslash \overline{s}^l$;}
\State{$\overline{s}^l=s_m$;}
\Else
\State{$\overline{s}^l=s_m$;}
\EndIf
\If{$\varphi(w,\overline{s}^r)\leq \varphi(w,s_{M-m+1})$}
\State{$\overline{\mathcal{S}}^{\prime\prime}_w=\overline{\mathcal{S}}^{\prime\prime}_w\backslash s_{M-m+1}$;}
\ElsIf{$\ell(\overline{s}^r)==\ell(s_{M-m+1})$}
\State{$\overline{\mathcal{S}}^{\prime\prime}_w=\overline{\mathcal{S}}^{\prime\prime}_w\backslash \overline{s}^r$;}
\State{$\overline{s}^r=s_{M-m+1}$;}
\Else
\State{$\overline{s}^r=s_{M-m+1}$;}
\EndIf
\EndFor
\State{$\underline{\mathcal{S}}_w=\underline{\mathcal{S}}^\prime_w \cup\underline{\mathcal{S}}^{\prime\prime}_w$, $\overline{\mathcal{S}}_w=\overline{\mathcal{S}}^\prime_w\cup\overline{\mathcal{S}}^{\prime\prime}_w$.}
\end{algorithmic}
\end{algorithm}

\subsection{Characterizing the distortion-cost region}
A semantic encoding scheme is a mapping from $\mathcal{W}$ to $\mathcal{S}$. To measure the expected distortion of mapping a meaning $w$ to a message $s$, we define
\begin{equation}
\varphi(w,s) \triangleq \sum_{\hat{s},\hat{w}} c(\hat{s}|s)q(\hat{w}|\hat{s})d(w,\hat{w}).
\end{equation}

The distortion of a semantic encoding scheme $\bm{U}$ can be written as
\begin{equation}
D_{\bm{U,Q}} = \sum_{w,s} p(w) u(s|w) \varphi(w,s). 
\end{equation}

To characterize the distortion-cost region for semantic encoding, we introduce a special class of semantic encoding schemes: deterministic semantic encoding.

\begin{defi}[Deterministic semantic encoding]
A semantic encoding scheme $\bm{U}$ is said to be deterministic if $\bm{U}$ maps each meaning to a single message deterministically. Each row of a deterministic encoding scheme $\bm{U}$ has only one non-zero element. We denote a deterministic semantic code $\bm{U}$ by $\bm{\Delta}_{i_1,i_2,...,i_N}$, where $i_1,i_2,...,i_N\in [M]$ are the column indexes of the non-zero elements in the rows of $\bm{U}$. That is,
\begin{equation}
u(s_m|w_j) = \begin{cases}
1, & \text{if}~m=i_j, \\
0, & \text{otherwise}.
\end{cases}
\end{equation}
\end{defi}

\begin{prop}\label{prop:timesharing}
A semantic distortion-cost pair $(L,D)$ can be achieved by a stochastic encoding scheme $\bm{U}$ if and only if $(L,D)$ can be achieved by time sharing among a set of deterministic encoding schemes in
\begin{equation*}
\Big\{\bm{\Delta}_{i_1,i_2,...,i_N}: \forall i_1,i_2,...,i_N\in [M] \Big\}.
\end{equation*}
\end{prop}

\begin{NewProof}
See Appendix \ref{sec:AppA}.
\end{NewProof}

Proposition~\ref{prop:timesharing} suggests two important properties of the distortion-cost region:
\begin{itemize}
    \item Any point in the distortion-cost region can be achieved by a set of deterministic semantic encoding schemes via time sharing.
    \item The distortion-cost region $R_{\text{enc}}$ is a convex set. The dis\-tor\-tion-cost function $D^*_{\bm{U,Q}}(L)$ is a convex function.
\end{itemize}

As a result, we only need to focus on the class of deterministic semantic encoding schemes to characterize the distortion-cost region.

\begin{defi}\label{defi:Sw}
For a meaning $w\in\mathcal{W}$, we define six subsets of $\mathcal{S}$: $\underline{\mathcal{S}}^\prime_w$,
$\underline{\mathcal{S}}^{\prime\prime}_w$, $\underline{\mathcal{S}}_w$,
$\overline{\mathcal{S}}^\prime_w$,
$\overline{\mathcal{S}}^{\prime\prime}_w$, $\overline{\mathcal{S}}_w\subseteq \mathcal{S}$:
\begin{enumerate}
\item $\underline{\mathcal{S}}^\prime_w$ is the smallest subset of $\mathcal{S}$ such that $\forall s\in\mathcal{S}$, there exists an $s^\prime\in \underline{\mathcal{S}}^\prime_w$ that satisfies
\begin{equation*}
\ell(s^\prime)\leq \ell(s),~~\varphi(w,s^\prime)\leq\varphi(w,s).
\end{equation*}
\item $\underline{\mathcal{S}}^{\prime\prime}_w$ is the smallest subset of $\mathcal{S}$ such that $\forall s\in\mathcal{S}$, there exists an $s^{\prime\prime}\in \underline{\mathcal{S}}^{\prime\prime}_w$ that satisfies
\begin{equation*}
\ell(s^{\prime\prime})\geq \ell(s),~~\varphi(w,s^{\prime\prime})\leq\varphi(w,s).
\end{equation*}
\item $\underline{\mathcal{S}}_w=\underline{\mathcal{S}}^\prime_w \cup\underline{\mathcal{S}}^{\prime\prime}_w $.
\item $\overline{\mathcal{S}}^\prime_w$ is the smallest subset of $\mathcal{S}$ such that $\forall s\in\mathcal{S}$, there exists an $s^\prime\in \overline{\mathcal{S}}^\prime_w$ that satisfies
\begin{equation*}
\ell(s^\prime)\leq \ell(s),~~\varphi(w,s^\prime)\geq\varphi(w,s).
\end{equation*}
\item $\overline{\mathcal{S}}^{\prime\prime}_w$ is the smallest subset of $\mathcal{S}$ such that $\forall s\in\mathcal{S}$, there exists an $s^{\prime\prime}\in \overline{\mathcal{S}}^{\prime\prime}_w$ that satisfies
\begin{equation*}
\ell(s^{\prime\prime})\geq \ell(s),~~\varphi(w,s^{\prime\prime})\geq\varphi(w,s).
\end{equation*}
\item $\overline{\mathcal{S}}_w=\overline{\mathcal{S}}^\prime_w \cup\overline{\mathcal{S}}^{\prime\prime}_w $.
\end{enumerate}
\end{defi}

For a $w\in\mathcal{W}$, the six subsets are defined based on the properties of $s$. Take the subset $\underline{\mathcal{S}}^\prime_w$ for example. Each of $s$ in $\underline{\mathcal{S}}^\prime_w$ has the property that there exists no $s^\prime\in\mathcal{S}\backslash\{s\}$ that is more accurate and cost-efficient than $s$ at the same time. It will be shown later that, {for each meaning $w$}, we only need to consider the messages in $\underline{\mathcal{S}}_w$ and $\overline{\mathcal{S}}_w$ to construct the deterministic encoding schemes and characterize the distortion-cost region of semantic encoding. The procedures to discover the six sets for a $w$ are described in Algorithm \ref{algo:sixsets}. The properties of the six subsets are summarized in Appendix \ref{sec:AppB}.

\begin{defi}\label{defi:G}
For a meaning $w\in\mathcal{W}$ and two messages $s,s^\prime\in\mathcal{S}$, $\ell(s)\neq\ell(s^\prime)$, we define a function
\begin{equation}
G(w,s,s^\prime)=\frac{\varphi(w,s^\prime)-\varphi(w,s)}{\ell(s^\prime)-\ell(s)}.
\end{equation}
\end{defi}
The function $G$ is important in that it guides us to discover a sequence of deterministic encoding schemes to characterize the boundary of the distortion-cost region. Theorem \ref{thm:region} below presents the main result of this section.

\begin{thm}[The distortion-cost region of semantic encoding]\label{thm:region}
Consider a semantic language $(\mathcal{W},\mathcal{S},\bm{P},\bm{Q})$, a semantic channel $\bm{C}$, a message cost function $\ell(s):\mathcal{S}\to\RR^+$, and a distortion measure $d(w,\hat{w}):\mathcal{W}\times\mathcal{W}\to\RR^+$. The boundary of the distortion-cost region of semantic encoding $R_{\text{enc}}$ is the piecewise linear connection of $\underline{T}+\overline{T}+2$ distortion-cost pairs $\left(L_{\underline{\bm{U}}^{(t)}},D_{\underline{\bm{U}}^{(t)},\bm{Q}} \right)$, $t=0,1,2,...,\underline{T}$ and $\left(L_{\overline{\bm{U}}^{(t)}},D_{\overline{\bm{U}}^{(t)},\bm{Q}} \right)$, $t=0,1,2,...,\overline{T}$, where $\left\{\underline{\bm{U}}^{(t)}:t=0,1,2,...,\underline{T}\right\}$ and $\left\{\overline{\bm{U}}^{(t)}:t=0,1,2,...,\overline{T}\right\}$ are deterministic semantic encoding schemes constructed in Algorithm~\ref{algo:region}. 
\end{thm}

\begin{NewProof}
See Appendix \ref{sec:AppC}.
\end{NewProof}

\begin{algorithm}[t]
\caption{Characterizing the distortion-cost region of semantic encoding $R_{\text{enc}}$.}
\label{algo:region}
\begin{algorithmic}[1]
\State{{\textbf{Input}}: $(\mathcal{W},\mathcal{S},\bm{P},\bm{Q})$, $\ell(s)$, $d(w,\widehat{w})$.}
\State{{\textbf{Output}}: $\left\{\underline{\bm{U}}^{(t)},\left(L_{\underline{\bm{U}}^{(t)}},D_{\underline{\bm{U}}^{(t)},\bm{Q}}\right):t=0,1,2,...,\underline{T}\right\}$;}
\State{\indent\indent$\left\{\overline{\bm{U}}^{(t)},\left(L_{\overline{\bm{U}}^{(t)}},D_{\overline{\bm{U}}^{(t)},\bm{Q}}\right):t=0,1,2,...,\overline{T}\right\}$.}
\algrule
\State{{\textbf{Step 1}}: Find $\underline{\mathcal{S}}_w$ and $\overline{\mathcal{S}}_w$ for each $w\in\mathcal{W}$.}
\For{$n=1,2,...,N$}
\State{$\underline{\mathcal{S}}_{w_n}, \overline{\mathcal{S}}_{w_n}=\textbf{Algorithm 1}(w_n,\mathcal{S})$.}
\EndFor
\State{{\textbf{Step 2}}: Construct a sequence of deterministic semantic source coding schemes $\underline{\bm{U}}^{(t)}=\bm{\Delta}_{\underline{i}_1^{(t)},\underline{i}_2^{(t)},...,\underline{i}_N^{(t)}}$, $t=0,1,2,...,\underline{T}$.}
\State{$t=0$;}
\State{Construct $\underline{\bm{U}}^{(0)}=\bm{\Delta}_{\underline{i}_1^{(0)},\underline{i}_2^{(0)},...,\underline{i}_N^{(0)}}$}, where $\underline{i}_n^{(0)}=\min\{m:s_m\in\underline{\mathcal{S}}_{w_n}\}$.
\While{$L_{\underline{\bm{U}}^{(t)}}<L_{\max}$}
\State{$t=t+1$;}
\State{$$\underline{n}^{(t)},\underline{m}^{(t)}=\argmin_{n,m:s_m\in\mathcal{S}_{w_n},m>\underline{i}^{(t-1)}_n}G\left(w_n,s_{\underline{i}^{(t-1)}_n},s_m\right).$$}
\State{Construct $\underline{\bm{U}}^{(t)}=\bm{\Delta}_{\underline{i}_1^{(t)},\underline{i}_2^{(t)},...,\underline{i}_N^{(t)}}$ such that
$$\underline{i}^{(t)}_n=\begin{cases}
\underline{i}^{(t-1)}_k, & \text{if}~n\neq \underline{n}^{(t)}; \\
\underline{m}^{(t)}, & \text{if}~n = \underline{n}^{(t)}.
\end{cases}$$}
\EndWhile
\State{$\underline{T}=t$.}
\State{{\textbf{Step 3}}: Construct a sequence of deterministic semantic source coding schemes $\overline{\bm{U}}^{(t)}=\bm{\Delta}_{\overline{i}_1^{(t)},\overline{i}_2^{(t)},...,\overline{i}_N^{(t)}}$, $t=0,1,2,...,\overline{T}$.}
\State{$t=0$;}
\State{Construct $\overline{\bm{U}}^{(0)}=\bm{\Delta}_{\overline{i}_1^{(0)},\overline{i}_2^{(0)},...,\overline{i}_N^{(0)}}$}, where $\overline{i}_n^{(0)}=\min\{m:s_m\in\overline{\mathcal{S}}_{w_n}\}$.
\While{$L_{\overline{\bm{U}}^{(t)}}<L_{\max}$}
\State{$t=t+1$;}
\State{$$\overline{n}^{(t)},\overline{m}^{(t)}=\argmax_{n,m:s_m\in\mathcal{S}_{w_n},m>\overline{i}^{(t-1)}_n}G\left(w_n,s_{\overline{i}^{(t-1)}_n},s_m\right).$$}
\State{Construct $\overline{\bm{U}}^{(t)}=\bm{\Delta}_{\overline{i}_1^{(t)},\overline{i}_2^{(t)},...,\overline{i}_N^{(t)}}$ such that
$$\overline{i}^{(t)}_n=\begin{cases}
\overline{i}^{(t-1)}_k, & \text{if}~n\neq \overline{n}^{(t)}; \\
\overline{m}^{(t)}, & \text{if}~n = \overline{n}^{(t)}.
\end{cases}$$}
\EndWhile
\State{$\overline{T}=t$.}
\end{algorithmic}
\end{algorithm}

Theorem~\ref{thm:region} states that the boundary of the distortion-cost region of semantic encoding is piecewise linear. The vertices of the boundary can be achieved by the deterministic encoding schemes constructed in Algorithm~\ref{algo:region}. Any non-vertex point on the boundary can be achieved by a stochastic encoding scheme or the time-sharing\footnote{Time sharing means that the system alternates between two deterministic encoding schemes over separate time intervals within the same communication session.} of two deterministic encoding schemes associated with the two closest vertices. The distortion-cost function of semantic encoding $D^*_{\bm{U,Q}}(L)$ is also piecewise linear and can be characterized by $\left\{\underline{\bm{U}}^{(t)}:t=0,1,2,...,\underline{T}\right\}$.

To conclude this section, we emphasize that the transmitter takes a proactive role in semantic communication, and hence, semantic encoding is often more efficient than semantic decoding as far as reducing distortion is concerned. In addition, the transmitter has the exact distribution of the intended meanings $p(w)$, while the receiver may only have an inaccurate prior $q(w)$. The semantic cost is fully determined by semantic encoding.

\section{Semantic Decoding and Combined Semantic Encoding and Decoding}\label{sec:dec}
In Section \ref{sec:dec}, we focused on semantic encoding from the transmitter's perspective. In this section, we study how to reduce semantic distortion from the receiver's perspective via semantic decoding. Furthermore, we discuss the combined semantic encoding and decoding in Section \ref{sec:decB}.

\subsection{Semantic decoding}\label{sec:decA}
In semantic decoding, the semantic encoder at the transmitter is dictated by the expression of the agreed language $\bm{P}$ and the receiver varies the mapping  $\bm{V}$ to optimize the semantic distortion
\begin{equation}
D_{\bm{P,V}}=\sum_{\hat{s},\hat{w}}v(\hat{w}|\hat{s})\psi_p(\hat{w},\hat{s}),
\end{equation}
where
\begin{equation}
\psi_p(\hat{w},\hat{s})\triangleq\sum_{w,s}p(w)p(s|w)c(\hat{s}|s)d(w,\hat{w}).
\end{equation}

We first characterize the semantic distortion-cost region that can be achieved by semantic decoding.
\begin{defi}[The distortion-cost region of semantic decoding]
A distortion-cost pair $(L,D)$ is achievable if there exists a semantic decoding scheme $\bm{V}$ such that $D_{\bm{P,V}}=D$ and $L_{\bm{P}}=L$. The distortion-cost region $R_{\text{dec}}$ is the set of all achievable distortion-cost pairs $(L,D)$.  
\end{defi}

Since the semantic encoder $\bm{P}$ is fixed, the achievable cost of $\bm{V}$ is also fixed, i.e.,
\begin{equation}
L_{\bm{P}} = \sum_w p(w)\sum_s p(w|s)\ell(s).
\end{equation}
Therefore, the distortion-cost region of semantic decoding is a vertical line in the distortion-cost space.

\begin{defi}[Deterministic semantic decoding]
A semantic decoding scheme $\bm{V}\in\RR^{M\times N}$ is said to be deterministic if $\bm{V}$ maps each message to a single meaning deterministically. We denote a deterministic semantic decoding scheme $\bm{V}$ by $\widetilde{\bm{\Delta}}_{n_1,n_2,...,n_M}$, where $n_1,n_2,...,n_M\in [N]$ are the column indexes of the non-zero elements in the rows of $\bm{V}$.
\end{defi}

\begin{thm}[The distortion-cost region of semantic decoding]\label{thm:decoding_region}
Consider a semantic language $(\mathcal{W},\mathcal{S},\bm{P},\bm{Q})$, a semantic channel $\bm{C}$, a message cost function $\ell(s):\mathcal{S}\to\RR^+$, and a distortion measure $d(w,\hat{w}):\mathcal{W}\times\mathcal{W}\to\RR^+$. The distortion-cost region of semantic decoding $R_{\text{dec}}$ is the vertical line between
\begin{equation*}
\Big(L_{\bm{P}}, D_{\bm{P},\widetilde{\bm{\Delta}}_{n^{\prime}_1,n^{\prime}_2,...,n^{\prime}_M}}\Big)~
\text{and}~
\Big(L_{\bm{P}}, D_{\bm{P},\widetilde{\bm{\Delta}}_{n^{\prime\prime}_1,n^{\prime\prime}_2,...,n^{\prime\prime}_M}}\Big),
\end{equation*}
where
\begin{eqnarray}
n^{\prime}_m \hspace{-0.2cm}&=&\hspace{-0.2cm} \argmin_n \psi_p(w_n,s_m), \\
n^{\prime\prime}_m \hspace{-0.2cm}&=&\hspace{-0.2cm} \argmax_n \psi_p(w_n,s_m).
\end{eqnarray}
\end{thm}

\begin{NewProof}
Omitted.
\end{NewProof}

The distortion-cost region of semantic decoding gives all achievable distortion-cost pairs with semantic decoding. However, unlike semantic encoding, the receiver may not have accurate prior information on $p(w)$. As a consequence, the receiver may not be able to find the optimal semantic decoding scheme $\widetilde{\bm{\Delta}}_{n^{\prime}_1,n^{\prime}_2,...,n^{\prime}_M}$ that minimizes the distortion.

Under a potentially inaccurate prior distribution $q(w)$, we are interested in the semantic decoding scheme employed by the receiver and the conditions under which this scheme is optimal. In the following, we shall use the subscripts `p' and `q' to denote the statistics obtained under prior distributions $p(w)$ and $q(w)$, respectively. For example, we have
\begin{equation}
\psi_q(\hat{w},\hat{s})\triangleq\sum_{w,s}q(w)p(s|w)c(\hat{s}|s)d(w,\hat{w}),
\end{equation}
under the prior $q(w)$.

\begin{prop}[Semantic decoding under inaccurate prior]\label{thm:dec}
Consider the expression $\bm{P}$ at the transmitter, a semantic channel $\bm{C}$, and a prior distribution $q(w)$ at the receiver. The best semantic decoding scheme perceived by the receiver is $\bm{V}^*_q = \widetilde{\bm{\Delta}}_{n^{(q)}_1,n^{(q)}_2,...,n^{(q)}_M}$, where
\begin{equation}
n^{(q)}_m = \argmin_n \psi_q(w_n,s_m).
\end{equation}
\end{prop}

To simplify notation, we denote by $\hat{w}_q(\hat{s})$ the meaning that a received message $\hat{s}$ will be mapped to under prior $q(w)$, i.e., $\hat{w}_q(\hat{s})=w_{n^{(q)}_m}$. The achieved distortion of $\bm{V}^*_q$ can be written as
\begin{equation}
D_{\bm{P},\bm{V}^*_q} = \sum_{\hat{s}}\psi_p(\hat{w}_q(\hat{s}),\hat{s}).
\end{equation}

On the other hand, if the true prior $p(w)$ is available at the receiver, the optimal decoding scheme is $\bm{V}^*_p=\widetilde{\bm{\Delta}}_{n^{\prime}_1,n^{\prime}_2,...,n^{\prime}_M}$ and the minimum distortion is
\begin{equation}
D_{\bm{P},\bm{V}^*_p} = \sum_{\hat{s}}\psi_p(\hat{w}_p(\hat{s}),\hat{s}).
\end{equation}

The condition under which semantic decoding achieves the optimal distortion $D_{\bm{P},\bm{V}^*_p}$ is non-trivial to establish for a general distortion function. In the following, we consider the Hamming distortion and study the condition under which $D_{\bm{P},\bm{V}^*_q}=D_{\bm{P},\bm{V}^*_p}$.

\begin{defi}[Hamming distortion]
Hamming distortion measures the probability of semantic error and is given by
\begin{equation}
d(w,\hat{w})=\begin{cases}
1, \text{if}~\hat{w}\neq w; \\
0, \text{if}~\hat{w}= w.
\end{cases}
\end{equation}
\end{defi}

\begin{prop}[Semantic decoding with the Hamming distortion]\label{thm:decHamming}
Consider semantic decoding with a prior $q(w)$ and Hamming distortion. Semantic decoding achieves the optimal distortion, i.e., $D_{\bm{P},\bm{V}^*_q}=D_{\bm{P},\bm{V}^*_p}$, if and only if
\begin{equation}\label{eq:deccondition}
\argmax_w q(w)p(\hat{s}|w) \subseteq \argmax_w p(w)p(\hat{s}|w),~\forall \hat{s},
\end{equation}
where $p(\hat{s}|w)\triangleq \sum_s p(s|w)c(\hat{s}|s)$.
\end{prop}

\begin{NewProof}
See Appendix \ref{sec:AppD}.
\end{NewProof}

\begin{figure}[t]
  \centering
  \includegraphics[width=0.9\linewidth]{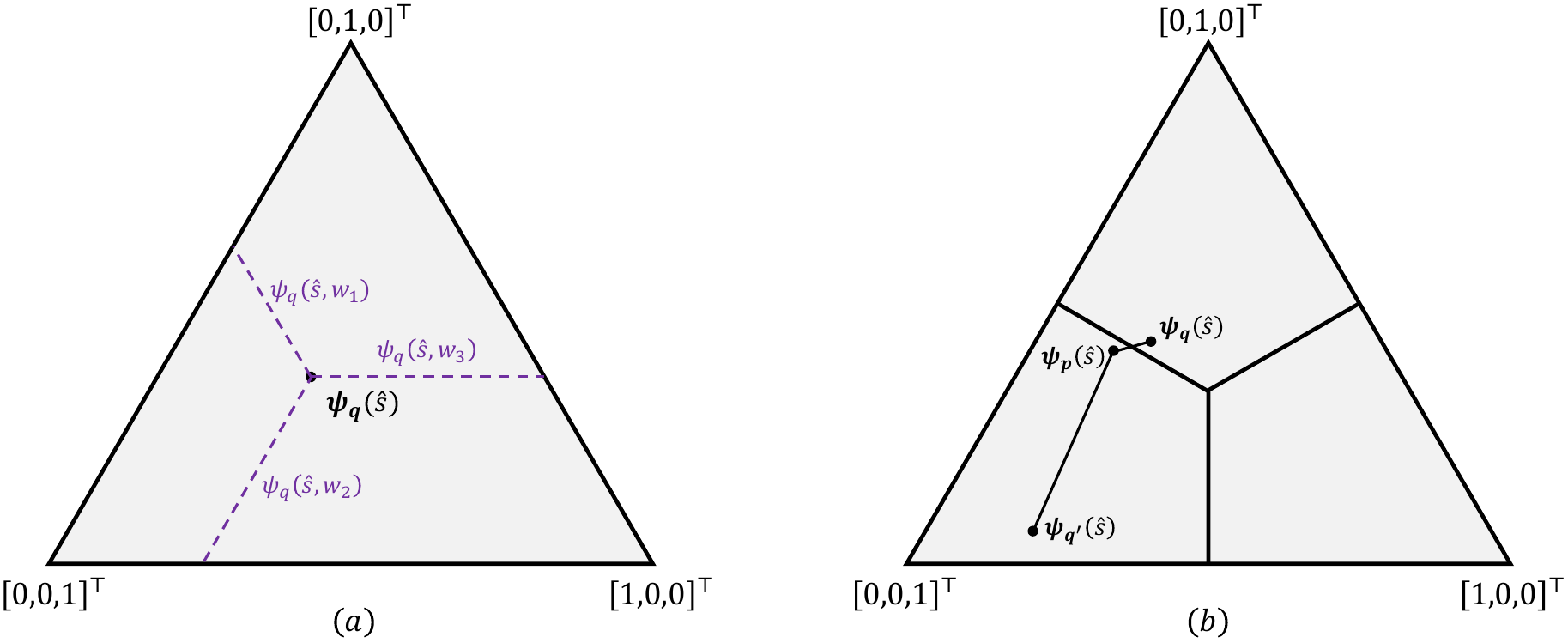}\\
  \caption{(a) Probability space in which $\bm{\alpha}_q(\hat{s})$ resides is an equilateral triangle when $N=3$. (b) We partition the equilateral triangle into three quadrilaterals. For a received message $\hat{s}$, the Hamming distortion of semantic decoding is determined by whether $\bm{\alpha}_p(\hat{s})$ and $\bm{\alpha}_q(\hat{s})$ fall into the same quadrilateral.}
\label{fig:geo}
\end{figure}

A geometric interpretation of \eqref{eq:deccondition} is given as follows. First, we normalize $q(w)p(\hat{s}|w)$ to a distribution $\bm{\alpha}_q(\hat{s})$ with each element being
\begin{equation}
\alpha_q(\hat{s},w)=\frac{q(w)p(\hat{s}|w)}{\sum_w q(w)p(\hat{s}|w)},~\forall w\in\mathcal{W}.
\end{equation}
The probability space in which $\bm{\alpha}_q(\hat{s})$ resides can be represented as a unit simplex. When $N=3$, for example, the probability space is an equilateral triangle, as shown in Fig.~\ref{fig:geo}(a). Based on the semantic decoding rule, we partition the probability space into $N$ regions, corresponding to the $N$ meanings in $\mathcal{W}$. When $N=3$, the equilateral triangle can be partitioned into three quadrilaterals, as shown in Fig.~\ref{fig:geo}(b). Under the prior $p(w)$ and $q(w)$, semantic decoding yields $M$ points $\left\{\bm{\alpha}_p(\hat{s}):\hat{s}\in\mathcal{S}\right\}$ and $\left\{\bm{\alpha}_q(\hat{s}):\hat{s}\in\mathcal{S}\right\}$, respectively.

In the above context, the geometric interpretation of \eqref{eq:deccondition} is that $\bm{\alpha}_p(\hat{s})$ and $\bm{\alpha}_q(\hat{s})$ fall into the same region of the probability space, $\forall \hat{s}\in\mathcal{S}$. That is, under the Hamming distortion, the optimality of semantic decoding is governed by a ``partition distance'' between $\bm{\alpha}_p(\hat{s})$ and $\bm{\alpha}_q(\hat{s})$, i.e., whether they fall into the same region, but not the conventional distance measure between two probability distributions, such as the L1 or Kullback-Leibler (KL) distance.

Proposition~\ref{thm:decHamming} suggests that the condition that semantic decoding achieves the optimal distortion can be very demanding. In actuality, semantic decoding $\bm{V}$ can yield even larger distortion than the original interpretation of the language $\bm{Q}$. This can be easily understood as the inaccurate prior information $q(w)$ may mislead the receiver. A real-life example of this phenomenon is that when you talk to people with prejudice, they immediately sound harsh, even if they do not mean it that way.

Considering the Hamming distortion, the original interpretation of the language $\bm{Q}$ yields a distortion of
\begin{equation*}
D_{\bm{P},\bm{Q}}=\sum_{\hat{s},\hat{w}}q(\hat{w}|\hat{s})\psi_p(\hat{w},\hat{s})
= 1 - \sum_{\hat{s},\hat{w}}p(\hat{w})p(\hat{s}|\hat{w})q(\hat{w}|\hat{s}).
\end{equation*}

The gap between $D_{\bm{P},\bm{Q}}$ and $D_{\bm{P},\bm{V}^*_q}$ can be written as
\begin{eqnarray}
&&\hspace{-0.5cm} D_{\bm{P},\bm{Q}}-D_{\bm{P},\bm{V}^*_q}= \\
&&\hspace{-0.5cm} \sum_{\hat{s}}\left(p\big(\hat{w}_q(\hat{s})\big)p\big(\hat{s}|\hat{w}_q(\hat{s})\big) - \sum_{\hat{w}}p(\hat{w})p(\hat{s}|\hat{w})q(\hat{w}|\hat{s}) \right).  \nonumber
\end{eqnarray}

If the prior information is accurate, i.e., $p(w)=q(w)$, we have
\begin{eqnarray}
&&\hspace{-0.5cm} D_{\bm{P},\bm{Q}}-D_{\bm{P},\bm{V}^*_q}= \\
&&\hspace{-0.5cm} \sum_{\hat{s}}\left(\max_w p(w)p(\hat{s}|w) - \sum_{\hat{w}}p(\hat{w})p(\hat{s}|\hat{w})q(\hat{w}|\hat{s}) \right)\geq 0.  \nonumber
\end{eqnarray}
and hence, $D_{\bm{P},\bm{V}^*_q}$ is strictly better than $D_{\bm{P},\bm{Q}}$. In general, however, $D_{\bm{P},\bm{Q}}-D_{\bm{P},\bm{V}^*_q}$ can be negative -- the wrong prior leads to a semantic decoding scheme $\bm{V}$ worse than $\bm{Q}$ and it is better to stick to the original interpretation of the language. In this case, we propose two refinements on $\bm{Q}$ below to construct an improved random semantic decoding scheme $\bm{V}$.

Considering a received message $\hat{s}$, we define two subsets $\widehat{\mathcal{W}}(\hat{s})\subseteq \mathcal{W}$ and $\mathcal{W}(\hat{s})\subseteq \mathcal{W}$, where
\begin{eqnarray}
\widehat{\mathcal{W}}(\hat{s}) \hspace{-0.2cm}& \triangleq &\hspace{-0.2cm} \big\{w: q(\hat{w}|\hat{s})>0 \big\}, \\
\mathcal{W}(\hat{s}) \hspace{-0.2cm}& \triangleq &\hspace{-0.2cm} \big\{w: p(s|w)c(\hat{s}|s)>0 \big\}.
\end{eqnarray}

\begin{prop}[Refining the interpretation]
A random semantic decoding scheme $\bm{V}$ with $D_{\bm{P},\bm{V}}\leq D_{\bm{P},\bm{Q}}$ can be constructed by refining the interpretation $\bm{Q}$ as follows. First, let $\bm{V=Q}$.

Refinement 1: If $\exists~w^{\prime}\in \widehat{\mathcal{W}}(\hat{s})$, s.t.,
\begin{equation}
    \min_{w\in\mathcal{W}(\hat{s})} d(w,w^{\prime}) \geq
    \max_{\substack{w\in\mathcal{W}(\hat{s}),\\ \widetilde{w}\in \widehat{\mathcal{W}}(\hat{s})\backslash\{w^{\prime}\}}} d(w,\widetilde{w}),
\end{equation}
we construct
\begin{equation}\label{eq:refine1}
v(w|\hat{s}) = \begin{cases}
0, & \text{if}~w=w^{\prime}, \\
\frac{q(w|\hat{s})}{1- q(w^{\prime}|\hat{s})}, & \text{if}~w\neq w^{\prime}.
\end{cases}
\end{equation}

Refinement 2: If $\exists~w^{\prime\prime}\in\widehat{\mathcal{W}}(\hat{s})$, s.t.,
\begin{equation}
    \max_{w\in\mathcal{W}(\hat{s})} d(w,w^{\prime\prime}) \leq
    \min_{\substack{w\in\mathcal{W}(\hat{s}),\\ \widetilde{w}\in \widehat{\mathcal{W}}(\hat{s})\backslash\{w^{\prime\prime}\}}} d(w,\widetilde{w}),
\end{equation}
we construct
\begin{equation}\label{eq:refine2}
v(w|\hat{s}) = \begin{cases}
1, & \text{if}~w=w^{\prime\prime}, \\
0, & \text{if}~w\neq w^{\prime\prime}.
\end{cases}
\end{equation}
\end{prop}

In the first refinement \eqref{eq:refine1}, there exists a $w^{\prime}$ that is strictly worse than any other $\widetilde{w}\in \widehat{\mathcal{W}}(\hat{s})$. This is possible because the interpretation of the language can be suboptimal for some distortion functions. In this case, we remove the mapping between $\hat{s}$ and $w^{\prime}$ and assign the probability $q(w^{\prime}|\hat{s})$ proportionally to other $\widetilde{w}\in \widehat{\mathcal{W}}(\hat{s})$. In the second refinement \eqref{eq:refine2}, there exists a $w^{\prime\prime}$ that is strictly better than any other $\widetilde{w}\in \widehat{\mathcal{W}}(\hat{s})$. As a result, we construct $v(w|\hat{s})$ such that $\hat{s}$ is deterministically decoded to $w^{\prime\prime}$ regardless of $q(w)$.


\subsection{Combined semantic encoding and decoding (CSED)}\label{sec:decB}
In the problem of language utilization, semantic encoding and decoding are two decoupled processes since no negotiation is allowed between the transmitter and receiver.  We have shown in Sections \ref{sec:enc} and \ref{sec:decA} that the semantic distortion can be reduced by either semantic encoding or decoding. A natural question arises: what if the transmitter performs semantic encoding and the receiver performs semantic decoding simultaneously?

To answer this question, let us first define combined semantic encoding and decoding (CSED) formally.

\begin{defi}[CSED]\label{defi:CSED}
In CSED, the transmitter adopts a semantic encoding scheme
\begin{equation}
\bm{U} = \sum_{i=0}^{\underline{T}}
\xi_i \underline{\bm{U}}^{(i)}
+ \sum_{i=\underline{T}+1}^{\underline{T}+\overline{T}+2}
\xi_{i-\underline{T}-1} \overline{\bm{U}}^{(i-\underline{T}-1)},
\end{equation}
where $\bm{\xi}\triangleq\{\xi_i\geq 0:\sum_{i=0}^{\underline{T}+\overline{T}+2}\xi_i=1\}$ is a probability distribution; $\left\{\underline{\bm{U}}^{(i)}:i=0,1,2,...,\underline{T}\right\}$ and $\left\{\overline{\bm{U}}^{(i)}:i=0,1,2,...,\overline{T}\right\}$ are deterministic semantic encoding schemes derived from Theorem~\ref{thm:region}. 
The receiver performs semantic decoding and obtains $\bm{V}^*_q=\widetilde{\bm{\Delta}}_{n^{(q)}_1,n^{(q)}_2,...,n^{(q)}_M}$ following Proposition \ref{thm:dec}. The semantic distortion and cost of CSED are given by
\begin{equation}
D_{\bm{U},\bm{V}^*_q}=\sum_{w,s,\hat{s},\hat{w}}p(w)u(s|w)c(\hat{s}|s)v(\hat{w}|\hat{s})d(w,\hat{w}).
\end{equation}
\begin{equation}
L_{\bm{U}}=\sum_{w,s}p(w)u(s|w)\ell(s).
\end{equation}
\end{defi}

CSED leads to a new distortion-cost region.

\begin{thm}[The distortion-cost region of CSED]\label{thm:CSEDregion}
The distortion-cost region of CSED $R_{\text{csed}}$, defined as the set of all achievable distortion-cost pairs $(L,D)$ via CSED, is a convex hull of the set of points:
\begin{equation*}
\left\{L_{\underline{\bm{U}}^{(i)}}, D_{\underline{\bm{U}}^{(i)},\bm{V}^*_q}:i=0,1,2,...,\underline{T} \right\},
\end{equation*}
\begin{equation*}
\left\{L_{\overline{\bm{U}}^{(i)}}, D_{\overline{\bm{U}}^{(i)},\bm{V}^*_q}:i=0,1,2,...,\overline{T} \right\}.
\end{equation*}
\end{thm}

\begin{NewProof}
Theorem \ref{thm:CSEDregion} follows from the time sharing argument that allows the system to interpolate among the performance characteristics of different deterministic encoding schemes, thereby enabling any point within the convex hull to be reachable.
\end{NewProof}

Correspondingly, we can define the distortion-cost function of CSED as the infimum of $R_{\text{csed}}$:
\begin{equation}
D^*_{\bm{U},\bm{V}^*_q}(L) \triangleq\inf \left\{D: (L,D)\in   R_{\text{csed}} \right\}.
\end{equation}

Given an average message cost $L$, the minimum distortion achieved by semantic encoding is $D^*_{\bm{U},\bm{Q}}(L)$. An interesting question is  whether $D^*_{\bm{U},\bm{V}^*_q}(L)$ is smaller than $D^*_{\bm{U},\bm{Q}}(L)$? In addition, when $L=L_{\bm{P}}$, whether $D^*_{\bm{U},\bm{V}^*_q}(L_{\bm{P}})$ is smaller than $D_{\bm{P},\bm{V}^*_q}$ or $D^*_{\bm{U},\bm{Q}}(L_{\bm{P}})$?
The answers to these questions are, unfortunately, negative. This can be easily understood via a counterexample.

\begin{figure}[t]
  \centering
  \includegraphics[width=0.5\linewidth]{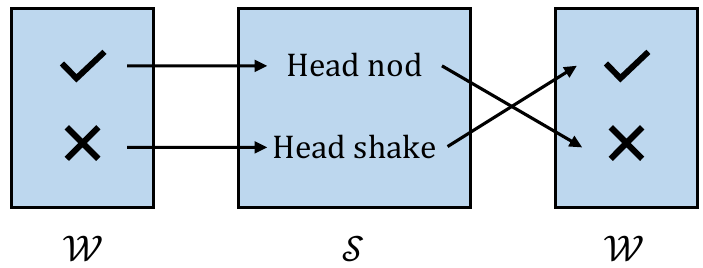}\\
  \caption{The nod-shake example. Both semantic encoding and semantic decoding is better than CSED.}
\label{fig:nodshake}
\end{figure}

\begin{example}[The node-shake example]
Suppose the meaning set is $\mathcal{W}=\{\text{yes},\text{no}\}$, the message set is $\mathcal{S}=\{\text{head nod},\text{head shake}\}$, and the costs of the two messages are the same. The expression and interpretation of the semantic language are quite opposite, as shown in Fig.~\ref{fig:nodshake}. At the transmitter, ``head nod'' stands for ``yes'' and ``head shake'' stands for ``no'', while at the receiver, ``head nod'' stands for ``no'' and ``head shake'' stands for ``yes''. Suppose the prior distribution $p(w)$ is known at the receiver and we consider the Hamming distortion.
\end{example}

In this example, the Hamming distortion can be reduced to $0$ with either semantic encoding or semantic decoding. However, when both semantic encoding and decoding are performed, the Hamming distortion is $1$. This suggests that CSED does not necessarily improve upon separate semantic encoding or decoding.

In what follows, we consider an error-free semantic channel and establish sufficient conditions under which CSED is better than semantic encoding and decoding.

\begin{thm}[CSED with an error-free semantic channel]\label{thm:csed}
Consider a semantic language $(\mathcal{W},\mathcal{S},\bm{P},\bm{Q})$, an error-free semantic channel $\bm{C}$, and a symmetric distortion measure $d(w,\hat{w})=d(\hat{w},w)$.

Under the conditions that
\begin{enumerate}
\item The receiver has perfect prior information, i.e., $p(w)=q(w)$.
\item The language is logically self-consistent, i.e., $q(w|s)=\frac{p(s|w)p(w)}{p(s)}$.
\item $\forall w,w^\prime\in\mathcal{W}$,
\begin{equation}
    \argmin_s \varphi(w,s)\cap\argmin_s \varphi(w^\prime,s)=\emptyset.
\end{equation}
\item $\forall s,s^\prime\in\big\{s:u(s|w)>0,s\in\mathcal{S},w\in\mathcal{W} \big\}$,
\begin{equation}
    \argmin_w \psi_q(w,s)\cap\argmin_w \psi_q(w,s^\prime)=\emptyset.
\end{equation}
\end{enumerate}

We have
\begin{equation}\label{eq:CSED}
    D^*_{\bm{U},\bm{V}^*_q}(L)=\min_{\bm{U}^\prime,\bm{V}^\prime} D_{\bm{U}^\prime,\bm{V}^\prime}(L),
\end{equation}
where $\bm{U}^\prime,\bm{V}^\prime$ can be any semantic encoding and decoding schemes. In this case, CSED is better than both semantic encoding and semantic decoding, i.e., $D^*_{\bm{U},\bm{V}^*_q}(L)\leq D^*_{\bm{U},\bm{Q}}(L)$ and $D^*_{\bm{U},\bm{V}^*_q}(L_{\bm{P}})\leq D_{\bm{P},\bm{V}^*_q}$.
\end{thm}

\begin{NewProof}
See Appendix \ref{sec:AppE}.
\end{NewProof}

\begin{rem}
The CSED scheme discussed in Theorem \ref{defi:CSED} consists of the semantic encoding scheme $\bm{U}$ and decoding scheme $\bm{V}$. In particular, $\bm{U}$ is optimized based on the interpretation $\bm{Q}$ of the language, and $\bm{V}$ is optimized based on the expression $\bm{P}$ of the language.

We have shown in the node-shake example that this can lead to a poorer performance than semantic encoding or decoding. However, what if the transmitter is clever enough and predicts that the receiver would improve its interpretation from $\bm{Q}$ to $\bm{V}$? In this case, the transmitter can optimize $\bm{P}$ to $\bm{U}^\prime$ based on $\bm{V}$, as opposed to $\bm{Q}$. In the node-shake example, $\bm{U}^\prime$ and $\bm{V}$ would lead to zero Hamming distortion.

Following this idea, the receiver can further predict that the transmitter would improve its expression from $\bm{P}$ to $\bm{U}^\prime$ and optimize $\bm{Q}$ based on $\bm{U}^\prime$ instead of $\bm{Q}$, and so on. The resulting Hamming distortion would alternate between $0$ and $1$, depending on how many times the transmitter and receiver have thought about each other's encoding and decoding schemes. This phenomenon resembles human communication -- two smart people may not be able to communicate efficiently as they would keep inferring each other's expression and interpretation and changing theirs continuously.
\end{rem}

\section{Semantic Communication: An Example}\label{sec:example}
This section gives a concrete example to illustrate semantic communication formulated in this paper. Based on the agreed semantic and technical languages, we show how semantic encoding, semantic decoding, and CSED operate, respectively, to reduce semantic distortion.

\subsection{The grid world and languages}

\begin{figure}[t]
  \centering
  \includegraphics[width=0.3\linewidth]{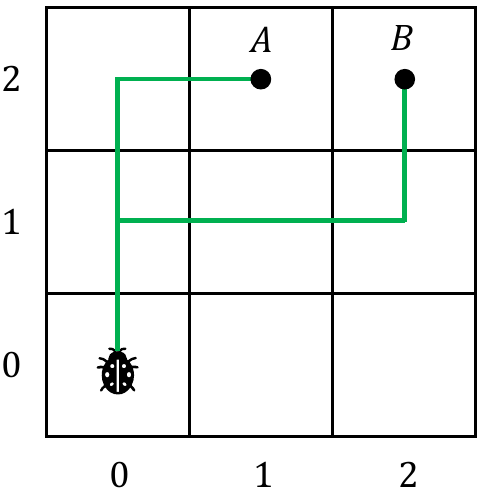}\\
  \caption{A bug walking in the grid world.}
\label{fig:exp3}
\end{figure}

\begin{example}[Grid world]
We consider a world of square grids, as shown in Fig.~\ref{fig:exp3}.
\begin{itemize}
\item The grids are indexed by $(i,j)$, $i,j\in\{0,1,2\}$. 
\item A bug sets out from $(0,0)$ and walks towards one of the two destinations A: $(1, 2)$ and B: $(2, 2)$. 
\item At each grid, the bug can only go up or right.
\item There is a transmitter and a receiver. Only the transmitter observes the trajectory of the bug.
\end{itemize}
\end{example}

The transmitter and receiver have agreed on semantic language and technical language.
\subsubsection{The message and meaning sets}
There are two words `$\mathcal{U}$' and `$\mathcal{R}$' in the semantic language, representing the actions of the bug (i.e., ``up'' and ``right''). A message is composed of a sequence of words, representing the sequential actions (i.e., a trajectory) of the bug starting from $(0, 0)$. The message `$\mathcal{URR}$', for example, means that the bug goes up, right, and right from $(0, 0)$. The maximal number of moves in the grid world is $4$; hence, the number of possible messages is $\sum_{i=0}^{4}2^i=31$. The message set can be written as
\begin{eqnarray*}
&&\hspace{-0.65cm}\mathcal{S}_{\text{all}}=\big\{\emptyset, \mathcal{U},\mathcal{R},
\mathcal{UU},\mathcal{UR},\mathcal{RU},\mathcal{RR},
\mathcal{UUU},\mathcal{UUR},\mathcal{URU}, \\
&&\hspace{-0.65cm}
\mathcal{URR}, \mathcal{RUU},\mathcal{RUR},\mathcal{RRU},\mathcal{RRR},
\mathcal{UUUU},\mathcal{UUUR},\mathcal{UURU}, \\
&&\hspace{-0.65cm} \mathcal{UURR},\mathcal{URUU},\mathcal{URUR},\mathcal{URRU},\mathcal{URRR},\mathcal{RUUU},\mathcal{RUUR},\\
&&\hspace{-0.65cm} \mathcal{RURU},\mathcal{RURR},\mathcal{RRUU},\mathcal{RRUR},\mathcal{RRRU},\mathcal{RRRR}
\big\}
\end{eqnarray*}
where $\emptyset$ means no message is transmitted.

Given the message set, the transmitter can convey various meanings such as the trajectory and the destination of the bug. In this example, we assume the receiver is only interested in the destination of the bug and define the meaning set as $\mathcal{W}=\{A,B\}$ and $p(A)=1/3$, $p(B)=2/3$. We assume the receiver has a prior distribution $q(A)=1/2$ and $q(B)=1/2$.

\subsubsection{Error-free technical communication}
For a given message, the technical language maps the words `$\mathcal{U}$' and `$\mathcal{R}$' to bit sequences `$0$' and `$10$', respectively. The set of channel symbols can be constructed by substituting `$\mathcal{U}$' and `$\mathcal{R}$' in $\mathcal{S}_{\text{all}}$ with `$0$' and `$10$', respectively.

The mappings between the messages and channel symbols are one-to-one. Throughout this section, we assume that the technical channel is error-free, and hence, the semantic channel is also error-free -- a transmitted message can be perfectly received by the receiver. The generalization to an error-prone technical channel is straightforward.

We consider the Hamming distortion. The cost function $\ell(s)$ is defined as the length of the bit sequence associated with $s$.

\subsubsection{Expression and interpretation}
\begin{figure}[t]
  \centering
  \includegraphics[width=0.9\linewidth]{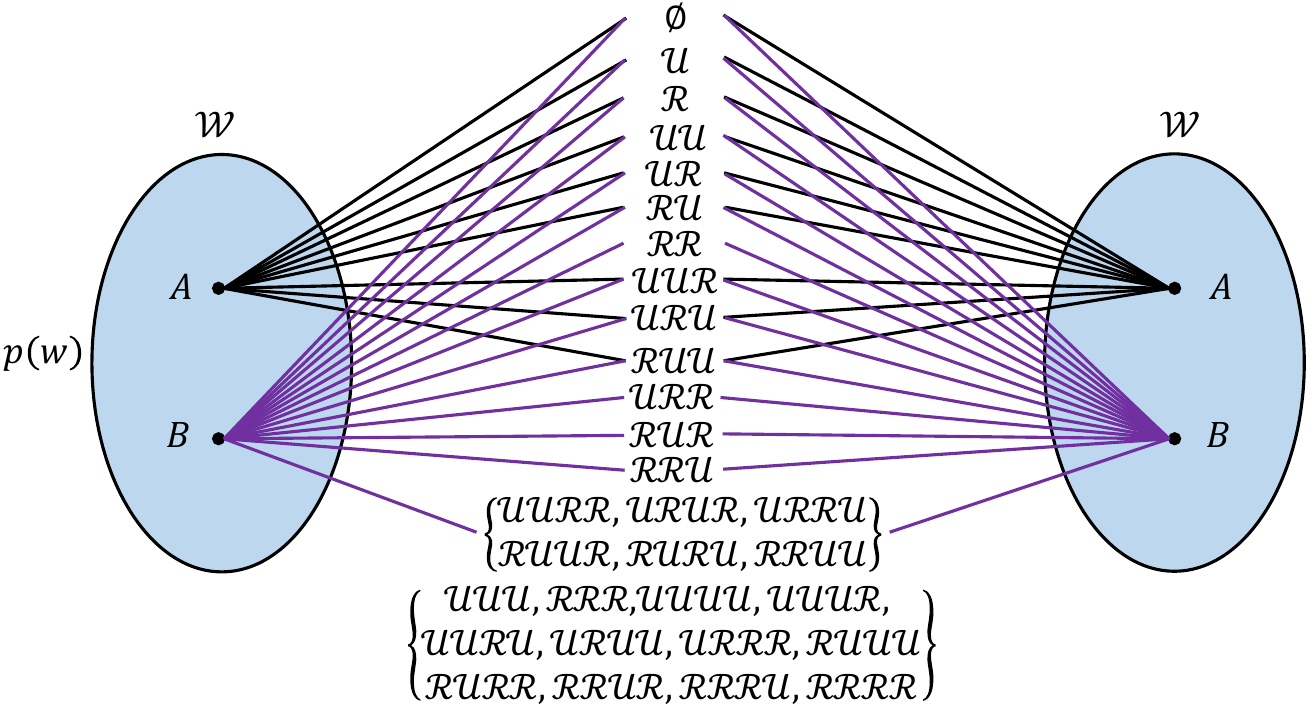}\\
  \caption{The expression and interpretation of the semantic language.}
\label{fig:exp3_mapping}
\end{figure}

Given the message and meaning sets, we illustrate the expression and interpretation of the semantic language in Fig.~\ref{fig:exp3_mapping}. As can be seen, to convey meanings in $\mathcal{W}$, the class of messages
\begin{eqnarray*}
&&\hspace{-0.5cm}\mathcal{S}_{\text{null}}=\big\{\mathcal{UUU},\mathcal{RRR}, \mathcal{UUUU},\mathcal{UUUR},\mathcal{UURU},\mathcal{URUU}, \\
&&\hspace{-0.5cm} \mathcal{URRR},\mathcal{RUUU},\mathcal{RURR},\mathcal{RRUR},\mathcal{RRRU},\mathcal{RRRR}
\big\}
\end{eqnarray*}
are illegitimate, and the class of messages 
\begin{eqnarray*}
\big\{\mathcal{UURR},\mathcal{URUR}, \mathcal{URRU},\mathcal{RUUR},\mathcal{RURU}, \mathcal{RRUU}
\big\}
\end{eqnarray*}
are equivalent as they have the same cost and lead to the same distortion. We can use a single message $\mathcal{UURR}$ to represent them in order to simplify notation. This gives us a refined message set 
\begin{eqnarray*}
&&\hspace{-0.5cm}\mathcal{S}=\mathcal{S}_{\text{all}}\backslash\mathcal{S}_{\text{null}}=\big\{\emptyset, \mathcal{U},\mathcal{R},
\mathcal{UU},\mathcal{UR},\mathcal{RU},\mathcal{RR},
\mathcal{UUR},\mathcal{URU}, \\
&&\hspace{-0.5cm}
\mathcal{RUU},\mathcal{URR},\mathcal{RUR},\mathcal{RRU},
\mathcal{UURR}
\big\},
\end{eqnarray*}
as shown in Fig.~\ref{fig:exp3_mapping}.
Note that we have arranged them in ascending order according to the message cost.
Next, we define the interpretation and expression, respectively.

Denoting by $\phi_A$ and $\phi_B$ the sets of trajectories that lead to $A$ and $B$, respectively, we have
\begin{equation*}
    \phi_A=\big\{\mathcal{UUR}, \mathcal{URU},\mathcal{RUU}  \big\},
\end{equation*}
\begin{equation*}
    \phi_B=\big\{\mathcal{UURR}, \mathcal{URUR},\mathcal{URRU} ,\mathcal{RRUU} ,\mathcal{RURU} ,\mathcal{RUUR}  \big\}.
\end{equation*}

The interpretation of the semantic language is defined as
\begin{equation}
q(\hat{w}=w|\hat{s})=\frac{\sum_{s\in\phi_w}\hat{s}\sqsubseteq s}{\sum_{w\in\mathcal{W}}\sum_{s\in\phi_w}\hat{s}\sqsubseteq s}
\end{equation}
where $\hat{s}\sqsubseteq s=1$ if $\hat{s}$ is a prefix of $s$, and $\hat{s}\sqsubseteq s=0$, otherwise. For example, suppose the received message $\hat{s}=\mathcal{U}$, we have $q(\hat{w}=A|\mathcal{U})=2/5$ because $\mathcal{U}$ is a prefix of two messages in $\phi_A$ and three messages in $\phi_B$. A complete list of $q(w|s)$ is given in Table~\ref{tab:language}.

\begin{table}[t]
\caption{Expression and interpretation of the semantic language.}
\setlength{\tabcolsep}{3.5mm} 
\label{tab:language}
\centering
\begin{tabular}{@{}cccccc@{}}
\toprule
$s$                             & $q(A|s)$ & $q(B|s)$ & $p(s|A)$ & $p(s|B)$ & $\ell(s)$ \\
\midrule
$\emptyset$      & $1/3$              & $2/3$              & $1/9$              & $1/19$             & $0$         \\
$\mathcal{U}$    & $2/5$              & $3/5$              & $1/9$              & $1/19$             & $1$         \\
$\mathcal{R}$    & $1/4$              & $3/4$              & $1/9$              & $1/19$             & $2$         \\
$\mathcal{UU}$   & $1/2$              & $1/2$              & $1/9$              & $1/19$             & $2$         \\
$\mathcal{UR}$   & $1/3$              & $2/3$              & $1/9$              & $1/19$             & $3$         \\
$\mathcal{RU}$   & $1/3$              & $2/3$              & $1/9$              & $1/19$             & $3$        \\
$\mathcal{RR}$   & $0$                & $1$                & $0$                & $1/19$             & $4$         \\
$\mathcal{UUR}$  & $1/2$              & $1/2$              & $1/9$              & $1/19$             & $4$         \\
$\mathcal{URU}$  & $1/2$              & $1/2$              & $1/9$              & $1/19$             & $4$         \\
$\mathcal{RUU}$  & $1/2$              & $1/2$              & $1/9$              & $1/19$             & $4$         \\
$\mathcal{URR}$  & $0$                & $1$                & $0$                & $1/19$             & $5$         \\
$\mathcal{RUR}$  & $0$                & $1$                & $0$                & $1/19$             & $5$         \\
$\mathcal{RRU}$  & $0$                & $1$                & $0$                & $1/19$             & $5$         \\
$\mathcal{UURR}$ & $0$                & $1$                & $0$                & $6/19$             & $6$   \\
\bottomrule
\end{tabular}
\end{table}

Correspondingly, for a meaning $w$, we define the expression of the language $p(s|w)$ as the uniform distribution over $\big\{s:q(w|s)>0\big\}$. Table~\ref{tab:language} lists $p(s|w)$, $\forall w,s$.

\subsubsection{Semantic encoding}
This subsection studies how to reduce semantic distortion from the transmitter's perspective via semantic encoding. We shall follow Theorem \ref{thm:region} and Algorithm \ref{algo:region} to characterize the semantic distortion-cost region with semantic encoding.

For an error-free semantic channel and Hamming distortion measure, the expected distortion of mapping a meaning $w$ to a message $s$ can be written as
\begin{equation*}
\varphi(w,s)=\sum_{\hat{s}}c(\hat{s}|s)\sum_{\hat{w}}q(\hat{w}|\hat{s})d(w,\hat{w})=1-q(\hat{w}=w|s).
\end{equation*}
A list of $\varphi(w,s)$ is given in Table~\ref{tab:phi}

\begin{table}[t]
\caption{$\varphi(w,s)$ and $\alpha_q(w,s)$.}
\setlength{\tabcolsep}{3.5mm} 
\label{tab:phi}
\centering
\begin{tabular}{@{}cccccc@{}}
\toprule
$s$              & $\varphi(A,s)$     & $\varphi(B,s)$    & $\alpha_q(A,s)$ & $\alpha_q(B,s)$  \\
\midrule
$\emptyset$      & $2/3$              & $1/3$              & $19/28$              & $9/28$           \\
$\mathcal{U}$    & $3/5$              & $2/5$              & $19/28$              & $9/28$            \\
$\mathcal{R}$    & $3/4$              & $1/4$              & $19/28$              & $9/28$         \\
$\mathcal{UU}$   & $1/2$              & $1/2$              & $19/28$              & $9/28$        \\
$\mathcal{UR}$   & $2/3$              & $1/3$              & $19/28$              & $9/28$            \\
$\mathcal{RU}$   & $2/3$              & $1/3$              & $19/28$              & $9/28$             \\
$\mathcal{RR}$   & $1$                & $0$                & $0$                  & $1$                  \\
$\mathcal{UUR}$  & $1/2$              & $1/2$              & $19/28$              & $9/28$                \\
$\mathcal{URU}$  & $1/2$              & $1/2$              & $19/28$              & $9/28$                \\
$\mathcal{RUU}$  & $1/2$              & $1/2$              & $19/28$              & $9/28$              \\
$\mathcal{URR}$  & $1$                & $0$                & $0$                & $1$               \\
$\mathcal{RUR}$  & $1$                & $0$                & $0$                & $1$                \\
$\mathcal{RRU}$  & $1$                & $0$                & $0$                & $1$              \\
$\mathcal{UURR}$ & $1$                & $0$                & $0$                & $1$       \\
\bottomrule
\end{tabular}
\end{table}

To characterize the semantic distortion-cost region, the first step is to construct the six subsets for $w\in\mathcal{W}$. According to Algorithm \ref{algo:sixsets}, the six subsets are given by
\begin{eqnarray*}
&&\hspace{-0.5cm} \underline{\mathcal{S}}^\prime(A) = \big\{\emptyset,\mathcal{U},\mathcal{UU}\big\},~~~
\underline{\mathcal{S}}^{\prime\prime}(A) = \big\{\mathcal{UUR},\mathcal{UURR}\big\}, \\
&&\hspace{-0.5cm} \underline{\mathcal{S}}(A) = \big\{\emptyset,\mathcal{U},\mathcal{UU},\mathcal{UUR},\mathcal{UURR}\big\}, \\
&&\hspace{-0.5cm} \overline{\mathcal{S}}^\prime(A) = \big\{\emptyset, \mathcal{RR}\big\},~~~~~
\overline{\mathcal{S}}^{\prime\prime}(A) = \big\{\mathcal{UURR}\big\},        \\
&&\hspace{-0.5cm} \overline{\mathcal{S}}(A)  = \big\{\emptyset, \mathcal{RR},\mathcal{UURR}\big\},      \\
&&\hspace{-0.5cm} \underline{\mathcal{S}}^\prime(B) = \big\{\emptyset, \mathcal{RR}\big\}, ~~~~~
\underline{\mathcal{S}}^{\prime\prime}(B) = \big\{\mathcal{UURR}\big\},       \\
&&\hspace{-0.5cm} \underline{\mathcal{S}}(B)   = \big\{\emptyset, \mathcal{RR},\mathcal{UURR}\big\},       \\
&&\hspace{-0.5cm} \overline{\mathcal{S}}^\prime(B) = \big\{\emptyset,\mathcal{U},\mathcal{UU}\big\},~~~
\overline{\mathcal{S}}^{\prime\prime}(B) = \big\{\mathcal{UUR},\mathcal{UURR}\big\},       \\
&&\hspace{-0.5cm} \overline{\mathcal{S}}(B)= \big\{\emptyset,\mathcal{U},\mathcal{UU},\mathcal{UUR},\mathcal{UURR}\big\},
\end{eqnarray*}

The second step constructs a sequence of deterministic encoding schemes that characterize the distortion-cost function. To simplify notation, we denote a deterministic encoding scheme by a set with two elements in this example, where the two elements correspond to the messages that $A$ and $B$ are mapped to. First, we construct $\underline{\bm{U}}^{(0)}=\big\{\emptyset,\emptyset \big\}$, where
\begin{eqnarray*}
D_{\underline{\bm{U}}^{(0)},\bm{Q}}\hspace{-0.2cm}&=&\hspace{-0.2cm} p(A)\varphi(A,\emptyset)+p(B)\varphi(B,\emptyset) = \frac{4}{9}, \\
L_{\underline{\bm{U}}^{(0)}}\hspace{-0.2cm}&=&\hspace{-0.2cm} p(A)\ell(\emptyset)+p(B)\ell(\emptyset) = 0.
\end{eqnarray*}

To construct $\underline{\bm{U}}^{(1)}$, we first compute the $G$ function and choose $\underline{w}^{(1)}$ and $\underline{s}^{(1)}$ that minimizes the $G$ function. In this case, we have
\begin{eqnarray*}
&&\hspace{-0.5cm} G(w=A,\emptyset,\mathcal{U}) =  -\frac{1}{15}, \\
&&\hspace{-0.5cm} G(w=A,\emptyset,\mathcal{UU}) =  -\frac{1}{12}, \\
&&\hspace{-0.5cm} G(w=B,\emptyset,\mathcal{RR}) =  -\frac{1}{12}.
\end{eqnarray*}
Therefore, we choose
\begin{equation}\label{eq:firstselection}
\underline{w}^{(1)}=A, ~~
\underline{s}^{(1)} = \mathcal{UU}, ~~
\underline{U}^{(1)} = \big\{\mathcal{UU},\emptyset\big\},
\end{equation}
and
\begin{eqnarray*}
D_{\underline{\bm{U}}^{(1)},\bm{Q}}\hspace{-0.2cm}&=&\hspace{-0.2cm} p(A)\varphi(A,\mathcal{UU})+p(B)\varphi(B,\emptyset) = \frac{7}{18}, \\
L_{\underline{\bm{U}}^{(1)}}\hspace{-0.2cm}&=&\hspace{-0.2cm} p(A)\ell(\mathcal{UU})+p(B)\ell(\emptyset) = \frac{2}{3}.
\end{eqnarray*}

\begin{figure*}[t]
  \centering
  \includegraphics[width=0.9\linewidth]{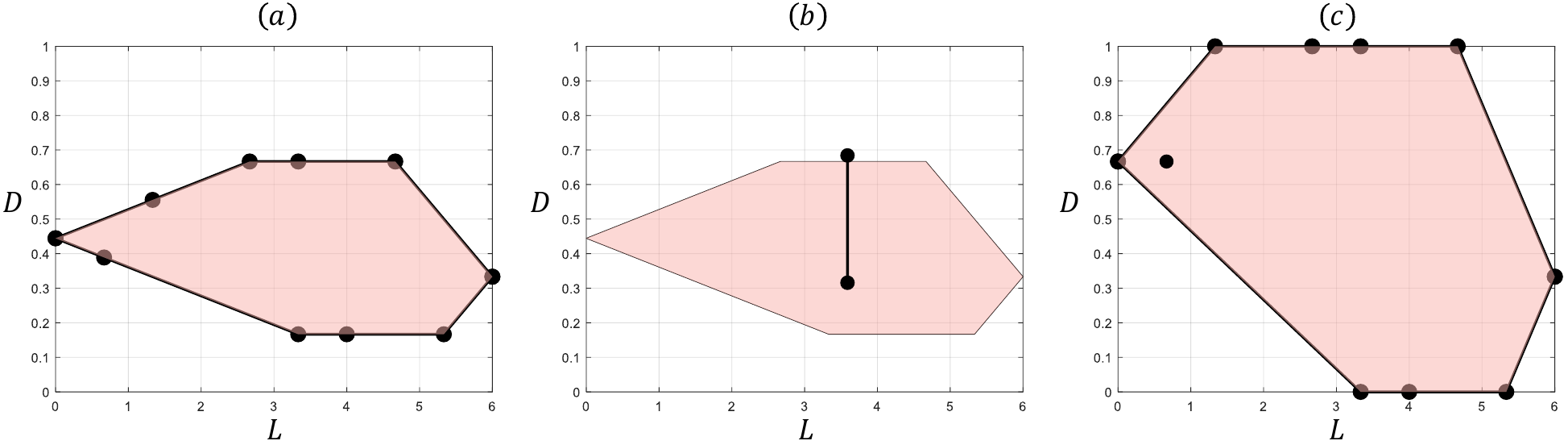}\\
  \caption{The distortion-cost regions of (a) semantic encoding $R_{\text{enc}}$, (b) semantic decoding $R_{\text{dec}}$, and (c) CSED $R_{\text{csed}}$.}
\label{fig:exp3_region}
\end{figure*}

Likewise, we can derive all the deterministic encoding schemes and the achieved distortion and cost. Here, we omit all the derivations and list the resulting encoding schemes in Table~\ref{tab:exp3_enc}, based on which the distortion-cost region of semantic encoding $R_{\text{enc}}$ is characterized in Fig.~\ref{fig:exp3_region}(a).

\begin{table}[]
\caption{The discovered encoding schemes.}
\setlength{\tabcolsep}{7mm} 
\label{tab:exp3_enc}
\centering
\begin{tabular}{lcc}
\toprule
Encoding schemes & $D$ & $L$ \\
\midrule
$\underline{U}^{(0)}=\big\{\emptyset,\emptyset\big\}$           & $4/9$  & $0$  \\
$\underline{U}^{(1)}=\big\{\mathcal{UU},\emptyset\big\}$        & $7/18$ & $2/3$  \\
$\underline{U}^{(2)}=\big\{\mathcal{UU},\mathcal{RR}\big\}$     & $1/6$  & $10/3$  \\
$\underline{U}^{(3)}=\big\{\mathcal{UUR},\mathcal{RR}\big\}$    & $1/6$  & $4$  \\
$\underline{U}^{(4)}=\big\{\mathcal{UUR},\mathcal{UURR}\big\}$  & $1/6$  & $16/3$  \\
$\underline{U}^{(5)}=\big\{\mathcal{UURR},\mathcal{UURR}\big\}$ & $1/3$  & $6$  \\
$\overline{U}^{(0)}=\big\{\emptyset,\emptyset\big\}$           & $4/9$  & $0$  \\
$\overline{U}^{(1)}=\big\{\mathcal{RR},\emptyset\big\}$        & $5/9$  & $4/3$ \\
$\overline{U}^{(2)}=\big\{\mathcal{RR},\mathcal{UU}\big\}$     & $2/3$  & $8/3$  \\
$\overline{U}^{(3)}=\big\{\mathcal{UURR},\mathcal{UU}\big\}$   & $2/3$  & $10/3$ \\
$\overline{U}^{(4)}=\big\{\mathcal{UURR},\mathcal{UUR}\big\}$  & $2/3$  & $14/3$ \\
$\overline{U}^{(5)}=\big\{\mathcal{UURR},\mathcal{UURR}\big\}$ & $1/3$  & $6$  \\
\bottomrule
\end{tabular}
\end{table}

\subsubsection{Semantic decoding and CSED}
Next, we study how to reduce the distortion from the receiver's perspective via semantic decoding. 
We first characterize the distortion-cost region of semantic decoding.

Under the Hamming distortion, we have $\psi_p(\hat{w},s)=p(s) - p(\hat{w})p(s|\hat{w})$. The semantic decoding schemes $\widetilde{\bm{\Delta}}_{n^{\prime}_1,n^{\prime}_2,...,n^{\prime}_M}$ and $\widetilde{\bm{\Delta}}_{n^{\prime\prime}_1,n^{\prime\prime}_2,...,n^{\prime\prime}_M}$ can be derived from Theorem \ref{thm:decoding_region}, as listed in Table \ref{tab:dec}, and
\begin{eqnarray*}
D_{\widetilde{\bm{\Delta}}_{n^{\prime}_1,n^{\prime}_2,...,n^{\prime}_M}} \approx 0.3158,&&\hspace{-0.4cm} D_{\widetilde{\bm{\Delta}}_{n^{\prime\prime}_1,n^{\prime\prime}_2,...,n^{\prime\prime}_M}} \approx 0.6842,  \\
&&\hspace{-1.5cm} L_{\bm{P}} \approx 3.5887.
\end{eqnarray*}

The distortion-cost region of semantic decoding $R_{\text{dec}}$ is plotted in Fig.~\ref{fig:exp3_region}(b).
In this example, the receiver has an inaccurate prior distribution. According to Proposition \ref{thm:dec}, we derive the optimal decoding scheme perceived by the receiver $\bm{V}^*_q$ in Table~\ref{tab:dec}. As can be seen, $\bm{V}^*_q$ matches $\widetilde{\bm{\Delta}}_{n^{\prime}_1,n^{\prime}_2,...,n^{\prime}_M}$. This is because Proposition \ref{thm:decHamming} is satisfied. In this case, semantic decoding with the inaccurate prior achieves the optimal distortion, i.e.,
\begin{equation*}
D_{\bm{P},\bm{V}^*_q}\approx 0.3158,
\end{equation*}
as shown in Fig.~\ref{fig:exp3_region}(b). It is worth noting that, in this example, the original interpretation of the language $\bm{Q}$ yields a distortion of
\begin{equation*}
D_{\bm{P},\bm{Q}}
= 1 - \sum_{s,w}p(w)p(s|\hat{w})q(w|s)
\approx 0.3262.
\end{equation*}

\begin{table}[t]
\caption{Semantic decoding schemes.}
\setlength{\tabcolsep}{3.5mm} 
\label{tab:dec}
\centering
\begin{tabular}{@{}cccccc@{}}
\toprule
$s$                             & $\widetilde{\bm{\Delta}}_{n^{\prime}_1,n^{\prime}_2,...,n^{\prime}_M}$ & $\widetilde{\bm{\Delta}}_{n^{\prime\prime}_1,n^{\prime\prime}_2,...,n^{\prime\prime}_M}$ & $\bm{V}^*_q$  \\
\midrule
$\emptyset$      & A & B & A                   \\
$\mathcal{U}$    & A & B & A                  \\
$\mathcal{R}$    & A & B & A                   \\
$\mathcal{UU}$   & A & B & A                   \\
$\mathcal{UR}$   & A & B & A                   \\
$\mathcal{RU}$   & A & B & A                  \\
$\mathcal{RR}$   & B & A & B                   \\
$\mathcal{UUR}$  & A & B & A                  \\
$\mathcal{URU}$  & A & B & A                  \\
$\mathcal{RUU}$  & A & B & A                  \\
$\mathcal{URR}$  & B & A & B                  \\
$\mathcal{RUR}$  & B & A & B                    \\
$\mathcal{RRU}$  & B & A & B                     \\
$\mathcal{UURR}$ & B & A & B               \\
\bottomrule
\end{tabular}
\end{table}

Finally, when combining both semantic encoding and decoding, the new distortion-cost region of CSED $R_{\text{csed}}$ can be obtained from Theorem \ref{thm:CSEDregion}, and we plot it in Fig.~\ref{fig:exp3_region}(c).

\begin{rem}[Impact of encoding scheme selection on the distortion-cost region of CSED]
When characterizing the distortion-cost function of semantic encoding, Algorithm \ref{algo:region} sequentially constructs $\underline{T}+1$ encoding schemes $\big\{\underline{\bm{U}}^{(t)}\big\}$. At a step $t$, it is possible to find several pairs of $\big(\underline{n}^{(t)},\underline{m}^{(t)}\big)$ that are equally optimal as they yield the minimum $G$ (see line 12 of Algorithm \ref{algo:region}). Likewise, when generating $\overline{\bm{U}}^{(t)}$, it is possible to find several pairs of $\big(\overline{n}^{(t)},\overline{m}^{(t)}\big)$ that are equally optimal as they yield the maximum $G$ (see line 20 of Algorithm \ref{algo:region}). Although these equally optimal pairs correspond to different encoding schemes, they finally yield the same distortion-cost region for semantic encoding. Therefore, we can randomly sample one from them in Algorithm \ref{algo:region}.
However, the selection does have an impact on the distortion-cost region of CSED.
According to Definition \ref{defi:CSED} and Theorem \ref{thm:CSEDregion}, CSED and its distortion-cost region are determined by the chosen semantic encoding schemes. Selecting different semantic encoding schemes in Algorithm \ref{algo:region} leads to different distortion-cost regions for CSED. For example, if we choose
\begin{equation}
\underline{w}^{(1)}=B, ~~
\underline{s}^{(1)} = \mathcal{RR}, ~~
\underline{U}^{(1)} = \big\{\emptyset,\mathcal{RR}\big\},
\end{equation}
in \eqref{eq:firstselection}, the distortion-cost pair $\left(\frac{2}{3},\frac{2}{3}\right)$ will change to $\left(\frac{2}{3},0\right)$, leading to reduced semantic distortion.
\end{rem}

\section{Extension}\label{sec:ext}
Our formulation of semantic communication in this paper can be extended in many ways. In the following, we highlight a few interesting problems worth further investigation.

\subsection{A technical comparison between language utilization and the rate-distortion theory}
Language utilization (semantic communication) and the rate-distortion theory (technical communication) have different roles in a broad communication process, as illustrated in Fig. \ref{fig:1}, but they can be compared from a technical perspective. The rate-distortion theory studies how to transmit a sequence of i.i.d. random variables $X_1$, $X_2$, …, $X_n$, $n\to\infty$, in a lossy fashion \cite{berger2003,coverBook}. To be comparable with language utilization,
\begin{enumerate}
\item We define a message set $\mathcal{S}$ that contains only typical sequences: each message $s\in\mathcal{S}$ is a typical sequence of infinite length and the cardinality of $\mathcal{S}$ is $2^{nH(X_n)}$.
\item We treat ``messages’’ and ``channel symbols’’ in the rate-distortion theory as ``meanings’’ and ``messages’’ in semantic communication, respectively. The semantic cost is the average length of channel symbols.
\end{enumerate}

In the above context, the rate-distortion theory provides the optimal design of semantic language (i.e., the mapping between meanings and messages) to trade-off distortion and cost, provided that codebook negotiation before transmission is allowed. In comparison, language utilization is about ``mismatch’’. For example, the codebooks possessed by the transmitter and receiver can mismatch, or the codebooks are not optimized for the optimal transmission of $\mathcal{S}$. In other words, language utilization exploits a semantic language that can be used for transmitting a variety of semantic sources robustly. When transmitting a specific kind of source, language utilization saves the cost of codebook negotiation, despite its suboptimal performance when compared with the rate-distortion theory (which tailors a codebook for the source).  An interesting problem is characterizing the trade-off between the suboptimality of language utilization and the additional cost one is willing to pay for codebook negotiation.

\subsection{Multiple Transmissions}
In the main body of this paper, we have primarily addressed the scenario of one-shot transmission from a transmitter to a receiver. An important and natural extension of this work involves considering multiple transmissions, which introduces additional complexity and opportunities for enhancing semantic communication. 
\begin{enumerate}
    \item Transmission of a single meaning. When transmitting a single meaning via multiple uses of the channel, the choice of semantic encoding strategy becomes crucial. One approach is repetition coding, where the same semantic message is encoded and transmitted multiple times. Alternatively, varying the encoding scheme across different channel uses may enhance robustness against noise and interference. The optimal strategy likely depends on the channel characteristics and the specific requirements for semantic fidelity. On the other hand, for decoding the transmitted meaning, several joint decoding strategies can be considered. Techniques such as majority voting or maximum ratio combining offer potential pathways to accurately infer the intended meaning by aggregating information across multiple transmissions. The effectiveness of each decoding approach will depend on factors such as channel reliability and the correlation between channel uses.
    \item Transmission of infinite meanings. With an infinite number of meanings to be transmitted, multiple transmissions provide a significant advantage by allowing the receiver to gather $p(\hat{s})$ for the received symbols, thereby facilitating the estimation of $p(s)$ and $p(w)$ via either direct inversion or regularization techniques, such as LASSO and Tikhonov regularization. In this scenario:
    i) Following the semantic decoding approach outlined in Proposition \ref{thm:dec}, we can achieve the optimal distortion $D_{\bm{P},\bm{V}^*_p}$, optimizing the fidelity of communicated meanings.
    ii) The receiver can leverage the concept of typicality in the transmitted sequence \cite{coverBook}, focusing on decoding among sequences that are ``typical'', given the observed transmissions. This approach takes advantage of statistical regularities in the sequence of transmitted meanings, enhancing decoding efficiency and accuracy.
\end{enumerate}


\subsection{Interaction-based transmission}
In addition to the feedforward link from the transmitter to the receiver, suppose there is a feedback link from the receiver to the transmitter. The feedback can be active, for which the receiver proactively feeds back some information to the transmitter.
The feedback can also be passive, for which the transmitter senses the environment (e.g., the change of the receiver’s behavior or the reflected waveform from the receiver) to collect feedback information.

Under this setup, the transmitter and receiver can communicate in an interactive fashion \cite{Schalkwijk:TIT:66,AttentionCode,Viswanathan:TIT:99,deepiot}.
\begin{enumerate}
\item  Feedback-aided transmission. The feedback from the receiver can be used to reduce the semantic distortion and cost for the feedforward transmission. This is more relevant to human conversations in real life. Consider that Einstein tries to explain the theory of relativity to a non-physicist. It is very likely that the listener does not understand anything after one-shot transmission. Nevertheless, through interactions, Einstein knows the concepts that the listener is confused about and explains these concepts in layman's terms. Finally, the receiver is able to understand the theory of relativity to some extent.

\item  Semantic communication with no language agreement. In the presence of the feedback link, communication is possible even if the transmitter and receiver have no agreed language, e.g., the transmitter does not know the interpretation of the receiver and the receiver does not know the expression of the transmitter. This can model the hypothetical situation of trying to converse with an alien.  Through interactions, they can gradually learn each other’s language. In particular, they can even form a new language through  interactions and it is interesting to characterize the expressibility and interpretability of the learned language as a form of the semantic capacity in conveying meanings.

\item  Effective and goal-oriented communication. Effective communication deals with the problem that what meaning to generate for the optimal operation of the cooperative task, considering the current states of the transmitter, receiver, and the progress of the task \cite{Tung:JSAC:21, Tung:Asilomar:21, gunduz2022survey}. In this sense, interaction-based communication is a natural fit for effective communication (and goal-oriented communication) as the transmitter has to know the current states of the transmitter, receiver, and the progress of the task through feedback and sensing.
\end{enumerate}

\subsection{Network semantic communication}
Beyond link-level communication, semantic communication can be extended to communication networks, such as broadcast channels, multiple-access channels, wiretap channels, relay channels, and mesh networks, wherein a wide range of challenging but intriguing problems can be foreseen. For example, in the case of a broadcast channel, the encoder may want to transmit a certain meaning to set of receivers, each with a different mapping to interpret its received message. This can model the scenario for a teacher teaching to a classroom of students with different levels, or a politician giving a speech to a large number of citizens from different backgrounds. In a network setting, we can also consider the extension of the feedback model in the previous subsection to a two-way communication system \cite{Shannnon:CP:93, Zhang:TIT:86}, where two agents try to communicate with each other simultaneously, with no or partial knowledge about each other's assumed languages.

\section{Conclusion}\label{sec:con}
This paper introduced a new framework for understanding semantic communication. We provided clarity on the essence of semantic communication and posed two fundamental challenges: language exploitation and language design. Notably, the language design problem aligns with the principles of the classical joint source-channel coding, where trade-offs between data transmission rates and versatile distortion metrics are of paramount importance. In contrast, this paper was dedicated to unraveling the intricacies of the language exploitation problem, which extends beyond the classical information theory.

The key distinction between language exploitation and established information theory lies in the existence of an undesignable, yet universally agreed-upon, semantic language shared among communication participants. Language exploitation leverages this common semantic ground to reduce misinterpretation during communication. To achieve this, we explored three key approaches: semantic encoding (pertaining to encoding intended meanings at the transmitter), semantic decoding (pertaining to reconstructing meaning from received messages), and the combined semantic encoding and decoding (CSED). Throughout the exploration, we meticulously characterized the distortion-cost region attainable through these strategies.

It is crucial to highlight that the notion of semantic language extends far beyond human languages or animal sounds and gestures. In a broader context, it encompasses any physical phenomena, ranging from seismic waves and solar flares to material responses to external stimuli. The process of language exploitation can be applied universally, as it involves the development of systems to harness, adapt to, or emulate these natural phenomena. Such processes include, for example, interacting with generative models like ChatGPT, simulating an individual's painting style, predicting earthquakes based on seismic wave properties, or creating anthropomorphic robots, to name a few. 

\appendices

\section{Proof of Proposition~\ref{prop:timesharing}}\label{sec:AppA}
We first prove the ``if" part: for any stochastic encoding scheme $\bm{U}$, there are a series of deterministic encoding schemes $\bm{U}_1,\bm{U}_2,...$ such that 
\begin{equation*}
	L_{\bm{U}}=\sum_i \lambda_i L_{\bm{U}_i},~~
	D_{\bm{U},\bm{Q}}=\sum_i \lambda_i D_{\bm{U}_i,\bm{Q}},
\end{equation*} 
where $\lambda_i\in[0,1]$ and $\sum_i \lambda_i=1$.
We rewrite $\bm{U}$ as
\begin{equation*}
    \bm{U}=\sum_{i_1,i_2,...,i_N} u(s_{i_1}|w_1)u(s_{i_2}|w_2)...u(s_{i_N}|w_N) {\bm{\Delta}}_{i_1,i_2,...,i_N}.
\end{equation*}

Let $\delta_{i_1,i_2,...,i_N}(s_m|w_n)$ be the element of ${\bm{\Delta}}_{i_1,i_2,...,i_N}$ on the $n$-th row and $m$-th column. We have
\begin{eqnarray*}
&&\hspace{-0.65cm} D_{\bm{U},\bm{Q}} =\sum_n p(w_n)\sum_{\hat{w}} d(w_n,\hat{w})\sum_m u(s_m|w_n)q(\hat{w}|s_m)\\
&&\hspace{-0.65cm}  =\sum_{i_1,i_2,...,i_N} u(s_{i_1}|w_1)u(s_{i_2}|w_2)...u(s_{i_N}|w_N)\\
&&\hspace{-0.65cm}  \times \sum_n p(w_n)\sum_{\hat{w}} d(w_n,\hat{w})\sum_m \delta_{i_1,i_2,...,i_N}(s_m|w_n)q(\hat{w}|s_m)\\
&&\hspace{-0.65cm} = \hspace{-0.2cm} \sum_{i_1,i_2,...,i_N} \hspace{-0.2cm}u(s_{i_1}|w_1)u(s_{i_2}|w_2)...u(s_{i_N}|w_N)D_{{\bm{\Delta}}_{i_1,i_2,...,i_N},\bm{Q}}.
\end{eqnarray*}
and 
\begin{eqnarray*}
&&\hspace{-0.65cm}  L_{\bm{U}} =\sum_n p(w_n)\sum_m u(s_m|w_n)\ell(s_m)\\
&&\hspace{-0.65cm} =\sum_{i_1,i_2,...,i_N} u(s_{i_1}|w_1)u(s_{i_2}|w_2)...u(s_{i_N}|w_N)\\
&&\hspace{-0.65cm} \times \sum_n p(w_n)\sum_m \delta_{i_1,i_2,...,i_N}(s_m|w_n)\ell(s_m)\\
&&\hspace{-0.65cm} =\sum_{i_1,i_2,...,i_N} \hspace{-0.2cm} u(s_{i_1}|w_1)u(s_{i_2}|w_2)...u(s_{i_N}|w_N)L_{{\bm{\Delta}}_{i_1,i_2,...,i_N}}.
\end{eqnarray*}
Thus, the distortion-cost pair $(L_{\bm{U}}, D_{\bm{U},\bm{Q}})$ can be achieved by deterministic encoding schemes in 
\begin{equation*}
\Big\{\bm{\Delta}_{i_1,i_2,...,i_N}:
        \forall i_1,i_2,...,i_N\in [M]\Big\}
\end{equation*}
via time sharing.

Now we prove the ``only if" part. Let
\begin{equation*}
\begin{split}
    \bm{A}&=\{{\bm{\Delta}}_1,{\bm{\Delta}}_2,...,{\bm{\Delta}}_{|\bm{A}|}\}\\
    &\subseteq \{{\bm{\Delta}}_{i_1,i_2,...,i_N}:
    \forall i_1,i_2,...,i_N\in [M]]\}
\end{split}
\end{equation*}
be an arbitrary but fixed set.
We define a distribution $(\lambda_1,\allowbreak\lambda_2,\allowbreak...,\lambda_{|\bm{A}|})$, where $\lambda_i\in[0,1]$ and $\sum_{i=1}^{|\bm{A}|} \lambda_i=1$.
Denote
\begin{equation*}
	L=\sum_{i=1}^{|\bm{A}|} \lambda_i L_{{\bm{\Delta}}_i},~~
	D=\sum_{i=1}^{|\bm{A}|} \lambda_i D_{{\bm{\Delta}}_i,\bm{Q}}.
\end{equation*}
Now we construct a stochastic encoding scheme $\bm{U}$ such that
 $L=L_{\bm{U}}$ and $D=D_{\bm{U},\bm{Q}}$.
By letting
\begin{equation}
    u(s|w)\triangleq \sum_{i=1}^{|\bm{A}|} \lambda_i{\bm{\Delta}}_i(s|w),
\end{equation}
for any $w,s$, we have $u(s|w)\ge 0$ and 
\begin{eqnarray*}
\sum_s u(s|w) \hspace{-0.2cm} & = &\hspace{-0.2cm} \sum_s\sum_{i=1}^{|\bm{A}|}\lambda_i{\bm{\Delta}}_i(s|w)=\sum_{i=1}^{|\bm{A}|}\lambda_i\sum_s{\bm{\Delta}}_i(s|w)\\
\hspace{-0.2cm} & = &\hspace{-0.2cm}
\sum_{i=1}^{|\bm{A}|}\lambda_i\times 1
    =1.
\end{eqnarray*}
Therefore, $\bm{U}$ is a legitimate encoding scheme.
Further,
\begin{equation*}
	\begin{split}
		L_{\bm{U}}
		&=\sum_n p(w_n)\sum_m u(s_m|w_n)\ell(s_m)\\
		&=\sum_n p(w_n)\sum_m \sum_{i=1}^{|\bm{A}|} \lambda_i{\bm{\Delta}}_i(s|w)\ell(s_m)\\
		&=\sum_{i=1}^{|\bm{A}|}\lambda_i \sum_n p(w_n)\sum_m  {\bm{\Delta}}_i(s|w)\ell(s_m)\\
		&=\sum_{i=1}^{|\bm{A}|}\lambda_i L_{{\bm{\Delta}}_i}=L\\
	\end{split}
\end{equation*}
and 
\begin{eqnarray*}
&&\hspace{-0.65cm} D_{\bm{U},\bm{Q}} =\sum_n p(w_n)\sum_{\hat{w}} d(w_n,\hat{w})\sum_m u(s_m|w_n)q(\hat{w}|s_m)\\
&&\hspace{-0.65cm} =\sum_n p(w_n)\sum_{\hat{w}} d(w_n,\hat{w})\sum_m \sum_{i=1}^{|\bm{A}|}\lambda_i{\bm{\Delta}}_i(s|w)q(\hat{w}|s_m)\\
&&\hspace{-0.65cm} =\sum_{i=1}^{|\bm{A}|}\lambda_i\sum_n p(w_n) \sum_{\hat{w}} d(w_n,\hat{w})\sum_m{\bm{\Delta}}_i(s|w)q(\hat{w}|s_m)\\
&&\hspace{-0.65cm} =\sum_{i=1}^{|\bm{A}|}\lambda_i D_{{\bm{\Delta}}_i,\bm{Q}}=D.
\end{eqnarray*}
The proposition is proved.

\section{Properties of the Six Subsets}\label{sec:AppB}
This appendix summarizes some properties of the six subsets in Definition 4.4.
\begin{propApp}[Orders of $\underline{\mathcal{S}}_w'$, ${\underline{\mathcal{S}}}_w''$ and ${\overline{\mathcal{S}}}_w'$, ${\overline{\mathcal{S}}}_w''$]\label{prop:1}
	For any meaning $w\in\mathcal{W}$, the sets ${\underline{\mathcal{S}}}_w'$, ${\underline{\mathcal{S}}}_w''$ and ${\overline{\mathcal{S}}}_w'$, ${\overline{\mathcal{S}}}_w''$ satisfy
	\begin{enumerate}
		\item $\forall\ s_{m_1},s_{m_2}\in{\underline{\mathcal{S}}}_w',m_1<m_2$, we have
		\begin{equation*}
			\ell\left(s_{m_1}\right)<\ell\left(s_{m_2}\right),\varphi\left(w,s_{m_1}\right)>\varphi\left(w,s_{m_2}\right)
		\end{equation*}
		\item $\forall\ s_{m_1},s_{m_2}\in{\underline{\mathcal{S}}}_w'',m_1<m_2$, we have
		\begin{equation*}
			\ell\left(s_{m_1}\right)<\ell\left(s_{m_2}\right),\varphi\left(w,s_{m_1}\right)<\varphi\left(w,s_{m_2}\right)
		\end{equation*}
		\item $\forall\ s_{m_1},s_{m_2}\in{\overline{\mathcal{S}}}_w',m_1<m_2$, we have
		\begin{equation*}
			\ell\left(s_{m_1}\right)<\ell\left(s_{m_2}\right),\varphi\left(w,s_{m_1}\right)<\varphi\left(w,s_{m_2}\right)
		\end{equation*}
		\item $\forall\ s_{m_1},s_{m_2}\in{\overline{\mathcal{S}}}_w'',m_1<m_2$, we have
		\begin{equation*}
			\ell\left(s_{m_1}\right)<\ell\left(s_{m_2}\right),\varphi\left(w,s_{m_1}\right)>\varphi\left(w,s_{m_2}\right).
		\end{equation*}
	\end{enumerate}
\end{propApp}
\begin{NewProof}
	We prove 1) by contradiction. For a give $w\in\mathcal{W}$, suppose that there exist $s_{m_1},s_{m_2}\in{\underline{\mathcal{S}}}_w',m_1<m_2$ such that one of the following two conditions holds.
	\begin{itemize}
		\item $\varphi\left(w,s_{m_1}\right)\le\varphi\left(w,s_{m_2}\right)$
		\item $\ell\left(s\right)\geq\ell\left(s'\right)$, $\varphi\left(w,s_{m_1}\right)>\varphi\left(w,s_{m_2}\right)$
	\end{itemize}	
	Note that $m_1<m_2$ implies that $\ell\left(s_{m_1}\right)\le\ell\left(s_{m_2}\right)$. The two conditions above are respectively equivalent to the two conditions below.
	\begin{enumerate}[label =(\alph{enumi})]
		\item $\ell\left(s\right)\le\ell\left(s'\right),\varphi\left(w,s_{m_1}\right)\le\varphi\left(w,s_{m_2}\right)$
		\item $\ell\left(s_{m_1}\right)=\ell\left(s_{m_2}\right)$, $\varphi\left(w,s_{m_1}\right)>\varphi\left(w,s_{m_2}\right)$
	\end{enumerate}

	If (a) holds, $s_{m_2}$ should be removed from ${\underline{\mathcal{S}}}_w'$. If (b) holds, $s_{m_1}$ should be removed from ${\underline{\mathcal{S}}}_w'$. Therefore, $\mathcal{S}_w'$ is not the smallest subset of $\mathcal{S}$ satisfying the conditions in Definition 4.4.
	The proofs of 2) to 4) are similar to 1) and are omitted.
\end{NewProof}

\begin{propApp}
For a given $w$, and the associated ${\underline{\mathcal{S}}}_w'$, ${\underline{\mathcal{S}}}_w''$ and ${\overline{\mathcal{S}}}_w'$, ${\overline{\mathcal{S}}}_w''$.
Define
\begin{eqnarray*}
&&\hspace{-0.65cm} {\underline{m}}_{w,1}=\min{\left\{m:s_m\in{\underline{\mathcal{S}}}_w'\right\}},{\underline{m}}_{w,2}=\max{\left\{m:s_m\in{\underline{\mathcal{S}}}_w'\right\}}, \\
&&\hspace{-0.65cm} {\underline{m}}_{w,3} =\min{\left\{m:s_m\in{\underline{\mathcal{S}}}_w''\right\}},{\underline{m}}_{w,4}=\max{\left\{m:s_m\in{\underline{\mathcal{S}}}_w''\right\}},\\
&&\hspace{-0.65cm} 
{\overline{m}}_{w,1}=\min{\left\{m:s_m\in{\overline{\mathcal{S}}}_w'\right\}},{\overline{m}}_{w,2}=\max{\left\{m:s_m\in{\overline{\mathcal{S}}}_w'\right\}},\\
&&\hspace{-0.65cm} 
{\overline{m}}_{w,3}=\min{\left\{m:s_m\in{\overline{\mathcal{S}}}_w''\right\}},{\overline{m}}_{w,4}=\max{\left\{m:s_m\in{\overline{\mathcal{S}}}_w''\right\}}.
\end{eqnarray*} 
It holds that
	\begin{enumerate}
		\item $\ell\left(s_{{\underline{m}}_{w,1}}\right)=L_{\textup{min}}$ and
		 \begin{equation*}
		 	\varphi\left(w,s_{{\underline{m}}_{w,1}}\right)=\min_{s':s'\in\mathcal{S},\ell\left(s'\right)=L_{\textup{min}}}{\varphi(w,s')}.
		 \end{equation*}
		 
		\item $\varphi\left(w,s_{{\underline{m}}_{w,2}}\right)=\min_{s'\in\mathcal{S}}{\varphi\left(w,s'\right)}$ and
		\begin{equation*}
			\ell\left(s_{{\underline{m}}_{w,2}}\right)=\min_{s':s'\in\mathcal{S},\varphi\left(w,s'\right)=\varphi\left(w,s_{{\underline{m}}_{w,2}}\right)}{\ell\left(s'\right)}.
		\end{equation*}
		 
		 \item $\varphi\left(w,s_{{\underline{m}}_{w,3}}\right)=\min_{s'\in\mathcal{S}}{\varphi\left(w,s'\right)}$ and
		 \begin{equation*}
		 	\ell\left(s_{{\underline{m}}_{w,3}}\right)=\max_{s':s'\in\mathcal{S},\varphi\left(w,s'\right)=\varphi\left(w,s_{{\underline{m}}_{w,3}}\right)}{\ell\left(s'\right)}.
		 \end{equation*}
		 
		 \item 	$\ell\left(s_{{\underline{m}}_{w,4}}\right)=L_{\textup{max}}$ and
		 \begin{equation*}
		 	\varphi\left(w,s_{{\underline{m}}_{w,4}}\right)=\min_{s':s'\in\mathcal{S},\ell\left(s'\right)=L_{\textup{max}}}{\varphi(w,s')}.
	 	\end{equation*}
		 
		 \item 	$\ell\left(s_{{\overline{m}}_{w,1}}\right)=L_{\textup{min}}$ and
		 \begin{equation*}
		 	\varphi\left(w,s_{{\overline{m}}_{w,1}}\right)=\max_{s':s'\in\mathcal{S},\ell\left(s'\right)=L_{\textup{min}}}{\varphi(w,s')}.
	 	\end{equation*}
		  
		  \item $\varphi\left(w,s_{{\overline{m}}_{w,2}}\right)=\max_{s'\in\mathcal{S}}{\varphi\left(w,s'\right)}$ and
		  \begin{equation*}
		  	\ell\left(s_{{\overline{m}}_{w,2}}\right)=\min_{s':s'\in\mathcal{S},\varphi\left(w,s'\right)=\varphi\left(w,s_{{\overline{m}}_{w,2}}\right)}{\ell\left(s'\right)}.
	  	\end{equation*}
		  
		  \item $\varphi\left(w,s_{{\overline{m}}_{w,3}}\right)=\max_{s'\in\mathcal{S}}{\varphi\left(w,s'\right)}$ and
		  \begin{equation*}
		  	\ell\left(s_{{\overline{m}}_{w,3}}\right)=\max_{s':s'\in\mathcal{S},\varphi\left(w,s'\right)=\varphi\left(w,s_{{\overline{m}}_{w,3}}\right)}{\ell\left(s'\right)}.
	  	\end{equation*}
		  
		  \item $\ell\left(s_{{\overline{m}}_{w,4}}\right)=L_{\textup{max}}$ and
		  \begin{equation*}
		  	\varphi\left(w,s_{{\overline{m}}_{w,4}}\right)=\max_{s':s'\in\mathcal{S},\ell\left(s'\right)=L_{\textup{max}}}{\varphi(w,s')}.
	  	\end{equation*}
	\end{enumerate}
\end{propApp}

\begin{NewProof}
	The proofs of the eight conditions are similar and thus we only prove 1) in the following.
	We first prove that $\ell\left(s_{{\underline{m}}_{w,1}}\right)=L_{\textup{min}}$. 
	It is sufficient to prove that
	\begin{equation*}
		\min{\left\{\ell\left(m\right):s_m\in{\underline{\mathcal{S}}}_w'\right\}}=L_{\textup{min}}.
	\end{equation*}
	Let ${\hat{\mathcal{S}}}_w\triangleq \left\{\hat{s}\in\mathcal{S}:\ell\left(\hat{s}\right)=L_{\textup{min}}\right\}$. Suppose  ${\hat{\mathcal{S}}}_w\cap{\underline{\mathcal{S}}}_w'=\emptyset$. 
	By Definition~\ref{defi:Sw}, for any $s'\in{\hat{\mathcal{S}}}_w\subseteq\mathcal{S}\setminus{\underline{\mathcal{S}}}_w'$, there exists an $s\in\mathcal{S}_w'$ such that
	\begin{equation*}
		\ell\left(s\right)\le\ell\left(s'\right),\varphi\left(w,s\right)\le\varphi\left(w,s'\right).
	\end{equation*}
	Thus, $L_{\textup{min}}\le\ell\left(s\right)\le\ell\left(s'\right)=L_{\textup{min}}$ which implies that $s\in{\hat{\mathcal{S}}}_w$. Thus, ${\hat{\mathcal{S}}}_w\cap{\underline{\mathcal{S}}}_w'\neq\emptyset$. On the other hand, $s\in{\underline{\mathcal{S}}}_w'$, i.e., $s\in{\hat{\mathcal{S}}}_w\cap{\underline{\mathcal{S}}}_w'\subseteq{\hat{\mathcal{S}}}_w\cap\left(\mathcal{S}\setminus{\underline{\mathcal{S}}}_w'\right)=\emptyset$, which does not hold. 
	Therefore,
	\begin{equation*}
		\ell\left(s_{\underline{m}_{w,1}}\right)=\min{\left\{\ell\left(m\right):s_m\in{\underline{\mathcal{S}}}_w'\right\}}=L_{\textup{min}}.
	\end{equation*}

	Now we prove that 
	\begin{equation*}
		\varphi\left(w,s_{{\underline{m}}_{w,1}}\right)=\min_{s':s'\in\mathcal{S},\ell\left(s'\right)=L_{\textup{min}}}{\varphi(w,s')}.
	\end{equation*}
	Proposition~\ref{prop:1} shows that for any $s_{m_1},s_{m_2}\in{\underline{\mathcal{S}}}_w'$, $m_1<m_2$, we have
	$\ell\left(s_{m_1}\right)<\ell\left(s_{m_2}\right)$.
	Thus, $\left|{\hat{\mathcal{S}}}_w\cap{\underline{\mathcal{S}}}_w'\right|=1$, i.e.,
	\begin{equation*}
		{\hat{\mathcal{S}}}_w\cap{\underline{\mathcal{S}}}_w'=\left\{s_{\underline{m}_{w,1}}\right\}. 
	\end{equation*}		
	If $|{\hat{\mathcal{S}}}_w|=1$, then clearly 1) holds. 
	Otherwise, by Definition~\ref{defi:Sw}, 
	$\forall\ s'\in{\hat{\mathcal{S}}}_w\setminus\left\{s_{m_{w,1}}\right\}\subseteq\mathcal{S}\setminus{\underline{\mathcal{S}}}_w'$, there exists an $s\in{\underline{\mathcal{S}}}_w'$ such that
	\begin{equation*}
		\ell\left(s\right)\le\ell\left(s'\right),\varphi\left(w,s\right)\le\varphi\left(w,s'\right).
	\end{equation*}		
	Since $\ell\left(s'\right)=L_{\textup{min}}$, we have $\ell\left(s\right)=L_{\textup{min}}$, i.e. $s=s_{m_{w,1}}$. Thus, $\varphi\left(w,s_{m_{w,1}}\right)\le\varphi\left(w,s'\right)$ for any $s'\in{\hat{\mathcal{S}}}_w$. 
\end{NewProof}

\begin{propApp}\label{prop:Pi}
	For any $i=1,2,3,4$, it holds that
	\begin{itemize}
		\item $\underline{\textsc{P}i}=(L_{{\underline{\bm{U}}}_i},D_{{\underline{\bm{U}}}_i,\bm{Q}})$, where ${\underline{\bm{U}}}_i=\bm{\Delta}_{{\underline{m}}_{w_1,i},{\underline{m}}_{w_2,i},\ldots,{\underline{m}}_{w_N,i}}$
		\item $\overline{\textsc{P}i}=(L_{{\overline{\bm{U}}}_i},D_{{\overline{\bm{U}}}_i,\bm{Q}})$, where ${\overline{\bm{U}}}_i=\bm{\Delta}_{{\overline{m}}_{w_1,i},{\overline{m}}_{w_2,i},\ldots,{\overline{m}}_{w_N,i}}$.
	\end{itemize}	
\end{propApp}

\begin{defiApp}\label{defi:psi}
We define 
\begin{equation*}
	\begin{split}
		\underline{\Psi}'&\triangleq\left\{\bm{\Delta}_{i_1,i_2\ldots,i_N}:\forall n,s_{i_n}\in{\underline{\mathcal{S}}}_{w_n}'\right\},\\
		\underline{\Psi}''&\triangleq\left\{\bm{\Delta}_{i_1,i_2\ldots,i_N}:\forall n,s_{i_n}\in{\underline{\mathcal{S}}}_{w_n}''\right\},\\
		\underline{\Psi}&\triangleq\left\{\bm{\Delta}_{i_1,i_2,\ldots,i_N}:\forall n,s_{i_n}\in{\underline{\mathcal{S}}}_{w_n}\right\},\\
		\overline{\Psi}'&\triangleq\left\{\bm{\Delta}_{i_1,i_2\ldots,i_N}:\forall n,s_{i_n}\in{\overline{\mathcal{S}}}_{w_n}'\right\},\\
		\overline{\Psi}''&\triangleq\left\{\bm{\Delta}_{i_1,i_2\ldots,i_N}:\forall n,s_{i_n}\in{\overline{\mathcal{S}}}_{w_n}''\right\},\\
		\overline{\Psi}&\triangleq\left\{\bm{\Delta}_{i_1,i_2,\ldots,i_N}:\forall n,s_{i_n}\in{\overline{\mathcal{S}}}_{w_n}\right\},\\
        \Psi&\triangleq\left\{\bm{\Delta}_{i_1,i_2,\ldots,i_N}:\forall n,s_{i_n}\in{\overline{\mathcal{S}}}_{w_n}\cup{\underline{\mathcal{S}}}_{w_n}\right\}.
	\end{split}
\end{equation*}
\end{defiApp}

It is easy to see that ${\underline{\Psi}}'\cup{\underline{\Psi}}''\subseteq\underline{\Psi}$ and ${\overline{\Psi}}'\cup\overline{\Psi}''\subseteq\overline{\Psi}$.

\begin{lemApp}\label{lm:4points}
	For any $\bm{U}=\bm{\Delta}_{i_1,i_2\ldots,i_N}$, the following four conditions hold.
	\begin{enumerate}
		\item $\exists\ \bm{U}_{\mathsf{l,d}}\in{\underline{\Psi}}'$ such that $D_{\bm{U},\bm{Q}}\geq D_{\bm{U}_{\mathsf{l,d}},\bm{Q}}$ and $L_{\bm{U}}\geq L_{\bm{U}_{\mathsf{l,d}}}$.
		\item $\exists\ \bm{U}_{\mathsf{r,d}}\in{\underline{\Psi}}''$ such that $D_{\bm{U},\bm{Q}}\geq D_{\bm{U}_{\mathsf{r,d}},\bm{Q}}$ and $L_{\bm{U}}\le L_{\bm{U}_{\mathsf{r,d}}}$.
		\item $\exists\ \bm{U}_{\mathsf{l,u}}\in{\overline{\Psi}}'$ such that $D_{\bm{U},\bm{Q}}\le D_{\bm{U}_{\mathsf{l,u}},\bm{Q}}$ and $L_{\bm{U}}\geq L_{\bm{U}_{\mathsf{l,u}}}$.
		\item $\exists\ \bm{U}_{\mathsf{r,u}}\in\overline{\Psi}''$ such that $D_{\bm{U},\bm{Q}}\le D_{\bm{U}_{\mathsf{r,u}},\bm{Q}}$ and $L_{\bm{U}}\le L_{\bm{U}_{\mathsf{r,u}}}$.
	\end{enumerate}
\end{lemApp}  
\begin{NewProof}
The proofs of the four conditions are similar. We only prove 1).
	By Definition~\ref{defi:Sw}, for any $n$, we can find an $i_n'$ such that $s_{i_n'}\in{\underline{\mathcal{S}}}_{w_n}'$ and 
	\begin{equation*}
		\ell\left(s_{i_n'}\right)\le\ell\left(s_{i_n}\right),\varphi\left(w,s_{i_n'}\right)\le\varphi\left(w,s_{i_n}\right).
	\end{equation*}
Let $\bm{U}_{\mathsf{l,d}}=\bm{\Delta}_{i_1',i_2'\ldots,i_N'}$, we have $D_{\bm{U}}\geq D_{\bm{U}_{\mathsf{l,d}},\bm{Q}}$ and $L_{\bm{U}}\geq L_{\bm{U}_{\mathsf{l,d}}}$.
\end{NewProof}

\section{Proof of Theorem~\ref{thm:region}}\label{sec:AppC}
Before proving Theorem \ref{thm:region}, we first give some properties of the $G$ function in Definition \ref{defi:G}.
\begin{lemApp}\label{lm:ab}
	For any $a_1,a_2,b_1,b_2\in\RR$, where $b_1,b_2>0$, the following three arguments are equivalent:
	\begin{enumerate}
		\item $\frac{a_1}{b_1}\geq\frac{a_2}{b_2}$.
		\item $\frac{a_1+a_2}{b_1+b_2}\geq\frac{a_2}{b_2}$.
		\item $\frac{a_1}{b_1}\geq\frac{a_1+a_2}{b_1+b_2}$.
	\end{enumerate}
\end{lemApp}

Define three vectors $\left[a_1,b_1\right]^\top$, $\left[a_2,b_2\right]^\top$ and $\left[a_1+a_2,b_1+b_2\right]^\top$. Lemma~\ref{lm:ab} follows directly from the relative magnitudes of the angles of the three vectors.

\begin{corApp}\label{cor:G}
	For any $n=1,2,\ldots,N$ and $s',s'',s'''\in S_{w_n}$, where $\ell\left(s'\right)<\ell\left(s''\right)<\ell\left(s'''\right)$, the following three arguments are equivalent.
	\begin{itemize}
		\item $G\left(w_n,s',s''\right)\le G(w_n,s',s''')$.
		\item $G\left(w_n,s',s''\right)\le G(w_n,s'',s''')$.
		\item $G\left(w_n,s',s'''\right)\le G(w_n,s'',s''')$.
	\end{itemize}
	The following three arguments are equivalent:
	\begin{itemize}
		\item $G\left(w_n,s',s''\right)\geq G(w_n,s',s''')$.
		\item $G\left(w_n,s',s''\right)\geq G(w_n,s'',s''')$.
		\item $G\left(w_n,s',s'''\right)\geq G(w_n,s'',s''')$.
	\end{itemize}
\end{corApp}

\begin{lemApp}\label{lm:abc}
	For any $a_1,a_2,..., a_K,b_1,b_2,\ldots,b_K,c\in\RR$ and $b_1,b_2,...,b_K>0$, where $K> 0$, we have
	\begin{enumerate}
		\item $\max\limits_{k\in[1,K]}\frac{a_k}{b_k}\geq\frac{\sum_{k}a_k}{\sum_{k}b_k}\geq\min\limits_{k\in[1,K]}\frac{a_k}{b_k}$.
		\item $\frac{a_k}{b_k}\geq c$, $\forall k$ implies that $\frac{\sum_{k}a_k}{\sum_{k}b_k}\geq c$. 
		\item $\frac{a_k}{b_k}\le c,\forall k$ implies that $\frac{\sum_{k}a_k}{\sum_{k}b_k}\le c$.
	\end{enumerate}
\end{lemApp}

Algorithm \ref{algo:region} constructs a sequence of deterministic encoding schemes that shape $R_{\text{enc}}$. In this process, two key steps are lines 2 and 20, which indicate the next encoding scheme at a time step $t$. Next, we analyze these two steps.

\begin{defiApp}
For $t=1,2,3,\ldots,\underline{T}$, we define
	\begin{equation*}
		{\underline{G}}^{(t)}\triangleq\min_{n,m:s_m\in\mathcal{S}_{w_n},m>{\underline{i}}_n^{(t-1)}}{G\left(w_n,s_{{\underline{i}}_n^{(t-1)}},s_m\right)}.
	\end{equation*}
For $t=1,2,3,\ldots,\overline{T}$, we define
	\begin{equation*}
		{\overline{G}}^{(t)}\triangleq\max_{n,m:s_m\in{\overline{\mathcal{S}}}_{w_n},m>{\overline{i}}_n^{\left(t-1\right)}}{G\left(w_n,s_{{\overline{i}}_n^{\left(t-1\right)}},s_m\right)}.
	\end{equation*}
\end{defiApp}

\begin{lemApp}\label{lm:G-is-increasing}
	${\underline{G}}^{(t)}$ is a non-decreasing function of $t=1,2,...,\underline{T}$ and ${\overline{G}}^{(t)}$ is a non-increasing function of $t=1,2,...,\overline{T}$.
\end{lemApp}
\begin{NewProof}
	We first prove that ${\underline{G}}^{(t)}$ is a non-decreasing function.
	It is sufficient to prove that ${\underline{G}}^{(t)}\le{\underline{G}}^{(t+1)}$ if for any $t=1,2,...,\underline{T}-1$. Recall that $s_{{\underline{m}}^{(t)}}\in{\underline{\mathcal{S}}}_{w_{{\underline{n}}^{(t)}}}$ and $s_{{\underline{m}}^{(t+1)}}\in{\underline{\mathcal{S}}}_{w_{{\underline{n}}^{(t+1)}}}$.
	If ${\underline{n}}^{(t+1)}\neq{\underline{n}}^{(t)}$, then by Algorithm~\ref{algo:region}, ${\underline{i}}_{{\underline{n}}^{(t+1)}}^{(t)}={\underline{i}}_{{\underline{n}}^{(t+1)}}^{\left(t-1\right)}$. Therefore, 
	\begin{equation*}
		\begin{split}
			{\underline{G}}^{(t)}&=\min_{n,m:s_m\in\mathcal{S}_{w_n},m>{\underline{i}}_n^{(t-1)}}{G\left(w_n,s_{{\underline{i}}_n^{(t-1)}},s_m\right)}\\
			&\le\min_{m:s_m\in\mathcal{S}_{w_{{\underline{n}}^{(t+1)}}},m>{\underline{i}}_{{\underline{n}}^{(t+1)}}^{\left(t-1\right)}}{G\left(w_{\underline{n}^{(t+1)}},s_{{\underline{i}}_{{\underline{n}}^{(t+1)}}^{\left(t-1\right)}},s_m\right)}\\
			&=\min_{m:s_m\in\mathcal{S}_{w_{{\underline{n}}^{(t+1)}}},m>{\underline{i}}_{{\underline{n}}^{(t+1)}}^{\left(t-1\right)}}{G\left(w_{\underline{n}^{(t+1)}},s_{{\underline{i}}_{{\underline{n}}^{(t+1)}}^{(t)}},s_m\right)}\\
			&={\underline{G}}^{(t+1)}.
		\end{split}
	\end{equation*}
	
	Now we consider the case when ${\underline{n}}^{(t+1)}={\underline{n}}^{(t)}$.  Since 
	\begin{equation*}
		\begin{split}
			{\underline{G}}^{(t)}&=G\left(w_{\underline{n}^{(t)}},s_{{\underline{i}}_{{\underline{n}}^{(t)}}^{\left(t-1\right)}},s_{{\underline{i}}_{{\underline{n}}^{(t)}}^{(t)}}\right)\\
			&\le G\left(w_{\underline{n}^{(t)}},s_{{\underline{i}}_{{\underline{n}}^{(t)}}^{\left(t-1\right)}},s_{{\underline{i}}_{{\underline{n}}^{(t)}}^{(t+1)}}\right),
		\end{split}
	\end{equation*}
	by Corollary~\ref{cor:G}, we have
	\begin{equation*}
		{\underline{G}}^{(t)}\le G\left(w_{\underline{n}^{(t)}},s_{{\underline{i}}_{{\underline{n}}^{(t)}}^{(t)}},s_{{\underline{i}}_{{\underline{n}}^{(t)}}^{(t+1)}}\right)={\underline{G}}^{(t+1)}.
	\end{equation*}
	
Likewise, it can be proven that ${\overline{G}}^{(t)}$ is a non-increasing function.
\end{NewProof}

\begin{lemApp}\label{lm:G^t}
	For any $n=1,2,...,N$, $t=1,2,...,\underline{T}$ and $s_m\in{\underline{\mathcal{S}}}_{w_n}$,
	\begin{enumerate}
		\item $m>{\underline{i}}_n^{(t)}$ implies that
		\begin{equation*}
			{\underline{G}}^{(t+1)}\le G\left(w_n,s_{{\underline{i}}_n^{(t)}},s_m\right).
		\end{equation*}
		\item $m<{\underline{i}}_n^{(t)}$ implies that
		\begin{equation*}
		{\underline{G}}^{(t)}\geq G\left(w_n,s_m,s_{{\underline{i}}_n^{(t)}}\right).
	\end{equation*}
	\end{enumerate}
\end{lemApp}
\begin{NewProof}
	Condition 1) holds since
	\begin{equation*}
		\begin{split}
			{\underline{G}}^{(t+1)}&=\min_{n',m':s_{m'}\in\mathcal{S}_{w_{n'}},m'>{\underline{i}}_{n'}^{(t)}}{G\left(w_{n'},s_{{\underline{i}}_{n'}^{(t)}},s_{m'}\right)}\\
			&\le\min_{m':s_{m'}\in\mathcal{S}_{w_{n}},m'>{\underline{i}}_{n}^{(t)}}{G\left(w_{n},s_{{\underline{i}}_{n}^{(t)}},s_{m'}\right)}\\
			&\le G\left(w_n,s_{{\underline{i}}_n^{(t)}},s_m\right).
		\end{split}
	\end{equation*}
	Now we prove that 2) holds. By letting \begin{equation*}
		\tilde{t}=\min\left\{t':n^{(t')}=n,m<{\underline{i}}_n^{(t')}\right\},
	\end{equation*}
	we have 
	\begin{equation*}
		{\underline{i}}_{n^{(t)}}^{(\tilde{t}-1)}\le m<{\underline{i}}_n^{(\tilde{t})}\le {\underline{i}}_n^{(t)}
	\end{equation*} \begin{equation*}
		{\underline{G}}^{(\tilde{t})}=G\left(w_n,s_{{\underline{i}}_n^{(\tilde{t}-1)}},s_{{\underline{i}}_n^{(\tilde{t})}}\right).
	\end{equation*}

Now we prove that 
\begin{equation}\label{eq:appen1}
	{\underline{G}}^{(\tilde{t})}\geq G\left(w_n,s_m, s_{{\underline{i}}_n^{(\tilde{t})}}\right).
\end{equation}
If $m={\underline{i}}_{{\underline{n}}^{(t)}}^{(\tilde{t}-1)}$, \eqref{eq:appen1} holds. Otherwise, if $m>{\underline{i}}_{{\underline{n}}^{(t)}}^{(\tilde{t}-1)}$, then by Condition 1) we just proved, it holds that
\begin{equation*}
G\left(w_n,s_{{\underline{i}}_n^{(\tilde{t}-1)}},s_{{\underline{i}}_n^{(\tilde{t})}}\right)=G^{(\tilde{t})}\le G\left(w_n,s_{{\underline{i}}_n^{(\tilde{t}-1)}},s_m\right).
\end{equation*}
Combining Corrollary~\ref{cor:G}, \eqref{eq:appen1} holds. Therefore,  \eqref{eq:appen1} holds for all $m$.

If $t=\tilde{t}$, then 2) holds. Otherwise, if $\tilde{t}<t$, then we define 
\begin{equation*}
		\mathbb{T}\triangleq\left\{t':n^{(t')}=n, \tilde{t}+1\le t'\le t\right\}.
	\end{equation*}
We rewrite $\mathbb{T}=\left\{t_1,t_2,\ldots,t_{|\mathbb{T}|}\right\}$, where 
\begin{equation*}
		\tilde{t}<t_1<t_2<\ldots<t_{\left|\mathbb{T}\right|}\le t.
	\end{equation*} 
By Lemma~\ref{lm:abc}, we have
\begin{equation*}
		{\underline{G}}^{(\tilde{t})}\le{\underline{G}}^{\left(t_1\right)}\le{\underline{G}}^{\left(t_2\right)}\le\ldots\le{\underline{G}}^{\left(t_{\left|\mathbb{T}\right|}\right)}\le {\underline{G}}^{(t)}.
	\end{equation*}
That is,
\begin{equation*}
	\begin{split}
{\underline{G}}^{(\tilde{t})}&\le\frac{\varphi\left(w_n,s_{{\underline{i}}_n^{\left(t_1\right)}}\right)-\varphi\left(w_n,s_{{\underline{i}}_n^{(t_1-1)}}\right)}{\ell\left(s_{{\underline{i}}_n^{\left(t_1\right)}}\right)-\ell\left(s_{{\underline{i}}_n^{(t_1-1)}}\right)}\\
&\le\frac{\varphi\left(w_n,s_{{\underline{i}}_n^{\left(t_2\right)}}\right)-\varphi\left(w_n,s_{{\underline{i}}_n^{(t_2-1)}}\right)}{\ell\left(s_{{\underline{i}}_n^{\left(t_2\right)}}\right)-\ell\left(s_{{\underline{i}}_n^{(t_2-1)}}\right)}\\
		&\le\ldots\\
		&\le\frac{\varphi\left(w_n,s_{{\underline{i}}_n^{\left(t_{\left|\mathbb{T}\right|}\right)}}\right)-\varphi\left(w_n,s_{{\underline{i}}_n^{(t_{\left|\mathbb{T}\right|}-1)}}\right)}{\ell\left(s_{{\underline{i}}_n^{\left(t_{\left|\mathbb{T}\right|}\right)}}\right)-\ell\left(s_{{\underline{i}}_n^{(t_{\left|\mathbb{T}\right|}-1)}}\right)}\\
		&\le {\underline{G}}^{(t)}
	\end{split}
\end{equation*}
Note that ${{\underline{i}}_n^{(t_1-1)}}={{\underline{i}}_n^{(\tilde{t})}}$, ${{\underline{i}}_n^{(t_{|\mathbb{T}|})}}={{\underline{i}}_n^{(t)}}$ and ${{\underline{i}}_n^{(t_\tau-1)}}={{\underline{i}}_n^{\left(t_{\tau-1}\right)}}$ for any $2<\tau<|\mathbb{T}|$. We have
\begin{equation}\label{eq:GG}
	\begin{split}
		{\underline{G}}^{(\tilde{t})}&\le\frac{\varphi\left(w_n,s_{{\underline{i}}_n^{\left(t_1\right)}}\right)-\varphi\left(w_n,s_{{\underline{i}}_n^{(\tilde{t})}}\right)}{\ell\left(s_{{\underline{i}}_n^{\left(t_1\right)}}\right)-\ell\left(s_{{\underline{i}}_n^{(\tilde{t})}}\right)}\\
		&\le\frac{\varphi\left(w_n,s_{{\underline{i}}_n^{\left(t_2\right)}}\right)-\varphi\left(w_n,s_{{\underline{i}}_n^{\left(t_1\right)}}\right)}{\ell\left(s_{{\underline{i}}_n^{\left(t_2\right)}}\right)-\ell\left(s_{{\underline{i}}_n^{\left(t_1\right)}}\right)}\\
		&\le\ldots\\
		&\le\frac{\varphi\left(w_n,s_{{\underline{i}}_n^{\left(t_{\left|\mathbb{T}\right|}\right)}}\right)-\varphi\left(w_n,s_{{\underline{i}}_n^{\left(t_{\left|\mathbb{T}\right|-1}\right)}}\right)}{\ell\left(s_{{\underline{i}}_n^{\left(t_{\left|\mathbb{T}\right|}\right)}}\right)-\ell\left(s_{{\underline{i}}_n^{\left(t_{\left|\mathbb{T}\right|-1}\right)}}\right)}\\
		&={\underline{G}}^{(t)}.
	\end{split}
\end{equation}
From Lemma~\ref{lm:abc} and \eqref{eq:GG}, we have
\begin{equation*}
		{\underline{G}}^{(\tilde{t})}\le\frac{\varphi\left(w_n,s_{{\underline{i}}_n^{\left(t_{\left|\mathbb{T}\right|}\right)}}\right)-\varphi\left(w_n,s_{{\underline{i}}_n^{(\tilde{t})}}\right)}{\ell\left(s_{{\underline{i}}_n^{\left(t_{\left|\mathbb{T}\right|}\right)}}\right)-\ell\left(s_{{\underline{i}}_n^{(\tilde{t})}}\right)}\le{\underline{G}}^{(t)}.
	\end{equation*}
 
Note that 
\begin{equation*}
	\begin{split}	\frac{\varphi\left(w_n,s_{{\underline{i}}_n^{\left(t_{\left|\mathbb{T}\right|}\right)}}\right)-\varphi\left(w_n,s_{{\underline{i}}_n^{(\tilde{t})}}\right)}{\ell\left(s_{{\underline{i}}_n^{\left(t_{\left|\mathbb{T}\right|}\right)}}\right)-\ell\left(s_{{\underline{i}}_n^{(\tilde{t})}}\right)}
		&=G\left(w_n,s_{{\underline{i}}_n^{(\tilde{t})}},s_{{\underline{i}}_n^{\left(t_{\left|\mathbb{T}\right|}\right)}}\right)\\
		&=G\left(w_n,s_{{\underline{i}}_n^{(\tilde{t})}},s_{{\underline{i}}_n^{(t)}}\right).
	\end{split}
\end{equation*}
We have
\begin{equation}\label{eq:appen2}
	{\underline{G}}^{(\tilde{t})}\le G\left(w_n,s_{{\underline{i}}_n^{(\tilde{t})}},s_{{\underline{i}}_n^{(t)}}\right)\le{\underline{G}}^{(t)}.
\end{equation}
By \eqref{eq:appen1} and \eqref{eq:appen2}, we have
\begin{equation*}
		G\left(w_n,s_m, s_{{\underline{i}}_n^{(\tilde{t})}}\right)\le G\left(w_n,s_{{\underline{i}}_n^{(\tilde{t})}},s_{{\underline{i}}_n^{(t)}}\right).
	\end{equation*}
Combining Corollary~\ref{cor:G}, it holds that
\begin{equation}\label{eq:appen3}
	G\left(w_n,s_m, s_{{\underline{i}}_n^{(t)}}\right)\le G\left(w_n,s_{{\underline{i}}_n^{(\tilde{t})}},s_{{\underline{i}}_n^{(t)}}\right).
\end{equation}
By \eqref{eq:appen2} and \eqref{eq:appen3}, we finally have
\begin{equation*}
		G\left(w_n,s_m, s_{{\underline{i}}_n^{(t)}}\right)\le{\underline{G}}^{(t)}.
	\end{equation*}
\end{NewProof}

\begin{corApp}
	For any $t=1,2,...,\overline{T}$ and any $s_m\in{\overline{S}}_{w_{{\overline{n}}^{(t)}}}$,
	\begin{enumerate}
		\item $m>{\overline{m}}^{(t)}$ implies that
		\begin{equation*}
		{\overline{G}}^{(t+1)}\geq G\left(w_{{\overline{n}}^{(t)}},s_{{\overline{m}}^{(t)}},s_m\right).
	\end{equation*}
		\item $m<{\overline{m}}^{(t)}$ implies that
		\begin{equation*}
		{\overline{G}}^{(t)}\le G\left(w_{{\overline{n}}^{(t)}},s_m,s_{{\overline{m}}^{(t)}}\right).
	\end{equation*}
	\end{enumerate}
\end{corApp}

To ease exposition, we next define a $J$ function. The connections between the $G$ and $J$ functions will be revealed later in Lemma \ref{lm:G=j}.
\begin{defiApp}
	For any ${\bm{U}}$ and ${\bm{U}}'$, where $L_{{\bm{U}}'}\neq L_{\bm{U}}$, we define
	\begin{equation*}
		\begin{split}
			&J_{\bm{Q}}\left({\bm{U}},{\bm{U}}'\right) \triangleq\frac{D_{{\bm{U}}',\bm{Q}}-D_{\bm{U},\bm{Q}}}{L_{{\bm{U}}'}-L_{\bm{U}}}.\\
			&{\underline{J}}^{(t)} \triangleq J_{\bm{Q}}\left({\underline{{\bm{U}}}}^{(t-1)},{\underline{{\bm{U}}}}^{(t)}\right).\\ 
   &{\overline{J}}^{(t)} \triangleq J_{\bm{Q}}\left({\overline{{\bm{U}}}}^{(t-1)},{\overline{{\bm{U}}}}^{(t)}\right).
		\end{split}
	\end{equation*}
\end{defiApp}

\begin{lemApp}\label{lm:G=j}
	For any $t=1,2,...,\underline{T}$, ${\underline{J}}^{(t)}={\underline{G}}^{(t)}$; for any $t=1,2,...,\overline{T}$, ${\overline{J}}^{(t)}={\overline{G}}^{(t)}$.
\end{lemApp}

\begin{lemApp}\label{lemma:C8}
	For any ${\bm{U}}={\bm{\Delta}}_{i_1,i_2\ldots,i_N}\in\Psi$, where $\Psi$ is defined in Definition \ref{defi:psi},
	\begin{enumerate}
		\item $L_{{\underline{{\bm{U}}}}^{(t)}}<L_{\bm{U}}$, $t=1,2,...,\underline{T}$, implies that
		\begin{equation*}
		{\underline{J}}^{(t+1)}\le J_{\bm{Q}}({\underline{{\bm{U}}}}^{(t)},{\bm{U}})
	\end{equation*}
		\item $L_{{\overline{{\bm{U}}}}^{(t)}}<L_{\bm{U}}$, $t=1,2,...,\overline{T}$, implies that
		\begin{equation*}
		{\overline{J}}^{(t+1)}\le J_{\bm{Q}}({\overline{{\bm{U}}}}^{(t)},{\bm{U}}).
	\end{equation*}
	\end{enumerate}
\end{lemApp}
\begin{NewProof}
	We first prove 1). Recall that ${\underline{{\bm{U}}}}^{(t)}={\bm{\Delta}}_{{\underline{i}}_1^{(t)},{\underline{i}}_2^{(t)}\ldots,{\underline{i}}_N^{(t)}}$ for any $t$. Then 
	\begin{equation*}
		\begin{split}
			&J_{\bm{Q}}\left({\underline{{\bm{U}}}}^{(t)},{\bm{U}}\right)\\
			=&\frac{\sum_{n} p\left(w_n\right)\varphi\left(w_n,s_{i_n}\right)-\sum_{n} p\left(w_n\right)\varphi\left(w_n,s_{{\underline{i}}_n^{(t)}}\right)}{\sum_{n}{p\left(w_n\right)\ell(s_{i_n})}-\sum_{n}{p\left(w_n\right)\ell(s_{{\underline{i}}_n^{(t)}})}}\\
			=&\frac{\sum_{n} p\left(w_n\right)\left(\varphi\left(w_n,s_{i_n}\right)-\varphi\left(w_n,s_{{\underline{i}}_n^{(t)}}\right)\right)}{\sum_{n} p\left(w_n\right)\left(\ell\left(s_{i_n}\right)-\ell\left(s_{{\underline{i}}_n^{(t)}}\right)\right)}.
		\end{split}
	\end{equation*}
	Denote
	\begin{equation*}
		\begin{split}
			I_0&=\left\{n:i_n={\underline{i}}_n^{(t)}\right\}.\\
			I_+&=\left\{n:i_n>{\underline{i}}_n^{(t)}\right\}.\\
			I_-&=\left\{n:i_n<{\underline{i}}_n^{(t)}\right\}.\\
			a_1&={\sum_{n} p\left(w_n\right)\left(\varphi\left(w_n,s_{i_n}\right)-\varphi\left(w_n,s_{{\underline{i}}_n^{(t)}}\right)\right)}.\\
			b_1&={\sum_{n} p\left(w_n\right)\left(\ell\left(s_{i_n}\right)-\ell\left(s_{{\underline{i}}_n^{(t)}}\right)\right)}.\\
			a_2&={\sum_{n\in I_-} p\left(w_n\right)\left(-\varphi\left(w_n,s_{i_n}\right)+\varphi\left(w_n,s_{{\underline{i}}_n^{(t)}}\right)\right)}.\\
			b_2&={\sum_{n\in I_-} p\left(w_n\right)\left(-\ell\left(s_{i_n}\right)+\ell\left(s_{{\underline{i}}_n^{(t)}}\right)\right)}.
		\end{split}
	\end{equation*}
	Then
	\begin{equation}\label{eq:JU}
		J_{\bm{Q}}\left({\underline{{\bm{U}}}}^{(t)},{\bm{U}}\right)=\frac{a_1}{b_1}.
	\end{equation}
	For any $n\in I_0$, we have $i_n={\underline{i}}_n^{(t)}$, and then \begin{equation*}
		\varphi\left(w_n,s_{i_n}\right)-\varphi\left(w_n,s_{{\underline{i}}_n^{(t)}}\right)=\ell\left(s_{i_n}\right)-\ell\left(s_{{\underline{i}}_n^{(t)}}\right)=0.
	\end{equation*}
	Note that 
	\begin{equation*}
		0<b_1+b_2={\sum_{n\in I_+} p\left(w_n\right)\left(-\ell\left(s_{i_n}\right)+\ell\left(s_{{\underline{i}}_n^{(t)}}\right)\right)}.
	\end{equation*}
	It holds that $I_+\neq \emptyset$.
	For any $n\in I_+$, we have $\ell\left(s_{i_n}\right)>\ell\left(s_{{\underline{i}}_n^{(t)}}\right)$. Then, by Lemma~\ref{lm:G^t}, 
	\begin{equation*}
		\begin{split}
			{\underline{G}}^{(t+1)}&\le G\left(w_n,s_{{\underline{i}}_n^{(t)}},s_{i_n}\right)\\
			&=\frac{\varphi\left(w_n,s_{i_n}\right)-\varphi\left(w_n,s_{{\underline{i}}_n^{(t)}}\right)}{\ell\left(s_{i_n}\right)-\ell\left(s_{{\underline{i}}_n^{(t)}}\right)}\\
			&=\frac{p\left(w_n\right)\left(\varphi\left(w_n,s_{i_n}\right)-\varphi\left(w_n,s_{{\underline{i}}_n^{(t)}}\right)\right)}{p\left(w_n\right)\left(\ell\left(s_{i_n}\right)-\ell\left(s_{{\underline{i}}_n^{(t)}}\right)\right)}.
		\end{split}
	\end{equation*}
	By Lemma~\ref{lm:abc}, we have
	\begin{equation}\label{eq:appen4+}
		\begin{split}
			{\underline{G}}^{(t+1)}&\le\frac{\sum_{n\in I_+} p\left(w_n\right)\left(\varphi\left(w_n,s_{i_n}\right)-\varphi\left(w_n,s_{{\underline{i}}_n^{(t)}}\right)\right)}{\sum_{n\in I_+} p\left(w_n\right)\left(\ell\left(s_{i_n}\right)-\ell\left(s_{{\underline{i}}_n^{(t)}}\right)\right)}\\
			&=\frac{a_1+a_2}{b_1+b_2}.
		\end{split}
	\end{equation}
	When $I_-= \emptyset$, 1) holds by \eqref{eq:JU} and \eqref{eq:appen4+}. Therefore, we only need to consider the case where $I_-\neq \emptyset$.
	For any $n\in I_-$, we have $\ell\left(s_{i_n}\right)>\ell\left(s_{{\underline{i}}_n^{(t)}}\right)$, and  by Lemma~\ref{lm:G^t}, 
	\begin{equation*}
		\begin{split}
			{\underline{G}}^{(t)}&\geq G\left(w_n,s_{{\underline{i}}_n^{(t)}},s_{i_n}\right)\\
			&=\frac{\varphi\left(w_n,s_{i_n}\right)-\varphi\left(w_n,s_{{\underline{i}}_n^{(t)}}\right)}{\ell\left(s_{i_n}\right)-\ell\left(s_{{\underline{i}}_n^{(t)}}\right)}\\
			&=\frac{p\left(w_n\right)\left(-\varphi\left(w_n,s_{i_n}\right)+\varphi\left(w_n,s_{{\underline{i}}_n^{(t)}}\right)\right)}{p\left(w_n\right)\left(-\ell\left(s_{i_n}\right)+\ell\left(s_{{\underline{i}}_n^{(t)}}\right)\right)}.
		\end{split}
	\end{equation*}
	By Lemma~\ref{lm:abc}, we have
	\begin{equation}\label{eq:appen4-}
		{\underline{G}}^{(t)}\geq\frac{a_2}{b_2}.
	\end{equation}
	By \eqref{eq:appen4+}, \eqref{eq:appen4-} and Lemma~\ref{lm:G-is-increasing}, we have
	\begin{equation*}
			\frac{a_2}{b_2}	\le {\underline{G}}^{(t)}\le {\underline{G}}^{(t+1)}\le \frac{a_1+a_2}{b_1+b_2}.
	\end{equation*}
	Finally, 1) holds by Lemma~\ref{lm:ab}.
	
The second part of Lemma \ref{lemma:C8} can be proved in a similar fashion.
\end{NewProof}

\begin{defiApp}
Theorem \ref{thm:region} characterizes a region using the distortion-cost pairs of $\underline{T}+\overline{T}+2$ semantic encoding schemes $\left\{\underline{\bm{U}}^{(t)}:t=0,1,2,...,\underline{T}\right\}$ and $\left\{\overline{\bm{U}}^{(t)}:t=0,1,2,...,\overline{T}\right\}$.
We denote the region by $\widetilde{R}_{\text{enc}}$, the piecewise linear connection of $\left(L_{\underline{\bm{U}}^{(t)}},D_{\underline{\bm{U}}^{(t)},\bm{Q}} \right)$, $t=0,1,2,...,\underline{T}$ by a function $\underline{D}(L)$,
and the piecewise linear connection of $\left(L_{\overline{\bm{U}}^{(t)}},D_{\overline{\bm{U}}^{(t)},\bm{Q}} \right)$, $t=0,1,2,...,\overline{T}$
by a function $\overline{D}(L)$.
\end{defiApp}

\begin{corApp}
	For any $L',L''\in [L_{\text{min}},L_{\text{max}}]$, where $L'<L''$, we have 
	\begin{equation*}
		{\underline{J}}^{({\underline{t}}')}\le\frac{\underline{D}(L'')-\underline{D}(L')}{L''-L'}\le{\underline{J}}^{({\underline{t}}'')},
	\end{equation*}
	\begin{equation*}
		{\overline{J}}^{(\overline{t}')}\geq\frac{\overline{D}(L'')-\overline{D}(L')}{L''-L'}\geq{\overline{J}}^{(\overline{t}'')},
	\end{equation*}
	where 
 \begin{eqnarray*}
&&\hspace{-0.65cm} {\underline{t}}'\triangleq\min\left\{t:L_{{\underline{{\bm{U}}}}^{(t)}}\geq L'\right\},
{\underline{t}}''\triangleq\min\left\{t:\allowbreak L_{{\underline{{\bm{U}}}}^{(t)}}\allowbreak\geq L''\right\}  \\
&&\hspace{-0.65cm}  \overline{t}'\triangleq\min\left\{t:L_{{\overline{{\bm{U}}}}^{\left({\overline{t}}'\right)}}\geq L'\right\},
\overline{t}''\triangleq\min\left\{t:L_{{\overline{{\bm{U}}}}^{(t)}}\geq L''\right\}.  \\
\end{eqnarray*}

\end{corApp}

\begin{NewProof}
	If $\underline{t}'=\underline{t}''$, it is straightforward that
	\begin{equation*}
		{\underline{J}}^{({\underline{t}}')}=\frac{\underline{D}(L'')-\underline{D}(L')}{L''-L'}={\underline{J}}^{({\underline{t}}'')}.
	\end{equation*}
	If $\underline{t}'<\underline{t}''$, then 
\begin{eqnarray*}
&&\hspace{-0.65cm} \underline{D}(L'') = \underline{D}(L')+\left(\underline{D}_{{\underline{{\bm{U}}}}^{(\underline{t}')}}-\underline{D}(L')\right) \\
&&\hspace{-0.65cm} +\sum_{t=\underline{t}'+1}^{\underline{t}''-1}\left({\underline{D}}_{{\underline{{\bm{U}}}}^{(t)}}-{\underline{D}}_{{\bm{U}}^{\left(t-1\right)}}\right)+\left(\underline{D}(L'')-{\underline{D}}_{{\underline{{\bm{U}}}}^{\left(\underline{t}''-1\right)}}\right) \\
&&\hspace{-0.65cm} = \underline{D}(L')+{\underline{J}}^{({\underline{t}}')}\left(L_{{\underline{{\bm{U}}}}^{(\underline{t}')}}-L'\right)\\
&&\hspace{-0.65cm} +\sum_{t=\underline{t}'+1}^{\underline{t}''-1}{{\underline{J}}^{(t)}\left(L_{{\underline{{\bm{U}}}}^{(t)}}-L_{{\underline{{\bm{U}}}}^{\left(t-1\right)}}\right)}+{\underline{J}}^{\left(\underline{t}''\right)}\left(L''-L_{{\underline{{\bm{U}}}}^{\left(\underline{t}''-1\right)}}\right).
\end{eqnarray*}
	By Lemma~\ref{lm:G-is-increasing} and Lemma~\ref{lm:G=j}, we have 
	\begin{equation*}
		\underline{J}^{(\underline{t}')}\le \underline{J}^{\left(\underline{t}'+1\right)}\le\ldots\le \underline{J}^{\left(\underline{t}''\right)}.
	\end{equation*}
Then,
\begin{eqnarray*}
&&\hspace{-0.65cm}  \underline{D}(L')+{\underline{J}}^{({\underline{t}}')}\left(L_{{\underline{{\bm{U}}}}^{(\underline{t}')}}-L'\right)\\
&&\hspace{-0.65cm} +\sum_{t=\underline{t}'+1}^{\underline{t}''-1}{{\underline{J}}^{\left({\underline{t}}'\right)}\left(L_{{\underline{{\bm{U}}}}^{(t)}}\!-\!L_{{\underline{{\bm{U}}}}^{\left(t-1\right)}}\right)}+{\underline{J}}^{\left({\underline{t}}'\right)}\left(L''\!-\!L_{{\underline{{\bm{U}}}}^{\left(\underline{t}''\!-\!1\right)}}\right)\\
&&\hspace{-0.65cm} \le \tilde{D}(L'')\\
&&\hspace{-0.65cm} \le \underline{D}(L')+{\underline{J}}^{(\underline{t}'')}\left(L_{{\underline{{\bm{U}}}}^{(\underline{t}')}}-L'\right)\\
&&\hspace{-0.65cm} +\sum_{t=\underline{t}'+1}^{\underline{t}''-1}{{\underline{J}}^{\left(\underline{t}''\right)}\left(L_{{\underline{{\bm{U}}}}^{(t)}}\!-\!L_{{\underline{{\bm{U}}}}^{\left(t-1\right)}}\right)}+{\underline{J}}^{\left(\underline{t}''\right)}\!\left(L''\!-\!L_{{\underline{{\bm{U}}}}^{\left(\underline{t}''\!-\!1\right)}}\right).
\end{eqnarray*}
That is,
	\begin{equation*}
		\underline{D}(L')+{\underline{J}}^{({\underline{t}}')}\left(L''\!-\!L'\right)\le\tilde{D}(L'')\le\underline{D}(L')+{\underline{J}}^{\left(\underline{t}''\right)}\left(L''\!-\!L'\right).
	\end{equation*}
Finally,
	\begin{equation*}
		\underline{J}^{(\underline{t}')}=\frac{\tilde{D}(L'')-\tilde{D}(L')}{L''-L'}=\underline{J}^{(\underline{t}'')}.
	\end{equation*}
The proof of the second part is similar.
\end{NewProof}

\begin{lemApp}\label{lemma:R~convex}
	$\underline{D}(L)$ is convex and $\overline{D}(L)$ is concave. 
	The region $\widetilde{R}_{\text{enc}}$ is convex.
\end{lemApp}
\begin{NewProof}
	We first prove that $\underline{D}(L)$ is convex. It is sufficient to prove that for any $L'$, $L''\in[L_{\text{min}},L_{\text{max}}]$ and $\lambda\in(0,1)$, we have 
	\begin{equation*}
		\underline{D}\left(\hat{L}\right)\le\lambda\underline{D}(L')+(1-\lambda)\underline{D}(L'')
	\end{equation*}
	where $\hat{L}=\lambda L'+(1-\lambda)L''$. Let $t'=\min\left\{t:L_{{\bm{U}}^{(t)}}\geq L'\right\}$, $t''=\min\left\{t:L_{{\bm{U}}^{(t)}}\geq L''\right\}$ and $\hat{t}=\min\left\{t:L_{{\bm{U}}^{(t)}}\geq\hat{L}\right\}$. If $L'=L''$, then the inequality holds. Otherwise, without loss of generality, assume $L'<L''$. Then, $L'<\hat{L}<L''$.
	By Corollary 3.11, we have
	\begin{equation*}
		\frac{\underline{D}\left(\hat{L}\right)-\underline{D}(L')}{\hat{L}-L'}\le J^{(\hat{t})}\le\frac{\underline{D}(L'')-\underline{D}\left(\hat{L}\right)}{L''-\hat{L}},
	\end{equation*}
	which implies that 
	\begin{equation*}
		\underline{D}\left(\hat{L}\right)\le\underline{D}(L')+{\underline{J}}^{\left(\hat{t}\right)}\left(\hat{L}-L'\right),
	\end{equation*}
	\begin{equation*}
		\underline{D}\left(\hat{L}\right)\le\underline{D}(L'')+J^{\left(\hat{t}\right)}\left(\hat{L}-L''\right).
	\end{equation*}
	Thus,
	\begin{equation*}
		\begin{split}
			\underline{D}\left(\hat{L}\right)&=\lambda\underline{D}\left(\hat{L}\right)+\left(1-\lambda\right)\underline{D}\left(\hat{L}\right)\\
			&\le\lambda\left(\underline{D}(L')+{\underline{J}}^{\left(\hat{t}\right)}\left(\hat{L}-L'\right)\right)\\
			&+\left(1-\lambda\right)\left(\underline{D}(L'')+{\underline{J}}^{\left(\hat{t}\right)}\left(\hat{L}-L''\right)\right)\\
			&=\lambda\underline{D}(L')+\left(1-\lambda\right)\underline{D}(L'')\\
			&+{\underline{J}}^{\left(\hat{t}\right)}\left(\hat{L}-\left({\lambda L}'+\left(1-\lambda\right)L''\right)\right)\\
			&=\lambda\underline{D}(L')+\left(1-\lambda\right)\underline{D}(L'').
		\end{split}
	\end{equation*}
	Likewise, it can be proven that $\overline{D}(L)$ is concave. Now we prove that the region $\widetilde{R}_{\text{enc}}$ is convex. It is sufficient to prove that, for any $(L',D'), (L'',D'')\in \widetilde{R}_{\text{enc}}$ and $\lambda\in(0,1)$, it holds that
	\begin{equation*}
		(\hat{L},\hat{D})\in\widetilde{R}_{\text{enc}}.
	\end{equation*}
	where $\hat{L}=\lambda L'+(1-\lambda)L''$ and $\hat{D}=\lambda D'+(1-\lambda)D''$. It is equivalent to proving that
	\begin{equation*}
		\underline{D}\left(\hat{L}\right)\le\hat{D}\le\overline{D}\left(\hat{L}\right).
	\end{equation*}
	Since $(L',D'), (L'',D'')\in\widetilde{R}_{\text{enc}}$, we have 
	\begin{equation*}
		\underline{D}(L')\le D'\text{ and }\underline{D}(L'')\le D''.
	\end{equation*}
	On the other hand, $\underline{D}(L)$ is convex, indicating that 
	\begin{equation*}
		\begin{split}
			\underline{D}\left(\hat{L}\right)&\le\lambda\underline{D}(L')+\left(1-\lambda\right)\underline{D}(L'')\\
			&\le\lambda D'+\left(1-\lambda\right)D''=\hat{D}.
		\end{split}
	\end{equation*}
	Likewise, we can also prove that $\hat{D}\le\overline{D}\left(\hat{L}\right)$. The lemma is proved.
\end{NewProof}

We are now ready to prove Theorem \ref{thm:region}.
The achievability can be easily established. According to Algorithm~\ref{algo:region}, for any $L\in [L_{\text{min}},L_{\text{max}}]$, $(L,\underline{D}(L))$ and $(L,\overline{D}(L))$ are in $R_{\text{enc}}$. Moreover, Proposition~\ref{prop:timesharing} shows that $R_{\text{enc}}$ is a convex region. Thus, for any $L\in [L_{\text{min}},L_{\text{max}}]$ and $D\in [\underline{D}(L), \overline{D}(L)]$, we have $(L,D)\in R_{\text{enc}}$. Therefore, $\widetilde{R}_{\text{enc}}\subseteq R_{\text{enc}}$.

Next, we prove the converse, i.e., $R_{\text{enc}}\subseteq\widetilde{R}_{\text{enc}}$. Assuming that there exists an encoding scheme ${\bm{U}^*}$ such that $\left(L_{\bm{U}^*},D_{\bm{U}^*,\bm{Q}}\right)\in R_{\text{enc}}$ and $\left(L_{\bm{U}^*},D_{\bm{U}^*,\bm{Q}}\right)\notin\widetilde{R}_{\text{enc}}$.
Proposition~\ref{prop:timesharing} indicates that $(L_{\bm{U}^*},D_{\bm{U}^*,\bm{Q}})$ can be achieved by the time sharing of a set of deterministic encoding schemes. Denoted by $\bm{A}$ the set containing all these encoding schemes.
Recall from Lemma~\ref{lemma:R~convex} that the region $\widetilde{R}_{\text{enc}}$ is convex.
Therefore, there exists at least one deterministic encoding scheme ${\bm{U}}'\in \bm{A}$ such that $\left(L_{{\bm{U}}'},D_{{\bm{U}}',\bm{Q}}\right)\in R_{\text{enc}}$ and $\left(L_{{\bm{U}}'},D_{{\bm{U}}',\bm{Q}}\right)\notin\widetilde{R}_{\text{enc}}$.  
Otherwise, $(L_{\bm{U}^*},D_{\bm{U}^*,\bm{Q}})\in \widetilde{R}_{\text{enc}}$, which contradicts the assumption.

By Lemma~\ref{lm:4points}, there exist four points ${\bm{U}}_{\mathsf{l,d}},{\bm{U}}_{\mathsf{r,d}},{\bm{U}}_{\mathsf{l,u}},{\bm{U}}_{\mathsf{r,u}}\in \Psi$ such that 
    \begin{itemize}
        \item $D_{\bm{U}',\bm{Q}}\geq D_{{\bm{U}}_{\mathsf{l,d}},\bm{Q}}$ and $L_{\bm{U}'}\geq L_{{\bm{U}}_{\mathsf{l,d}}}$;
		\item $D_{\bm{U}',\bm{Q}}\geq D_{{\bm{U}}_{\mathsf{r,d}},\bm{Q}}$ and $L_{\bm{U}'}\le L_{{\bm{U}}_{\mathsf{r,d}}}$;
		\item $D_{\bm{U}',\bm{Q}}\le D_{{\bm{U}}_{\mathsf{l,u}},\bm{Q}}$ and $L_{\bm{U}'}\geq L_{{\bm{U}}_{\mathsf{l,u}}}$;
		\item $D_{\bm{U}',\bm{Q}}\le D_{{\bm{U}}_{\mathsf{r,u}},\bm{Q}}$ and $L_{\bm{U}'}\le L_{{\bm{U}}_{\mathsf{r,u}}}$.
    \end{itemize}
Hence, there exists at least one encoding scheme
\begin{equation*}
	{\bm{\Delta}}\in \left\{{\bm{U}}_{\mathsf{l,d}},{\bm{U}}_{\mathsf{r,d}},{\bm{U}}_{\mathsf{l,u}},{\bm{U}}_{\mathsf{r,u}}\right\}\subseteq \Psi
\end{equation*}
such that one of the following two statements holds.
\begin{enumerate}
	\item $D_{\bm{\Delta},\bm{Q}}<\underline{D}(L_{{\bm{\Delta}}})$;
	\item $D_{\bm{\Delta},\bm{Q}}>\overline{D}(L_{{\bm{\Delta}}})$.
\end{enumerate}
Otherwise, $\left(L_{{\bm{U}}'},D_{{\bm{U}}',\bm{Q}}\right)\in\widetilde{R}_{\text{enc}}$.

We first assume that statement 1) holds. 
Proposition~\ref{prop:Pi} shows that $\left(L_{{\underline{{\bm{U}}}}^{\left(0\right)}},D_{{\underline{{\bm{U}}}}^{\left(0\right)},\bm{Q}}\right)=\underline{\text{P1}}$. Thus $L_{\text{min}}<L_{{\bm{\Delta}}}$. Otherwise, $D_{\bm{\Delta},\bm{Q}}\ge \underline{D}(L_{{\bm{\Delta}}})$, which contradicts statement 1).
Let $t=\max\{t':L_{{\underline{{\bm{U}}}}^{(t')}}<L_{{\bm{\Delta}}}\}$. Thus, $L_{{\underline{{\bm{U}}}}^{(t)}}<L_{{\bm{\Delta}}}\le L_{{\underline{{\bm{U}}}}^{(t+1)}}$. 

Let $\lambda=\frac{L_{{\bm{\Delta}}}-L_{{\underline{{\bm{U}}}}^{(t)}}}{L_{{\underline{{\bm{U}}}}^{(t+1)}}-L_{{\underline{{\bm{U}}}}^{(t)}}}$, we have $\lambda\in(0,1]$ and 
\begin{equation*}
	L_{{\bm{\Delta}}}=\left(1-\lambda\right)L_{{\bm{U}}^{(t)}}+\lambda L_{{\bm{U}}^{(t+1)}}.
\end{equation*}

From Lemma 3.10, it holds that 
\begin{equation*}
	{\underline{J}}^{(t+1)}\le J_{\bm{Q}}({\bm{U}}^{(t)},{\bm{\Delta}}).
\end{equation*}
As a result,
\begin{equation*}
		D_{\bm{\Delta},\bm{Q}}\ge \left(1-\lambda\right)D_{{\underline{{\bm{U}}}}^{(t)},\bm{Q}}+\lambda D_{{\underline{{\bm{U}}}}^{(t+1)},\bm{Q}}
		=\underline{D}\left(L_{{\bm{\Delta}}}\right).
\end{equation*}
This contradicts statement 1). Therefore, statement 1) does not hold. 
Similarly, it can be proven that statement 2) does not hold either. 

Overall, any encoding scheme ${\bm{U}^*}$ with $\left(L_{\bm{U}^*},D_{\bm{U}^*,\bm{Q}}\right)\in R_{\text{enc}}$ must satisfy $\left(L_{\bm{U}^*},D_{\bm{U}^*,\bm{Q}}\right)\in\widetilde{R}_{\text{enc}}$. The region characterized by Theorem \ref{thm:region} is exactly the distortion-cost region of semantic encoding $R_{\text{enc}}$.

\section{Proof of Proposition~\ref{thm:decHamming}}\label{sec:AppD}
Under the Hamming distortion measure, $\psi_q(\hat{w},\hat{s})$ can be refined as
\begin{eqnarray*}
\psi_q(\hat{w},\hat{s})
\hspace{-0.2cm}&=&\hspace{-0.2cm}
\sum_{w,s}q(w)p(s|w)c(\hat{s}|s)d(w,\hat{w}) \\
\hspace{-0.2cm}&=&\hspace{-0.2cm}
\sum_{w,s}q(w)p(s|w)c(\hat{s}|s) - q(\hat{w})\sum_{s}p(s|\hat{w})c(\hat{s}|s) \\
\hspace{-0.2cm}&=&\hspace{-0.2cm}
q(\hat{s}) - q(\hat{w})p(\hat{s}|\hat{w}),
\end{eqnarray*}
where $q(\hat{s})\triangleq \sum_{w,s}q(w)p(s|w)c(\hat{s}|s)$.

Likewise, we have $\psi_p(\hat{w},\hat{s})= p(\hat{s}) - p(\hat{w})p(\hat{s}|\hat{w})$, where $p(\hat{s})\triangleq \sum_{w,s}p(w)p(s|w)c(\hat{s}|s)$.

Following Proposition \ref{thm:dec}, semantic decoding gives us
\begin{equation}\label{eq:wq}
\hat{w}_q(\hat{s})=
\argmin_{\hat{w}} \psi_q(\hat{w},\hat{s})=
\argmax_{w} q(w)p(\hat{s}|w),
\end{equation}
\begin{equation}\label{eq:dvq}
D_{\bm{P},\bm{V}^*_q}=\sum_{\hat{s}}\psi_p\big(\hat{w}_q(\hat{s}),\hat{s}\big) = 1-\sum_{\hat{s}}p\big(\hat{w}_q(\hat{s})\big)p\big(\hat{s}|\hat{w}_q(\hat{s})\big).
\end{equation}

An intuitive explanation of \eqref{eq:wq} and \eqref{eq:dvq} is as follows. Under the prior distribution $q(w)$, \eqref{eq:wq} is simply the maximum a posteriori (MAP) decoding for a given $\hat{s}$. When the decoding scheme is determined, a received message $\hat{s}$ is decoded to $\hat{w}_q(\hat{s})$. As a result, the probability of successful decoding is the summation $\sum_{\hat{s}}p\big(\hat{w}_q(\hat{s})\big)p\big(\hat{s}|\hat{w}_q(\hat{s})\big)$ with each term being the probability of transmitting $\hat{w}_q(\hat{s})$ and receiving $\hat{s}$. The Hamming distortion, i.e., the semantic error, is then given by \eqref{eq:dvq}.

Likewise, for the true prior p(w), we would have
\begin{equation}\label{eq:wp}
\hat{w}_p(\hat{s})=
\argmax_{w} p(w)p(\hat{s}|w),
\end{equation}
\begin{equation}\label{eq:dvp}
D_{\bm{P},\bm{V}^*_p}= 1-\sum_{\hat{s}}p\big(\hat{w}_p(\hat{s})\big)p\big(\hat{s}|\hat{w}_p(\hat{s})\big).
\end{equation}

Comparing \eqref{eq:dvp} and \eqref{eq:dvq}, $D_{\bm{P},\bm{V}^*_q}=D_{\bm{P},\bm{V}^*_p}$ if and only if \eqref{eq:deccondition} holds. Note that $\hat{w}_q(\hat{s})$ and $\hat{w}_p(\hat{s})$ in \eqref{eq:wp} and \eqref{eq:wq} are the solutions to the left- and right-hand sides of \eqref{eq:deccondition}, respectively, but they are not necessarily the only solutions. Therefore, \eqref{eq:deccondition} does not imply $\hat{w}_q(\hat{s})=\hat{w}_p(\hat{s})$.

\section{Proof of Theorem \ref{thm:csed}}\label{sec:AppE}
The distortion-cost function of CSED  $D^*_{\bm{U},\bm{V}^*_q}(L)$ can be achieved by the deterministic encoding schemes $\big\{ \underline{\bm{U}}^{(i)}:i=0,1,2,...,\underline{T} \big\}$ and the linear combination of two contiguous $\underline{\bm{U}}^{(i)}$ and $\underline{\bm{U}}^{(i+1)}$.
Thus, \eqref{eq:CSED} holds for any point on $D^*_{\bm{U},\bm{V}^*_q}(L)$ provided that we can prove \eqref{eq:CSED} for all $\underline{\bm{U}}^{(i)}$.

Under conditions 1) and 2), we have
\begin{equation*}
\varphi(w,s) = \frac{\psi_p(\hat{w},s)}{p(s)}.
\end{equation*}

When operated with CSED, for a meaning $w\in\mathcal{W}$, we define the encoded message $s$ and the decoded message $\hat{w}$ under encoding and decoding schemes $\bm{U}\in\big\{ \underline{\bm{U}}^{(i)}:i=0,1,2,...,\underline{T} \big\}$ and $\bm{V}^*_q$, respectively, as
\begin{equation*}
[s|w] = \argmin_s \varphi(w,s),
\end{equation*}
\begin{equation*}
[\hat{w}|s] = \argmin_{w^\prime} \psi_q(w^\prime,s)= \argmin_{w^\prime} \varphi(w^\prime,s).
\end{equation*}
We further define the subset of messages that $\mathcal{W}$ is mapped to as
\begin{equation*}
[\mathcal{S}|\mathcal{W}] = \left\{[s|w]:w\in\mathcal{W} \right\}.
\end{equation*}

From conditions 3) and 4), we first prove that the mappings between $\mathcal{W}$ and $[\mathcal{S}|\mathcal{W}]$ and that between $[\mathcal{S}|\mathcal{W}]$ and $\mathcal{W}$ are one-to-one.

We prove this by contradiction. Suppose the mappings between $\mathcal{W}$ and $[\mathcal{S}|\mathcal{W}]$ are not one-to-one, condition 3) gives us $\left|[\mathcal{S}|\mathcal{W}] \right|>N$. On the other hand, from 4), we have $N=\left|\mathcal{W} \right|\geq [\mathcal{S}|\mathcal{W}]$. Contradiction. The mappings between $\mathcal{W}$ and $[\mathcal{S}|\mathcal{W}]$ are thus one-to-one. Likewise, it can be proven that the mappings between $[\mathcal{S}|\mathcal{W}]$ and $\mathcal{W}$ are also one-to-one.

Next, we prove
\begin{equation}\label{eq:csed1}
d\big(w,\big[\hat{w}|[s|w] \big]\big)
=\min_{w^\prime} d(w,w^\prime),~\forall w\in\mathcal{W}.
\end{equation}

First, we sort $\mathcal{W}=\big\{w_{(n)}:n=1,2,...,N \big\}$ according to $\min_{s} \varphi(w,s)$ such that $w_{(n_1)}\leq w_{(n_2)}$ implies that $\min_{s} \varphi(w_{(n_1)},s)\leq \min_{s} \varphi(w_{(n_2)},s)$. Eq. \eqref{eq:csed1} can be proved by induction.

For $w_{(1)}$, $\bm{U}$ maps it to
\begin{equation*}
[s|w_{(1)}] = \argmin_s \varphi(w,s),
\end{equation*}
and $\bm{V}^*_q$ maps $[s|w_{(1)}]$ to
\begin{equation*}
w_{(\beta)} = \argmin_{w} \varphi\left(w,[s|w_{(1)}]\right).
\end{equation*}
where $\beta\geq 1$. If $\beta>1$,
\begin{equation*}
\varphi\left(w_{(\beta)},[s|w_{(1)}]\right)\leq
\varphi\left(w_{(1)},[s|w_{(1)}]\right).
\end{equation*}

Since $\varphi\left(w_{(\beta)},[s|w_{(1)}]\right)=
\varphi\left(w_{(1)},[s|w_{(1)}]\right)$ contradicts the condition 4), we have
\begin{equation*}
\varphi\left(w_{(\beta)},[s|w_{(1)}]\right)<
\varphi\left(w_{(1)},[s|w_{(1)}]\right),
\end{equation*}
which means $\beta\leq 1$, and hence, $\beta =1$. This proves that
\begin{equation*}
w_{(1)}=\left[\hat{w}\left|\left[s|w_{(1)}\right]\right.\right].
\end{equation*}

For any $1<n<N$, we assume $w_{(n)}=\left[\hat{w}\left|\left[s|w_{(n)}\right]\right.\right]$ and prove $w_{(n+1)}=\left[\hat{w}\left|\left[s|w_{(n+1)}\right]\right.\right]$ in the following.

For $w_{(n+1)}$, let
\begin{equation*}
w_{(\beta)}= \left[\hat{w}\left|\left[s|w_{(n+1)}\right]\right.\right]=\argmin_{w} \varphi\left(w, \left[s|w_{(n+1)}\right]\right),
\end{equation*}
we prove below that $\beta=n+1$.

First, $\beta<n+1$ contradicts 4). If $\beta>n+1$,
\begin{equation*}
\varphi\left(w_{(\beta)},[s|w_{(n+1)}]\right)\leq
\varphi\left(w_{(n+1)},[s|w_{(n+1)}]\right),
\end{equation*}
and
\begin{equation*}
\min_s \varphi\left(w_{(n+1)},s\right) \leq
\min_s \varphi\left(w_{(\beta)},s \right).
\end{equation*}

Since $\min_s \varphi\left(w_{(n+1)},s \right) =\varphi\left(w_{(n+1)},\left[s|w_{(n+1)}\right] \right)$, we have
\begin{eqnarray*}
&&\hspace{-1.5cm} \min_s \varphi\left(w_{(n+1)},s \right) =
\min_s \varphi\left(w_{(\beta)},s \right) = \\
&&\hspace{-0.5cm}  \varphi\left(w_{(\beta)}, \left[s|w_{(n+1)}\right] \right)=
\varphi\left(w_{(n+1)},\left[s|w_{(n+1)}\right] \right).
\end{eqnarray*}

As a result, 
\begin{equation*}
\left[s|w_{(n+1)}\right] \in
\argmin_s \varphi\left(w_{(n+1)}, s \right) \cap
\argmin_s \varphi\left(w_{(\beta)}, s \right).
\end{equation*}
This contradicts 3).

Therefore, $\beta=n+1$ and $w_{(n+1)}=\left[\hat{w}\left|\left[s|w_{(n+1)}\right]\right.\right]$.

In conclusion,  we have $d\big(w, \left[\hat{w}\left|\left[s|w\right]\right.\right]\big) =\min_{w^\prime} d(w,w^\prime)$, $\forall w$, and hence,
\begin{eqnarray*}
&&\hspace{-0.5cm} D_{\bm{U},\bm{V}^*_q}(L)=\sum_{w,s, \hat{w}}p(w)u(s|w)v(\hat{w}|s)d(w,\hat{w}) \\
&&\hspace{-0.5cm}
=\sum_{w}p(w)d\big(w, \left[\hat{w}\left|\left[s|w\right]\right.\right]\big) =
\sum_{w}p(w) \min_{w^\prime} d(w,w^\prime) \\
&&\hspace{-0.5cm} \leq
\min_{\bm{U}^\prime,\bm{V}^\prime} D_{\bm{U}^\prime,\bm{V}^\prime}(L),
\end{eqnarray*}
for $\underline{\bm{U}}^{(i)}$, $i=0,1,2,...,\underline{T}$.

\bibliographystyle{IEEEtran}
\bibliography{References}

\end{document}